\documentclass[10pt]{article}
\usepackage{latexsym,amsmath,amscd,amssymb,graphics}
\usepackage[backend=bibtex,sorting=none]{biblatex}
\usepackage{enumerate}
\usepackage[]{mcode}
\usepackage[toc,page]{appendix}
\usepackage[a4paper,margin=.75in]{geometry}
\usepackage{array}
\usepackage[section]{placeins}
\usepackage{afterpage}
\usepackage{verbatim}

\usepackage{graphicx}
\usepackage{subfig}
\usepackage{framed}
\usepackage[colorlinks]{hyperref}
\usepackage{url}
\usepackage{mathrsfs}
\usepackage{epstopdf}

\usepackage[all]{xy}
\usepackage{indentfirst}
\usepackage[mathscr]{euscript}
\usepackage{algorithm}
\usepackage{algpseudocode}
\usepackage{authblk}

\makeatletter
\renewcommand*{\@fnsymbol}[1]{\ensuremath{\ifcase#1\or \dagger\or *\or \ddagger\or
		\mathsection\or \mathparagraph\or \|\or **\or \dagger\dagger
		\or \ddagger\ddagger \else\@ctrerr\fi}}
\makeatother

\makeatletter

\@addtoreset{figure}{section}
\def\thefigure{\thesection.\@arabic\c@figure}
\def\fps@figure{h, t}
\@addtoreset{table}{bsection}
\def\thetable{\thesection.\@arabic\c@table}
\def\fps@table{h, t}
\@addtoreset{equation}{section}

\makeatother


\newcommand{\unaryminus}{\scalebox{0.75}[1.0]{\( - \)}}

\numberwithin{theorem}{subsection}

\def\bea{\begin{eqnarray}}
\def\eea{\end{eqnarray}}
\def\ba{\begin{array}}
\def\ea{\end{array}}

\def\bomega{\boldsymbol{\omega}}
\def\bOm{\boldsymbol{\Omega}}

\def\diag{\mathrm{ \textbf{diag}}}

\def\bx{{\boldsymbol {x} }}

\let\<\langle
\let \>\rangle
\newcommand{\rem}[1]{}

\newcommand{\bu}{\boldsymbol{u}}

\newcommand{\bv}{\boldsymbol{v}}
\newcommand{\bz}{\boldsymbol{z}}

\newcommand{\bGam}{\boldsymbol{\Gamma}}
\newcommand{\bGamma}{\boldsymbol{\Gamma}}
\newcommand{\bzeta}{\boldsymbol{\zeta}}

\newcommand{\bs}{\mathbf{s}}

\newcommand{\bkappa}{\boldsymbol{\kappa}}

\newcommand{\bchi}{\boldsymbol{\chi}} 
\newcommand{\btheta}{\boldsymbol{\theta}}
\newcommand{\inertia}{\mathbb{I}}

\newcommand{\dd}[2]{\frac{\mathrm{d} #1}{\mathrm{d} #2}}

\newcommand{\fricudirbase}{\sigma}
\newcommand{\fricudir}{\boldsymbol{\fricudirbase}}

\newcommand{\dprime}{\prime \prime}

\newcommand{\todo}[1]{\vspace{5 mm}\par \noindent
\framebox{\begin{minipage}[c]{0.95 \textwidth}
\tt #1 \end{minipage}}\vspace{5 mm}\par}
\newcommand{\revision}[2]{#2} 

\newcommand{\revisionS}[2]{#2} 

\usepackage[normalem]{ulem}

\graphicspath{ {./} {./figures_gen/}  {./figures_disk/} {./figures_disk_cp/} {./figures_ball/} {./figures_ball_cp/} {./figures_ball_detach/}  }

\title{On the Normal Force and Static Friction Acting on a Rolling Ball Actuated by Internal Point Masses}
\author[1]{Vakhtang Putkaradze\thanks{Email address: \texttt{putkarad@ualberta.ca}}}
\author[2]{Stuart Rogers\thanks{Email address: \texttt{srogers@umn.edu}}}
\affil[1]{Department of Mathematical and Statistical Sciences, Faculty of Science, University of Alberta, CAB 632, Edmonton, AB T6G 2G1, Canada}
\affil[2]{Institute for Mathematics and its Applications, College of Science and Engineering, University of Minnesota, 207 Church Street SE, 306 Lind Hall, Minneapolis, MN 55455, USA}
\date{\today}

\providecommand{\keywords}[1]{\textbf{\textit{Keywords}:} #1}
\begin{document}

\maketitle

\abstract{\noindent 
The goal of this paper is to investigate the normal and tangential forces  acting  at the point of contact between a horizontal surface and a  rolling ball actuated by internal  point  masses moving in the ball's frame of reference. The normal force and   static friction are derived from the equations of motion for a rolling ball  actuated   by internal point masses that move inside the ball's frame of reference, and, as a  special case, a  rolling  disk  actuated  by internal point masses. The masses may  move along one-dimensional trajectories fixed in the ball's and disk's frame. The dynamics of a ball and disk actuated by masses moving along one-dimensional trajectories are simulated numerically and the minimum coefficients of static friction required to prevent slippage are computed.}   
\\
\\
\tableofcontents 

\section{Introduction} \label{sec_introduction}
Internally actuated rolling ball robots hold great promise for environmental data collection, surveillance, and observation, such as is required for meteorology, law enforcement, security, defense, crop management, pollution detection, planetary exploration, etc. Many actuation mechanisms and control algorithms have been proposed for locomoting these robots, such as discussed in \cite{BoKiMa2012,BoKiMa2013} (internal rotors), 
\cite{burkhardt2016reduced} (internal magnets), \cite{kilin2015spherical} (internal gyroscopic pendulum), \cite{GaBa:RNC3457} (internal pendulum and yoke), \cite{burkhardt2016reduced,das2001design,mojabi2002introducing,shen2008controllability} (internal masses moving along linear trajectories), \cite{ilin2017dynamics,Putkaradze2018dynamicsP} (internal masses moving along    more general trajectories).
When detachment from or slip at the surface occurs, the actuation mechanisms of rolling ball robots become inefficient. Moreover, from the perspective of theoretical mechanics, the consideration of the exact dynamics at the moment of  slippage is quite difficult, as discussed in \cite{FrictionForceDisscussion2009}. Some previous works have discussed these dynamics for non-actuated rolling ball robots. For example, progress has been made in the case of continuous slippage \cite{IvPi2014}, but in general, it is always desirable to prevent dynamics that cause detachment or slippage. Thus, for example, the paper \cite{VaNa2012} investigates the magnitudes of the normal and tangential forces at the contact point   for the non-actuated ball, in order to  enforce  the  no-slip  postulate  and the main assumptions of nonholonomic mechanics. In this short paper, we will show how to calculate the normal force and static friction for a rolling ball actuated by moving internal point masses, so that the assumptions of no-detachment and no-slip at the contact point made in \cite{Putkaradze2018dynamicsP} may be readily checked. The expressions for the normal and tangential forces at the contact point obtained here, we hope, will facilitate practical implementations of rolling ball robots that obey the performance envelope defined by the no-detachment and no-slip conditions.

\revision{R2Q1}{There are 3 regimes for the dynamics of a ball actuated by moving internal point masses:}
\begin{enumerate}
	\item \textbf{Rolling without Slipping} These dynamics and their associated contact point forces are the main focus of this paper and are derived in Section~\ref{sec_3d_traj}.
	\item \textbf{Detachment} These dynamics are derived in Appendix~\ref{app_detachment}. 
	\item  \revisionS{RR2Q2}{\textbf{Sliding Friction and the Painlev\'e Paradox}  This paradox corresponds to the impossibility to uniquely continue the solution past certain boundaries in phase space if dry friction at the contact point is assumed. }
	\end{enumerate}
\revisionS{RR2Q2}{ The second item deals with the detachment dynamics when the ball loses contact with and leaves the surface. In that case, the forces applied on the ball at the moment of detachment are discontinuous and the numerical solution of the problem is challenging. The third item above, the description of a system experiencing dry sliding friction at the contact point, leads to the Painlev\'e paradox. Even for simple dynamical systems such as a falling rigid rod whose end slides on the plane with friction, there is the impossibility of continuing the dynamics using both the condition of contact and the laws of sliding friction, provided that the coefficient of static friction $\mu_\mathrm{s}$ is large enough. For the implementation of the nonholonomic constraint, that coefficient is taken to be very large (or even infinite).  For discussion and resolution of these highly complex issues of contact dynamics, see \cite{wagner2013analysis,ivanova2016painleve,ivanov2008detachment,ivanov2008geometric}. Our mechanical system is substantially more complex than the one considered by Painlev\'e, and thus careful treatment of the dynamics' continuation through detachment is beyond the scope of this paper.} 

In order for a ball to roll without slipping on a horizontal surface so that the rolling constraint is in effect, the magnitude $N$ of the normal force of the surface acting at the ball's contact point must be positive  (i.e. the normal force direction must oppose gravity's direction) so that  
\begin{equation}
\label{eq_no_detachment}
N > 0.
\end{equation}
In addition, to prevent slipping, an inequality constraint  due to  dry  static friction must be satisfied:  
\begin{equation}
\label{Amontons_law}
\mu_\mathrm{s} N \ge f_\mathrm{s},
\end{equation}
where $\mu_\mathrm{s}$ is the coefficient of static friction and $f_\mathrm{s}$ is the magnitude of the static friction.  $\mu_\mathrm{s}$ is positive and depends on the material properties of the ball and surface  and possibly on other environmental factors. The  condition \eqref{Amontons_law} for dry  static friction acting at the contact point is a simplified model, following from Amontons' laws \cite{blau2008friction}. While finer aspects of the behavior of dry static friction acting at the contact point are certainly known, we shall use condition \eqref{Amontons_law}  to enforce the no-slip dynamics, as it is the  most widely used and accepted. \revision{R1Q9\\R2Q3}{In reality, the true relationship between the normal and tangential forces is substantially more complicated than \eqref{Amontons_law} and is still up for considerable debate. The transition from no-slip to slipping motion is rather complex, and we do not attempt to study it here. We refer the reader to recent papers by V.V. Kozlov \cite{Kozlov2011DryFriction,Kozlov2011PainleveFriction}, which explain the complexity of the mechanism of sliding friction and treat associated paradoxes arising from naive applications of dry friction laws.}

Therefore, any numerical simulation of the dynamics of a rolling ball, especially one actuated by an internal mechanism such as moving internal point masses, must verify that $N>0$ to ensure that the rolling constraint is indeed always in effect. Moreover, if $N>0$ for $a \le t \le b$ and if $\mu_\mathrm{s}$ is unknown, it is also useful to compute 
\begin{equation}
\hat \mu_\mathrm{s} \equiv \max_{a \le t \le b} \frac{f_\mathrm{s}}{N},
\end{equation}
which is the minimum coefficient of static friction permitted before slippage occurs. If $\mu_\mathrm{s}$ is known and if the ball has an internal actuation mechanism that may be controlled, in order to construct a control for the ball such that the ball rolls without slipping, it is necessary to include the constraints $N>0$ and $\mu_\mathrm{s} N \ge f_\mathrm{s}$ in conjunction with the no-slip dynamics. In order to enable these computations, this paper derives the normal force and static friction acting on a ball actuated by internal point masses, whose dynamics were investigated in \cite{Putkaradze2018dynamicsP}, assuming that this ball rolls without slipping. In the process, the ball's equations of motion are derived via Newton's laws, validating a previous derivation via Lagrange-d'Alembert's principle in \cite{Putkaradze2018dynamicsP}. \revision{R1Q3}{We shall note that reference \cite{balandin2013motion} derived the normal force acting on a ball actuated by a single spherical pendulum. The present work generalizes the computation of the normal force to the case when the ball is actuated by masses moving along arbitrary trajectories.}

\section{Rolling Ball with 3-d Parameterizations of the Point Mass Trajectories} \label{sec_3d_traj}
  This section pedagogically derives   the equations of motion of a rolling ball,  defining the coordinate systems, notation, and  variables,  generalizing the derivations in \cite{ilin2017dynamics}, and validating the derivation in \cite{Putkaradze2018dynamicsP}  obtained via Lagrange-d'Alembert's principle.  
  
Consider a rigid ball of radius $r$ containing some static internal structure as well as $n \in \mathbb{N}^0$ point masses  which are free to move inside the ball, where $\mathbb{N}^0$ denotes the set of nonnegative integers. This ball rolls without slipping on a horizontal  surface in the presence of a uniform gravitational field. The ball with its static internal structure has mass $m_0$ and the $i^\mathrm{th}$ point mass has mass $m_i$ for $1 \le i \le n$. Let $M = \sum_{i=0}^n m_i$ denote the mass of the total system. The total mechanical system consisting of the ball with its static internal structure and the $n$ point masses is referred to as the ball or the rolling ball, the ball with its static internal structure but without the $n$  point masses may also be referred to as $m_0$, and the $i^\mathrm{th}$  point mass may also be referred to as $m_i$ for $1 \le i \le n$. Note that the dynamics of this system are equivalent to that of the Chaplygin ball \cite{Ho2011_pII,Putkaradze2018dynamicsP}, equipped with point masses. 

\begin{figure}[h]
	\centering
	\includegraphics[width=0.5\linewidth]{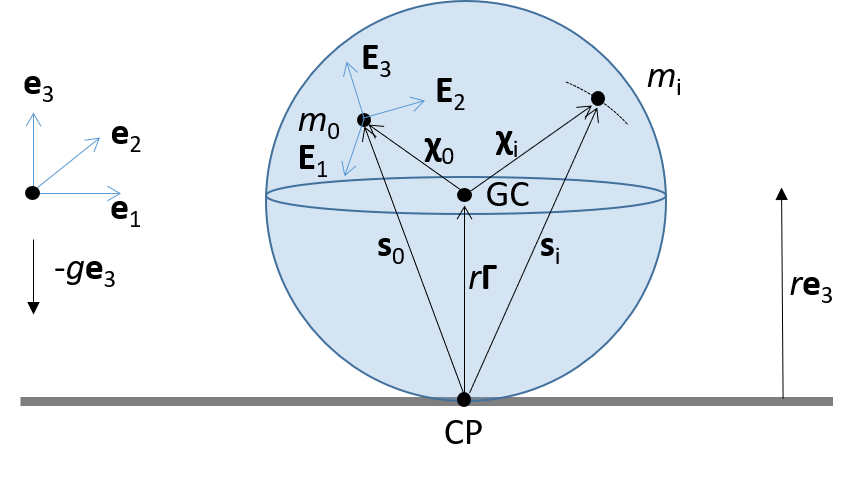}
	\caption{A ball of radius $r$ and mass $m_0$ rolls without slipping on a horizontal surface in the presence of a uniform gravitational field of magnitude $g$. The ball's geometric center, center of mass, and contact point with the horizontal surface are denoted by GC, $m_0$, and CP, respectively. The ball's motion is actuated by $n$ point masses, each of mass $m_i$, $1 \le i \le n$, that move inside the ball. The spatial frame has origin located at height $r$ above the horizontal surface and orthonormal axes $\mathbf{e}_1$, $\mathbf{e}_2$, and $\mathbf{e}_3$. The body frame has origin located at the ball's center of mass (denoted by $m_0$) and orthonormal axes $\mathbf{E}_1$, $\mathbf{E}_2$, and $\mathbf{E}_3$. All vectors inside the ball are expressed with respect to the body frame, while all vectors outside the ball are expressed with respect to the spatial frame. }
	\label{fig:detailed_rolling_ball}
\end{figure}

Two coordinate systems, or frames of reference, will be used to describe the motion of the rolling ball, an inertial spatial coordinate system and a body coordinate system in which each particle within the ball is always fixed. For brevity, the spatial coordinate system will be referred to as the spatial frame and the body coordinate system will be referred to as the body frame. These two frames are depicted in Figure~\ref{fig:detailed_rolling_ball}. The spatial frame has orthonormal axes $\mathbf{e}_1$, $\mathbf{e}_2$, $\mathbf{e}_3$, such that the $\mathbf{e}_1$-$\mathbf{e}_2$ plane is parallel to the horizontal surface and passes through the ball's geometric center (i.e. the $\mathbf{e}_1$-$\mathbf{e}_2$ plane is a height $r$ above the horizontal surface), such that $\mathbf{e}_3$ is vertical (i.e. $\mathbf{e}_3$ is perpendicular to the horizontal surface) and points ``upward" and away from the horizontal surface, and such that $\left(\mathbf{e}_1, \mathbf{e}_2, \mathbf{e}_3 \right)$ forms a right-handed coordinate system. For simplicity, the spatial frame axes are chosen to be
\begin{equation}
\mathbf{e}_1 = \begin{bmatrix} 1 & 0 & 0 \end{bmatrix}^\mathsf{T}, \quad \mathbf{e}_2 = \begin{bmatrix} 0 & 1 & 0 \end{bmatrix}^\mathsf{T}, \quad \mathrm{and} \quad \mathbf{e}_3 = \begin{bmatrix} 0 & 0 & 1 \end{bmatrix}^\mathsf{T}.
\end{equation}
The acceleration due to gravity in the uniform gravitational field is $\mathfrak{g} = -g \mathbf{e}_3  = \begin{bmatrix} 0 & 0 & -g  \end{bmatrix}^\mathsf{T}$ in the spatial frame.

The body frame's origin is chosen to coincide with the position of $m_0$'s center of mass. The body frame has orthonormal axes $\mathbf{E}_1$, $\mathbf{E}_2$, and $\mathbf{E}_3$, chosen to coincide with $m_0$'s principal axes, in which $m_0$'s inertia tensor $\inertia$ is diagonal, with corresponding principal moments of inertia $d_1$, $d_2$, and $d_3$. That is, in this body frame the inertia tensor is the diagonal matrix $\inertia = \diag \left( \begin{bmatrix} d_1 & d_2 & d_3 \end{bmatrix} \right)$.
\rem{\begin{equation}
	\inertia_0 = \begin{bmatrix} d_1 & 0 & 0 \\ 0 & d_2 & 0 \\ 0 & 0 & d_3 \end{bmatrix}.
	\end{equation}}
Moreover, $\mathbf{E}_1$, $\mathbf{E}_2$, and $\mathbf{E}_3$ are chosen so that $\left(\mathbf{E}_1, \mathbf{E}_2, \mathbf{E}_3 \right)$ forms a right-handed coordinate system. For simplicity, the body frame axes are chosen to be
\begin{equation}
\mathbf{E}_1 = \begin{bmatrix} 1 & 0 & 0 \end{bmatrix}^\mathsf{T}, \quad \mathbf{E}_2 = \begin{bmatrix} 0 & 1 & 0 \end{bmatrix}^\mathsf{T}, \quad \mathrm{and} \quad \mathbf{E}_3 = \begin{bmatrix} 0 & 0 & 1 \end{bmatrix}^\mathsf{T}.
\end{equation}

In the spatial frame, the body frame is the moving frame $\left(\Lambda \left(t\right) \mathbf{E}_1, \Lambda \left(t\right) \mathbf{E}_2, \Lambda \left(t\right) \mathbf{E}_3  \right)$, where $\Lambda \left(t\right) \in SO(3)$ defines the orientation (or attitude) of the ball at time $t$ relative to its reference configuration, for example at some initial time. 

For $0 \le i \le n$, let $\mathbf{z}_i(t)$ denote the position of $m_i$'s center of mass in the spatial frame. Let $\bchi_i(t)$ denote the body frame vector from the ball's geometric center to $m_i$'s center of mass. Then for $m_0$, $\bchi_0$ is the constant (time-independent) vector from the ball's geometric center to $m_0$'s center of mass. Note that the position of $m_i$'s center of mass in the body frame is $\bchi_i(t) -\bchi_0$ and in the spatial frame is $\mathbf{z}_i(t)=\mathbf{z}_0(t)+\Lambda(t) \left[\bchi_i(t)-\bchi_0\right]$. In general, a particle with position $\mathbf{w}(t)$ in the body frame has position $\mathbf{z}(t) = \mathbf{z}_0(t)+\Lambda(t) \mathbf{w}(t)$ in the spatial frame and has position $\mathbf{w}(t)+\bchi_0$ in the body frame translated to the ball's geometric center. In addition, suppose a time-varying external force $\mathbf{F}_\mathrm{e}(t)$ acts at the ball's geometric center.  Note that
$\mathbf{F}_\mathrm{e}(t)$ does not involve the  static  friction  induced by  the surface to enforce   the no-slip constraint. Instead, it involves forces due to other environmental factors such as  air resistance (i.e. drag) and wind force.  

\revision{R1Q4 \\ R2Q4\\R1Q6}{To obtain the dynamics of this rolling ball, it is assumed that the trajectories $\left\{\bchi_i(t)\right\}_{i=1}^n$ of the $n$ point masses are prescribed, in which case the dynamics can be obtained more efficiently by considering a single point mass of mass $M-m_0$ and whose trajectory is $\frac{1}{M-m_0}\sum_{i=1}^n m_i \bchi_i(t)$, the center of mass of the trajectories of the $n$ point masses. In subsequent work \cite{putkaradze2017optimal}, we consider the control of this rolling ball, in which case it is desirable to have $n$ degrees of freedom instead of a single degree of freedom. In the current work, the ball's motion is not controlled. }

\begin{figure}[h] 
\centering
\includegraphics[width=0.4\textwidth]{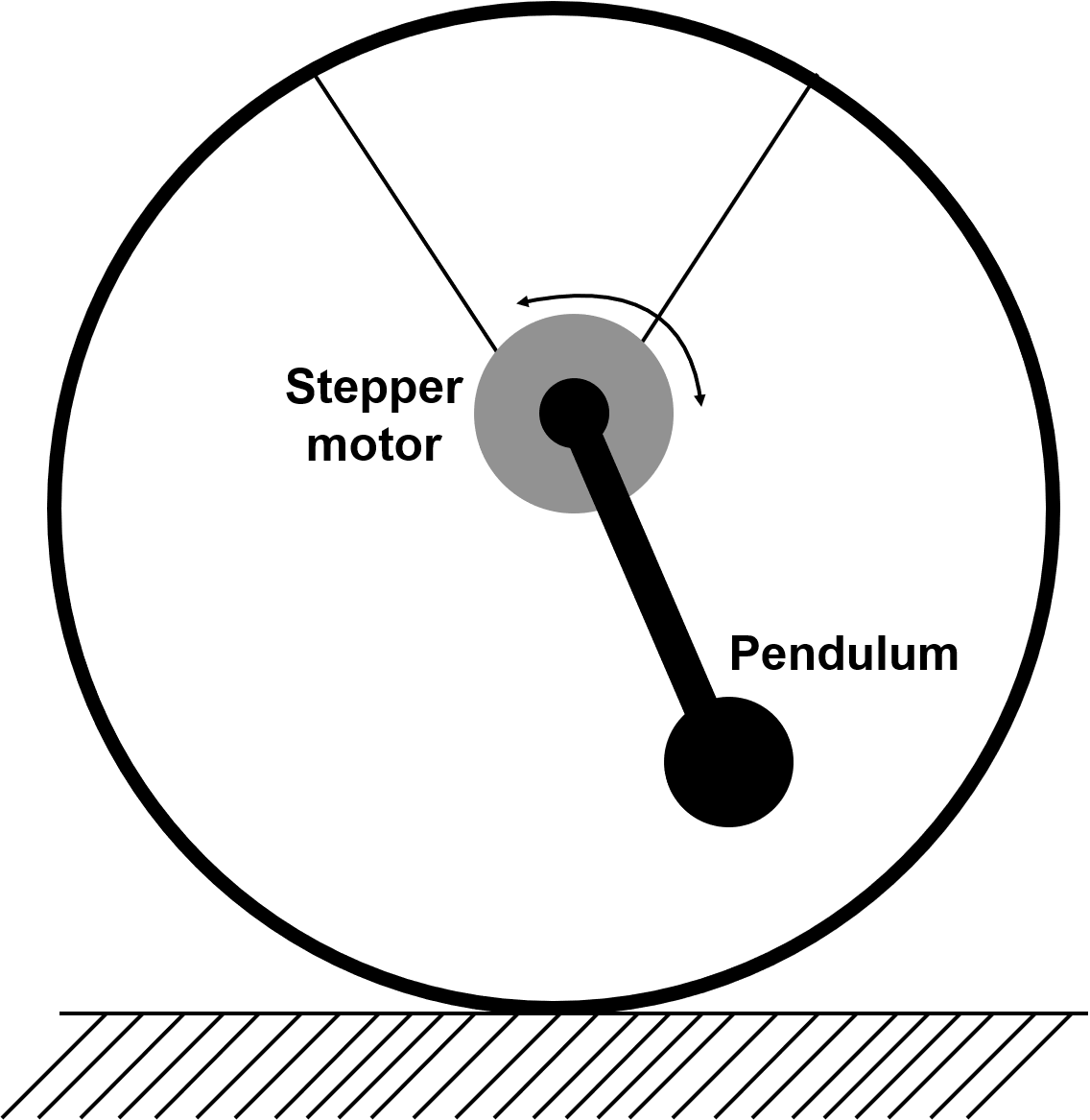}
\hspace{5mm} 
\includegraphics[width=0.4\textwidth]{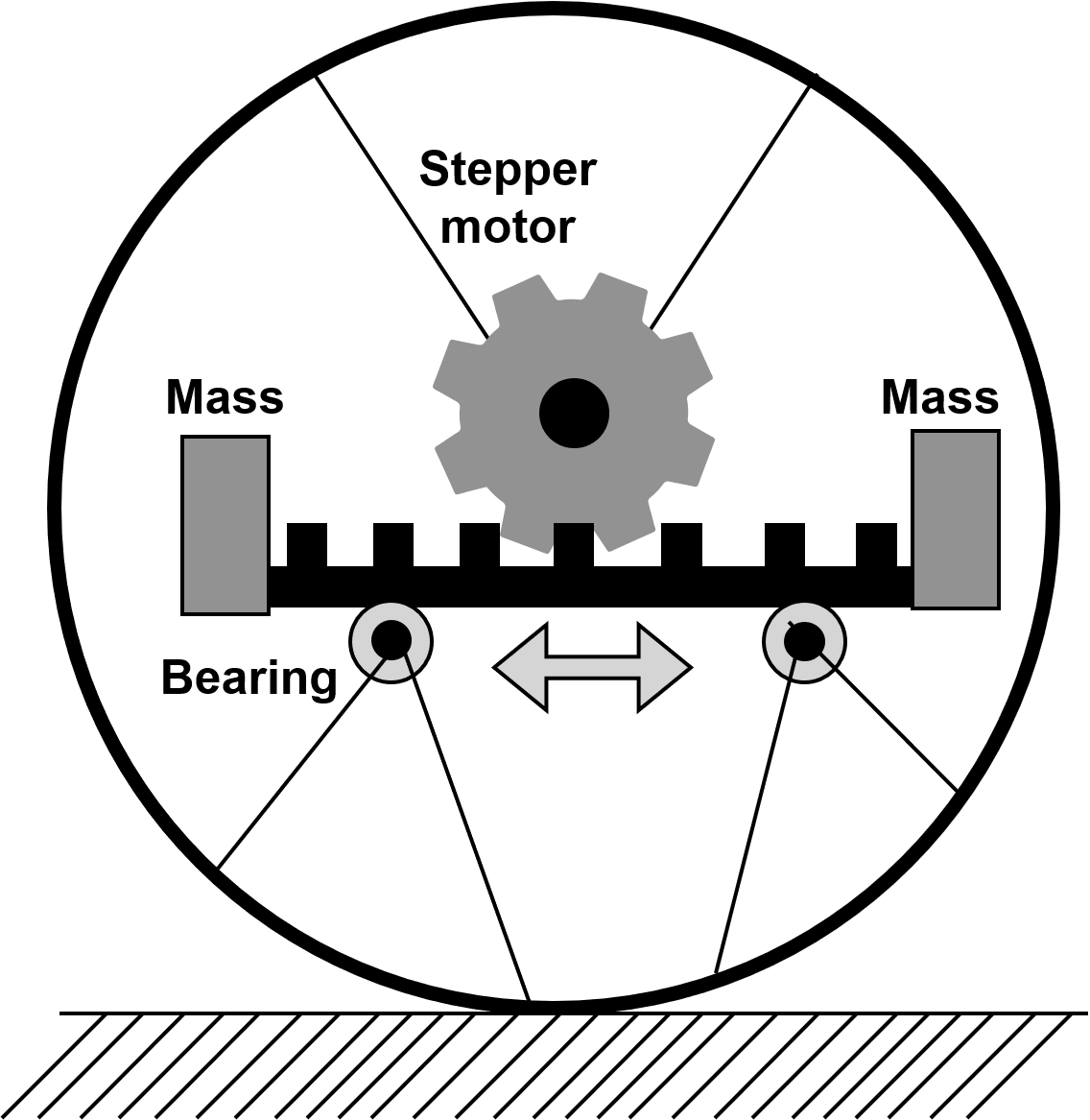} 
\caption{ 
\label{fig:stepper_motor} 
Examples of driving mechanisms actuated by a stepper motor. In both cases, the motor is rigidly attached to the internal frame of the ball, with the angle of rotation of the motor's shaft prescribed as a function of time. Left: a stepper motor swings a pendulum. Right: A stepper motor uses a sprocket to translate a rigid toothed rod with masses attached.}
\end{figure}

\revision{R1Q5}{It is also worth noting the validity of the assumption that $\bchi_i(t)$ is a prescribed function of time for $1 \le i \le n$.  } 
\rem{ %%%BEGIN REM 
\revisionS{RR1Q1\\RR1Q2}{ Let us consider a simplified system where the $i^\mathrm{th}$ mass moves parallel to a given direction $\mathbf{k}$, i.e. $\bchi_i(t)=\chi_i(t) \mathbf{k} $.Suppose also that the platform of mass $m_0$ serves as a base for the actuators, has position $X$, and moves, for simplicity, along the same direction $\mathbf{k}$.  Then, $\left\{ \chi_i(t) \right\}_{i=1}^n$ and $X(t)$ satisfy the following equations of motion involving inertia, friction, potential ($F_\mathrm{p}$), and external motor driving ($F_{\mathrm{e},i}$) forces: 
\begin{equation} 
\label{chi_i_eq} 
\begin{aligned} 
m_i \left( \ddot \chi_i + \ddot X \right) + \Gamma_i \dot \chi_i &=F_\mathrm{p}\left(\chi_i \right)+ F_{\mathrm{e},i}\left(\chi_i,t \right) \, 
\\
M \ddot X + \sum_{i=1}^n m_i \ddot \chi_i &= F_{\mathrm{e},0} .
\end{aligned}
\end{equation} 
Thus, in general, the motion $\left\{ \chi_i \right\}_{i=1}^n$ of the masses is coupled with the motion $X$ of the platform, especially if the forces or torques applied by the motor are prescribed as given functions of time \cite{BaSvYa2018}.  In our paper, however, we envision a different actuation mechanism, which specifies the positions $\left\{ \chi_i \right\}_{i=1}^n$. This can be achieved in practice, for example, by using stepper motors that, at any given time, can be set to move the mass $m_i$ close to a certain desired position $\chi_{i,0}$. We will now quantify under what physical assumptions this driving mechanism leads to the prescribed trajectories $\left\{ \chi_i(t) \right\}_{i=1}^n$. 
 }
} %%%END REM 
\revisionS{RR1Q1\\RR1Q2}{ In general, for internal masses actuated by motors with a given torque, the motion of the masses as a function of time cannot be prescribed a priori, but instead must be solved for in conjunction with the ball's motion \cite{BaSvYa2018}. 
We envision a different driving mechanism based on a stepper motor which is rigidly attached to the internal frame of the rolling ball. Unlike a regular electric motor which generates a given torque based on the input voltage/current, a stepper motor is a device which rotates the motor's shaft by a given amount measured in a discrete number of steps, where each step is typically 1-2 degrees depending on the motor's design, with a typical error of 1-2\% of the step angle.  The maximum achievable rotation speed is dependent on the motor's type and the masses involved. Modern stepper motors are capable of turning quite rapidly, at least several full revolutions per second and possibly more depending on the torques applied to the shaft. Thus, as long as the motor used is capable of supplying the torques required, we can assume that the rotation of the motor's shaft with respect to the ball can be specified within a given accuracy as a function of time, independent of the motion of the ball itself. This rotation of the shaft can then be used to drive masses along different trajectories fixed in the ball's frame. Some examples of driving mechanisms of this type are illustrated in Figure~\ref{fig:stepper_motor}. In the left panel of Figure~\ref{fig:stepper_motor}, the stepper motor swings a pendulum, so that the trajectory of the mass is a circle. In the right panel of Figure~\ref{fig:stepper_motor}, the stepper motor translates a rod with masses attached along a line. It is possible to create more complex trajectories, for example, by using a curved toothed rod instead of the straight one depicted in the right panel of Figure~\ref{fig:stepper_motor}. }

\rem{ %%%%BEGIN REM 
\revision{R1Q5}{The first assumption states that the internal dissipation in the driving mechanism is fast compared to the time scale of the mass's motion. If the typical time scale of the motion is $T_i$, then the rapid dissipation condition can be formulated  as
$\frac{T_i \Gamma_i}{m_i} \gg 1$. 
The second assumption states that given $t$, the driving mechanism resists possible deformations from the given state $\chi_{i,0}(t)$. In other words, given $\chi_{i,0}(t)$, locally, we have $F_{\mathrm{e},i}\left(\chi_i,t\right)= - P \left(\chi_{i,0}(t) - \chi_i(t) \right)$ for some rapidly changing function $P\left(\xi \right)$ where $P(0)=0$, e.g. $F_{\mathrm{e},i}\left(\chi_i,t\right)= - K \left(\chi_{i,0}(t) - \chi_i(t) \right)$ where $K$ is a very large stiffness coefficient. Given a desired accuracy $\Delta \chi_i$, the condition that $\bchi_i(t)$ is prescribed is formulated as $\left|F_\mathrm{p} \right| \ll \left|P \left(\Delta \chi_i \right) \right|$, so the forces provided by the motor and the stiffness of the attachment to the driving mechanism overcome any potential forces acting on each mass.
}
} %%%END REM 

Let us turn to the dynamical description of the ball's motion. For conciseness, the ball's geometric center is often denoted GC, $m_0$'s center of mass is often denoted CM, and the ball's contact point with the surface is often denoted CP. The GC is located at $\mathbf{z}_\mathrm{GC}(t) = \mathbf{z}_0(t)-\Lambda(t) \bchi_0$ in the spatial frame, at $-\bchi_0$ in the body frame, and at $\mathbf{0}$ in the body frame translated to the GC. The CM is located at $\mathbf{z}_0(t)$ in the spatial frame, at $\mathbf{0}$ in the body frame, and at $\bchi_0$ in the body frame translated to the GC. The CP is located at $\mathbf{z}_\mathrm{CP}(t) = \mathbf{z}_0(t)-\Lambda(t) \left[r\bGam(t)+\bchi_0 \right]$ in the spatial frame, at $-\left[r\bGam(t)+\bchi_0 \right]$ in the body frame, and at $-r\bGam(t)$ in the body frame translated to the GC, where $\bGamma(t) \equiv \Lambda^{-1}(t) \mathbf{e}_3$. Since the third spatial coordinate of the ball's GC is always $0$ and of the ball's CP is always $-r$, only the first two spatial coordinates of the ball's GC and CP, denoted by $\bz(t)$, are needed to determine the spatial location of the ball's GC and CP.

For succintness, the explicit time dependence of variables is often dropped. That is, the orientation of the ball at time $t$ is denoted simply $\Lambda$ rather than $\Lambda(t)$, the position of $m_i$'s center of mass in the spatial frame at time $t$ is denoted $\mathbf{z}_i$ rather than $\mathbf{z}_i(t)$, the position of $m_i$'s center of mass in the body frame translated to the GC at time $t$ is denoted $\bchi_i$ rather than $\bchi_i(t)$, the spatial $\mathbf{e}_1$- and $\mathbf{e}_2$-components of the ball's GC and CP at time $t$ are denoted $\bz$ rather than $\bz(t)$, and the external force is denoted $\mathbf{F}_\mathrm{e}$ rather than $\mathbf{F}_\mathrm{e}(t)$. 

\begin{figure}[h]
	\centering
	\includegraphics[width=0.5\linewidth]{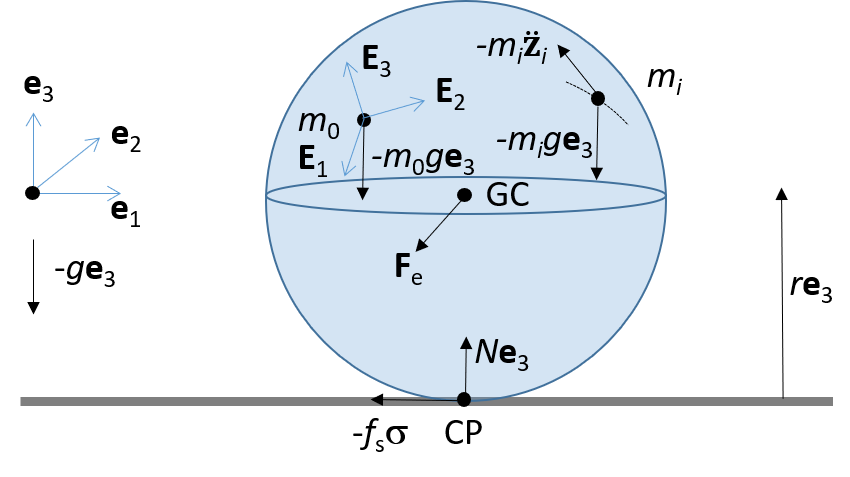}
	\caption{Free body diagram showing all the forces that act on the ball depicted in Figure~\ref{fig:detailed_rolling_ball}. }
	\label{fig:free_body_diagram_ball}
\end{figure}

Recall that $N$ denotes the magnitude of the normal force acting at the ball's CP. Assume that the ball rolls without slipping so that $N>0$ and 
\begin{equation} \label{eq_RC}
{\dot {\mathbf{z}}}_\mathrm{GC}=  \Lambda \bOm \times r \mathbf{e}_3 = \Lambda \left[\bOm \times r\bGamma \right],
\end{equation}
where $\bOm \equiv \left( \Lambda^{-1} \dot \Lambda \right)^\vee$ is the ball's body angular velocity and $\bGamma \equiv \Lambda^{-1} \mathbf{e}_3$. Recall that $f_\mathrm{s}$ denotes the magnitude of the static friction acting at the ball's CP and let $\fricudir$ denote the unit-length direction antiparallel to the static friction. Note that $\fricudir$ is parallel to the surface and therefore orthogonal to $\mathbf{e}_3$. Newton's laws for linear motion state that the time derivative of the ball's spatial linear momentum equals the sum of the forces exerted on the ball. Figure~\ref{fig:free_body_diagram_ball} illustrates the free body diagram depicting all the forces acting on the ball. Since the ball of mass $m_0$ is acted upon by gravity at the ball's CM, by an external force $\mathbf{F}_\mathrm{e}$ at the ball's GC, by a normal force $N \mathbf{e}_3$ at the ball's CP, and by a static friction $-f_\mathrm{s}\fricudir$ at the ball's CP and since each point mass $m_i$, for $1 \le i \le n$, is acted upon by gravity and has spatial acceleration ${\ddot {\mathbf{z}}}_i$, Newton's laws for linear motion give the time derivative of the ball's spatial linear momentum as
\begin{equation} \label{eq_Newton}
\dd{}{t} \left( m_0 {\dot {\mathbf{z}}}_0 \right) = \left(N-Mg\right)\mathbf{e}_3+\mathbf{F}_\mathrm{e}-f_\mathrm{s}\fricudir-\sum_{i=1}^n m_i {\ddot {\mathbf{z}}}_i.
\end{equation}
Since $m_0$ is constant, $\dd{}{t} \left( m_0 {\dot {\mathbf{z}}}_0 \right) = m_0 {\ddot {\mathbf{z}}}_0$ and \eqref{eq_Newton} simplifies to
\begin{equation} \label{eq_Newton2}
\mathbf{0} = \left(N-Mg\right)\mathbf{e}_3+\mathbf{F}_\mathrm{e}-f_\mathrm{s}\fricudir-\sum_{i=0}^n m_i {\ddot {\mathbf{z}}}_i.
\end{equation}
For $0 \le i \le n$, recall that
\begin{equation} \label{eq_z_i}
\mathbf{z}_i = \mathbf{z}_\mathrm{GC}+\Lambda \bchi_i.
\end{equation}
Differentiating \eqref{eq_z_i} with respect to time, using the rolling constraint \eqref{eq_RC}, and recalling that $\dot \Lambda = \Lambda \widehat \bOm$ (since $\bOm \equiv \left( \Lambda^{-1} \dot \Lambda \right)^\vee$) and $\mathbf{s}_i \equiv r \bGamma +\bchi_i$ yield
\begin{equation} \label{eq_dot_z_i}
{\dot {\mathbf{z}}}_i = {\dot {\mathbf{z}}}_\mathrm{GC}+\dot \Lambda \bchi_i+\Lambda {\dot \bchi}_i = \Lambda \left[\bOm \times r\bGamma \right]+\Lambda \widehat \bOm \bchi_i + \Lambda {\dot \bchi}_i = \Lambda \left[\bOm \times \mathbf{s}_i + {\dot \bchi}_i  \right].
\end{equation}
Differentiating \eqref{eq_dot_z_i} with respect to time, using $\dot \Lambda = \Lambda \widehat \bOm$, $\mathbf{s}_i \equiv r \bGamma +\bchi_i$, and ${\dot {\mathbf{s}}}_i = r \dot \bGamma +{\dot \bchi}_i$, and recalling that $\dot \bGamma=\bGamma \times \bOm$ (since $\dot \bGamma=\left(\dd{}{t} \Lambda^{-1}\right)\mathbf{e}_3=-\Lambda^{-1} \dot \Lambda \Lambda^{-1} \mathbf{e}_3=-\widehat \bOm \bGamma=-\bOm \times \bGamma=\bGamma \times \bOm$) yield
\begin{equation} \label{eq_ddot_z_i}
\begin{split}
{\ddot {\mathbf{z}}}_i &= \Lambda \left[\dot \bOm \times \mathbf{s}_i+ \bOm \times {\dot {\mathbf{s}}}_i + {\ddot \bchi}_i  \right] + \dot \Lambda \left[\bOm \times \mathbf{s}_i + {\dot \bchi}_i  \right] \\
&= \Lambda \left[\dot \bOm \times \mathbf{s}_i+ \bOm \times {\dot {\mathbf{s}}}_i + {\ddot \bchi}_i  \right] + \Lambda \widehat \bOm \left[\bOm \times \mathbf{s}_i + {\dot \bchi}_i  \right] \\
&= \Lambda \left[\dot \bOm \times \mathbf{s}_i+ \bOm \times \left(\bOm \times \mathbf{s}_i+{\dot \bchi}_i+{\dot {\mathbf{s}}}_i \right) + {\ddot \bchi}_i  \right] \\
&= \Lambda \left[\dot \bOm \times \mathbf{s}_i+ \bOm \times \left(\bOm \times \left(r \bGamma +\bchi_i\right)+{\dot \bchi}_i+r \dot \bGamma +{\dot \bchi}_i \right) + {\ddot \bchi}_i  \right] \\
&= \Lambda \left[\dot \bOm \times \mathbf{s}_i+ \bOm \times \left(\bOm \times \bchi_i+2{\dot \bchi}_i \right) + {\ddot \bchi}_i  \right].
\end{split}
\end{equation}
Substituting \eqref{eq_ddot_z_i} into \eqref{eq_Newton2} gives
\begin{equation} \label{eq_Newton3}
\mathbf{0} =  \left(N-Mg\right)\mathbf{e}_3+\mathbf{F}_\mathrm{e}-f_\mathrm{s}\fricudir- \Lambda \sum_{i=0}^n m_i \left[\dot \bOm \times \mathbf{s}_i + \bOm \times \left(\bOm \times \bchi_i +2 {\dot \bchi}_i \right) +{\ddot \bchi}_i \right].
\end{equation}
Dotting both sides of \eqref{eq_Newton3} with $\mathbf{e}_3$, recalling that  $\fricudir$ is orthogonal to $\mathbf{e}_3$, and solving for $N$ gives
\begin{equation} \label{eq_normal}
N = Mg+\left<\sum_{i=0}^n m_i \left[\dot \bOm \times \mathbf{s}_i + \bOm \times \left(\bOm \times \bchi_i +2 {\dot \bchi}_i \right) +{\ddot \bchi}_i \right],\bGamma\right>-F_{\mathrm{e},3}.
\end{equation} 
Solving \eqref{eq_Newton3} for $-f_\mathrm{s} \fricudir$ and substituting the formula for $N$ given by \eqref{eq_normal} yield 
\begin{equation} \label{eq_static_friction}
-f_\mathrm{s} \fricudir = \begin{bmatrix}
\left(\Lambda \sum_{i=0}^n m_i \left[\dot \bOm \times \mathbf{s}_i + \bOm \times \left(\bOm \times \bchi_i +2 {\dot \bchi}_i \right) +{\ddot \bchi}_i \right] - \mathbf{F}_\mathrm{e} \right)_{12}\\
0
\end{bmatrix}.
\end{equation}

Newton's laws for angular motion state that the time derivative of the ball's spatial angular momentum, computed about the ball's CM, equals the sum of the torques exerted on the ball about the ball's CM. Equating the time derivative of the ball's spatial angular momentum, computed about the ball's CM, to the sum of the torques about the ball's CM yields
\begin{equation} \label{eq_rot_Newton}
\dd{}{t} \left( \inertia_\mathrm{S} \bomega \right) = -\Lambda \mathbf{s}_0 \times \left(N \mathbf{e}_3 - f_\mathrm{s} \fricudir \right) - \Lambda \bchi_0 \times \mathbf{F}_\mathrm{e} +\sum_{i=1}^n \left(\mathbf{z}_i - \mathbf{z}_0 \right) \times m_i \left(- {\ddot {\mathbf{z}}}_i -g \mathbf{e}_3 \right),
\end{equation}
where $\inertia_\mathrm{S}$ is the ball's spatial moment of inertia and $\bomega \equiv \left[ \dot{\Lambda} \Lambda^{-1} \right]^\vee$ is the ball's spatial angular velocity. By definition, the ball's body moment of inertia is $\inertia \equiv \Lambda^{-1} \inertia_\mathrm{S} \Lambda$, so that $\inertia_\mathrm{S} = \Lambda \inertia \Lambda^{-1}$. By definition, $\bomega \equiv \left[ \dot{\Lambda} \Lambda^{-1} \right]^\vee = \Lambda \bOm $, so that $\dot \bomega = \dot \Lambda \bOm + \Lambda \dot \bOm = \Lambda \widehat \bOm \bOm + \Lambda \dot \bOm = \Lambda \left( \bOm \times \bOm \right) + \Lambda \dot \bOm = \Lambda \dot \bOm$, recalling that $\dot \Lambda = \Lambda \widehat \bOm$. Also recall that $\dot \Lambda^{-1} = - \Lambda^{-1} \dot \Lambda \Lambda^{-1}$. Using these facts, the time derivative of the ball's spatial angular momentum, computed about the ball's CM, may be simplified:
\begin{equation} \label{eq_lhs_rot}
\begin{split}
\dd{}{t} \left( \inertia_\mathrm{S} \bomega \right) &= \dot \inertia_\mathrm{S} \bomega + \inertia_\mathrm{S} \dot \bomega = \left(\dot \Lambda \inertia \Lambda^{-1} + \Lambda \inertia \dot \Lambda^{-1}   \right)\Lambda \bOm + \Lambda \inertia \Lambda^{-1} \Lambda \dot \bOm \\
&= \left(\dot \Lambda \inertia \Lambda^{-1} - \Lambda \inertia \Lambda^{-1} \dot \Lambda \Lambda^{-1}   \right)\Lambda \bOm + \Lambda \inertia \dot \bOm \\
&= \left(\Lambda \widehat \bOm \inertia - \Lambda \inertia \Lambda^{-1} \Lambda \widehat \bOm \right)\bOm + \Lambda \inertia \dot \bOm = \Lambda \left(\bOm \times \inertia \bOm + \inertia \dot \bOm \right).
\end{split}
\end{equation}
Substituting \eqref{eq_lhs_rot} into \eqref{eq_rot_Newton} yields
\begin{equation} \label{eq_rot_Newton2}
\Lambda \left(\bOm \times \inertia \bOm + \inertia \dot \bOm \right) = -\Lambda \mathbf{s}_0 \times \left(N \mathbf{e}_3 - f_\mathrm{s} \fricudir \right) - \Lambda \bchi_0 \times \mathbf{F}_\mathrm{e} -\sum_{i=1}^n \left(\mathbf{z}_i - \mathbf{z}_0 \right) \times m_i \left( {\ddot {\mathbf{z}}}_i +g \mathbf{e}_3 \right).
\end{equation}
Multiplying \eqref{eq_rot_Newton2} by $\Lambda^{-1}$, using \eqref{eq_z_i}, and recalling that $\mathbf{s}_0 \equiv r \bGamma +\bchi_0$ and $\tilde \bGamma \equiv \Lambda^{-1} \mathbf{F}_\mathrm{e}$ yield
\begin{equation} \label{eq_rot_Newton3}
\bOm \times \inertia \bOm + \inertia \dot \bOm  = -\left(r \bGamma+\bchi_0 \right) \times \Lambda^{-1} \left(N \mathbf{e}_3 - f_\mathrm{s} \fricudir \right) - \bchi_0 \times \tilde \bGamma -\sum_{i=1}^n \left(\bchi_i - \bchi_0 \right) \times m_i \left(\Lambda^{-1} {\ddot {\mathbf{z}}}_i +g \bGamma \right).   
\end{equation}
Solving \eqref{eq_Newton2} for $N \mathbf{e}_3-f_\mathrm{s}\fricudir$, which is the net force exerted by the surface on the ball at the CP, yields
\begin{equation} \label{eq_CP_force}
N \mathbf{e}_3-f_\mathrm{s}\fricudir = Mg \mathbf{e}_3-\mathbf{F}_\mathrm{e}+\sum_{i=0}^n m_i {\ddot {\mathbf{z}}}_i = \sum_{i=0}^n m_i \left( {\ddot {\mathbf{z}}}_i +g \mathbf{e}_3 \right) - \mathbf{F}_\mathrm{e}.
\end{equation}
Multiplying \eqref{eq_CP_force} by $\Lambda^{-1}$ yields
\begin{equation} \label{eq_CP_force2}
\Lambda^{-1} \left( N \mathbf{e}_3-f_\mathrm{s}\fricudir \right) = \sum_{i=0}^n m_i \left(\Lambda^{-1} {\ddot {\mathbf{z}}}_i +g \bGamma \right) - \tilde \bGamma.
\end{equation}
Substituting \eqref{eq_CP_force2} into \eqref{eq_rot_Newton3} yields 
\begin{equation} \label{eq_rot_Newton4}
\begin{split}
\bOm \times \inertia \bOm + \inertia \dot \bOm  &= -\left(r \bGamma+\bchi_0 \right) \times \left[\sum_{i=0}^n m_i \left(\Lambda^{-1} {\ddot {\mathbf{z}}}_i +g \bGamma \right)- \tilde \bGamma \right]- \bchi_0 \times \tilde \bGamma \\
&\hphantom{=}  -\sum_{i=1}^n \left(\bchi_i - \bchi_0 \right) \times m_i \left(\Lambda^{-1} {\ddot {\mathbf{z}}}_i +g \bGamma \right) \\
\rem{ %%%BEGIN REM 
&= -\left(r \bGamma+\bchi_0 \right) \times \left[\sum_{i=0}^n m_i \left(\Lambda^{-1} {\ddot {\mathbf{z}}}_i +g \bGamma \right) - \tilde \bGamma \right] - \bchi_0 \times \tilde \bGamma \\
&\hphantom{=} -\sum_{i=0}^n \left(\bchi_i - \bchi_0 \right) \times m_i \left(\Lambda^{-1} {\ddot {\mathbf{z}}}_i +g \bGamma \right) \\
}%%%END REM 
&=  - r \bGamma \times \left[\sum_{i=0}^n m_i \left(\Lambda^{-1} {\ddot {\mathbf{z}}}_i +g \bGamma \right) - \tilde \bGamma \right] -\sum_{i=0}^n \bchi_i \times m_i \left(\Lambda^{-1} {\ddot {\mathbf{z}}}_i +g \bGamma \right) \\
&= - \sum_{i=0}^n m_i \bs_i \times \left(\Lambda^{-1} {\ddot {\mathbf{z}}}_i +g \bGamma \right) + r \bGamma \times \tilde \bGamma.
\end{split} 
\end{equation}
Substituting \eqref{eq_ddot_z_i} into \eqref{eq_rot_Newton4} yields
\begin{equation} \label{eq_rot_Newton5}
\begin{split}
\bOm \times \inertia \bOm + \inertia \dot \bOm  &=
- \sum_{i=0}^n m_i \bs_i \times \left\{ \dot \bOm \times \mathbf{s}_i+ \bOm \times \left(\bOm \times \bchi_i+2{\dot \bchi}_i \right) + {\ddot \bchi}_i +g \bGamma \right\} + r \bGamma \times \tilde \bGamma.
\end{split} 
\end{equation}
Solving \eqref{eq_rot_Newton5} for $\dot \bOm$ yields
\begin{equation} \label{eq_Om}
\dot \bOm = \left[\sum_{i=0}^n m_i \widehat{\mathbf{s}_i}^2  -\inertia \right]^{-1}  \left[\bOm \times \inertia \bOm+r \tilde \bGamma \times \bGamma+ \sum_{i=0}^n m_i \mathbf{s}_i \times  \left\{ g \bGamma+ \bOm \times \left(\bOm \times \bchi_i +2 {\dot \bchi}_i \right) +{\ddot \bchi}_i \right\}  \right],
\end{equation}
which agrees with the result obtained via Lagrange-d'Alembert's principle in \cite{Putkaradze2018dynamicsP}. 

\paragraph{Rolling Ball with Static Internal Structure} \revision{R1Q8}{ By setting the number of point masses $n$ to 0, the dynamics and contact point forces for the rolling ball with static internal structure are readily obtained.} Letting $n=0$, \eqref{eq_Om} simplifies to \cite{Putkaradze2018dynamicsP}
\begin{equation} \label{eq_dot_Omega_static}
\dot \bOm = \left[ m_0 \widehat{\mathbf{s}_0}^2  -\inertia \right]^{-1} \left[\bOm \times \inertia \bOm+r \tilde \bGamma \times \bGamma+  m_0 \mathbf{s}_0 \times  \left\{ g \bGamma+ \bOm \times \left(\bOm \times \bchi_0 \right) \right\}  \right],
\end{equation}
\eqref{eq_normal} simplifies to
\begin{equation} \label{eq_normal_static}
N = m_0 \left(g+\left< \dot \bOm \times \mathbf{s}_0 + \bOm \times \left(\bOm \times \bchi_0 \right),\bGamma\right> \right)-F_{\mathrm{e},3},
\end{equation}
and \eqref{eq_static_friction} simplifies to
\begin{equation} \label{eq_static_friction_static}
-f_\mathrm{s} \fricudir = \begin{bmatrix}
\left(m_0 \Lambda \left[\dot \bOm \times \mathbf{s}_0 + \bOm \times \left(\bOm \times \bchi_0 \right) \right] - \mathbf{F}_\mathrm{e} \right)_{12}\\
0
\end{bmatrix}.
\end{equation}
\revisionS{RR2Q1}{For the case of the Routh sphere, i.e. a ball such that the line joining the ball's CM and GC forms an axis of mass distribution symmetry, \eqref{eq_dot_Omega_static} may be integrated by quadratures \cite{chaplygin2002motion} so that \eqref{eq_normal_static} and \eqref{eq_static_friction_static} may be analyzed analytically \cite{rozenblat2007separation}.}

\section{Rolling Ball with 1-d Parameterizations of the Point Mass Trajectories} \label{sec_1d_traj}

\begin{figure}[h]
	\centering
	\includegraphics[width=0.5\linewidth]{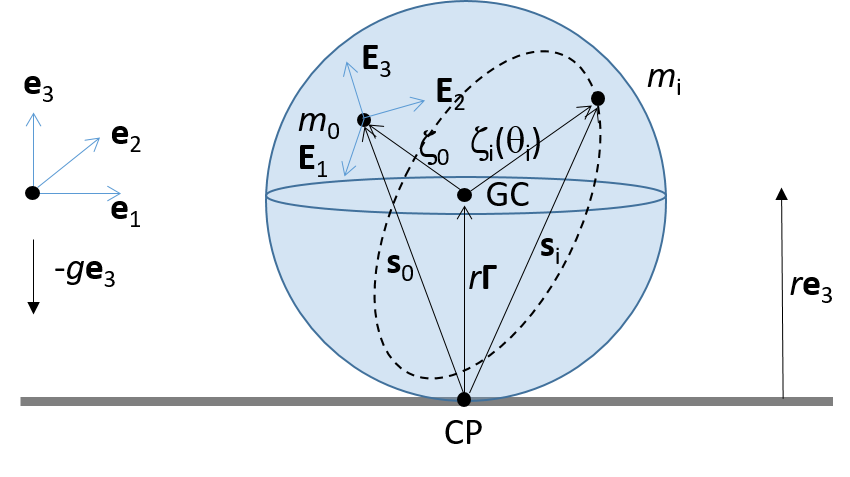}
	\caption{Each  point mass, denoted by $m_i$, $1 \le i \le n$, moves along a rail fixed inside the ball depicted here by the dashed hoop. The trajectory of the rail is denoted by $\bzeta_i$ and is parameterized by $\theta_i$. }
	\label{fig:detailed_1dparam_rolling_ball}
\end{figure}

For $1 \le i \le n$, assume now that the trajectory  $\bchi_i$ of the $i^\mathrm{th}$  point mass is required to move along a 1-d rail, like a circular hoop.
Moreover, for $1 \le i \le n$, assume that the $i^\mathrm{th}$  rail is parameterized by a 1-d parameter $\theta_i$, so that the trajectory $\bzeta_i$ of the $i^\mathrm{th}$  rail, in the body frame translated to the ball's geometric center, as a function of $\theta_i$ is $\bzeta_i(\theta_i)$. Thus, the trajectory of the $i^\mathrm{th}$   point mass as a function of time $t$ is $\bchi_i(t) \equiv \bzeta_i(\theta_i (t))$, $1 \le i \le n$. Refer to Figure~\ref{fig:detailed_1dparam_rolling_ball} for an illustration. To make notation consistent, define $\bzeta_0(\theta_0) \equiv \bchi_0$, so that the constant (time-independent) vector  $\bchi_0 = \bchi_0(t) \equiv \bzeta_0(\theta_0 (t))$ for any scalar-valued, time-varying function $\theta_0(t)$. By the chain rule and using the notation \textvisiblespace$^\cdot$ to denote differentiation with respect to time $t$ and $\bzeta_i^\prime$ to denote differentiation of $\bzeta_i$ with respect to $\theta_i$, for $0 \le i \le n$,
\begin{equation} \label{eq_1d_param}
\begin{split}
\bchi_i(t) &\equiv \bzeta_i(\theta_i (t)) = \bzeta_i, \\
\dot \bchi_i(t) &= \dd{\bzeta_i }{\theta_i}(\theta_i (t)) \dot \theta_i(t) = \bzeta_i^{\prime} (\theta_i (t))  \dot \theta_i(t) =  \bzeta_i^{\prime} \dot \theta_i  =  \dot \theta_i \bzeta_i^{\prime}, \\
\ddot \bchi_i(t) &=\frac{d^2 \bzeta_i }{d \theta_i^2}(\theta_i (t)) \dot \theta_i^2(t) + \dd{\bzeta_i }{\theta_i}(\theta_i (t)) \ddot \theta_i(t) \\
& =\bzeta_i^{\dprime}(\theta_i (t)) \dot \theta_i^2(t) + \bzeta_i^{\prime} (\theta_i (t))  \ddot \theta_i(t)=\bzeta_i^{\dprime} \dot \theta_i^2 + \bzeta_i^{\prime} \ddot \theta_i = \dot \theta_i^2 \bzeta_i^{\dprime} + \ddot \theta_i \bzeta_i^{\prime}. 
\end{split}
\end{equation}
With this new notation, $\mathbf{s}_i \equiv r \bGamma +\bchi_i = r \bGamma +\bzeta_i$ for $0\le i\le n$.

Plugging the formulas for $\bchi_i$, $\dot \bchi_i$, and $\ddot \bchi_i$ given in \eqref{eq_1d_param} into the relevant formulas in Section~\ref{sec_3d_traj} yields the equations of motion, normal force, and static friction for a rolling ball with 1-d parameterizations of the point mass trajectories. \eqref{eq_Om} becomes \cite{Putkaradze2018dynamicsP}
\begin{equation} \label{eq_Om_1d}
\begin{split} 
\dot \bOm &= \left[\sum_{i=0}^n m_i \widehat{\mathbf{s}_i}^2  -\inertia \right]^{-1}  \Bigg[\bOm \times \inertia \bOm+r \tilde \bGamma \times \bGamma 
\\ &\hphantom{=\left[\sum_{i=0}^n m_i \widehat{\mathbf{s}_i}^2  -\inertia \right]^{-1}  \Bigg[} + \sum_{i=0}^n m_i \mathbf{s}_i \times  \left\{ g \bGamma+ \bOm \times \left(\bOm \times \bzeta_i +2 \dot \theta_i \bzeta_i^{\prime} \right) + \dot \theta_i^2 \bzeta_i^{\dprime} + \ddot \theta_i  \bzeta_i^{\prime} \right\}  \Bigg],
\end{split}
\end{equation}
\eqref{eq_normal} becomes
\begin{equation} \label{eq_normal_1d}
N = Mg+\left<\sum_{i=0}^n m_i \left[\dot \bOm \times \mathbf{s}_i + \bOm \times \left(\bOm \times \bzeta_i +2 \dot \theta_i \bzeta_i^{\prime} \right) + \dot \theta_i^2 \bzeta_i^{\dprime} + \ddot \theta_i \bzeta_i^{\prime} \right],\bGamma\right>-F_{\mathrm{e},3},
\end{equation}
and \eqref{eq_static_friction} becomes
\begin{equation} \label{eq_static_friction_1d}
-f_\mathrm{s} \fricudir = \begin{bmatrix}
\left(\Lambda \sum_{i=0}^n m_i \left[\dot \bOm \times \mathbf{s}_i +  \bOm \times \left(\bOm \times \bzeta_i +2 \dot \theta_i \bzeta_i^{\prime} \right) + \dot \theta_i^2 \bzeta_i^{\dprime} + \ddot \theta_i \bzeta_i^{\prime} \right] - \mathbf{F}_\mathrm{e} \right)_{12}\\
0
\end{bmatrix}.
\end{equation}
Equations \eqref{eq_Om_1d}, \eqref{eq_normal_1d}, and \eqref{eq_static_friction_1d}  and their subsequent analysis  constitute  the main focus of this paper. 
\section{Rolling Disk with 1-d Parameterizations of the Point Mass Trajectories} \label{sec_disk}
\begin{figure}[h]
	\centering
	\includegraphics[width=0.6\linewidth]{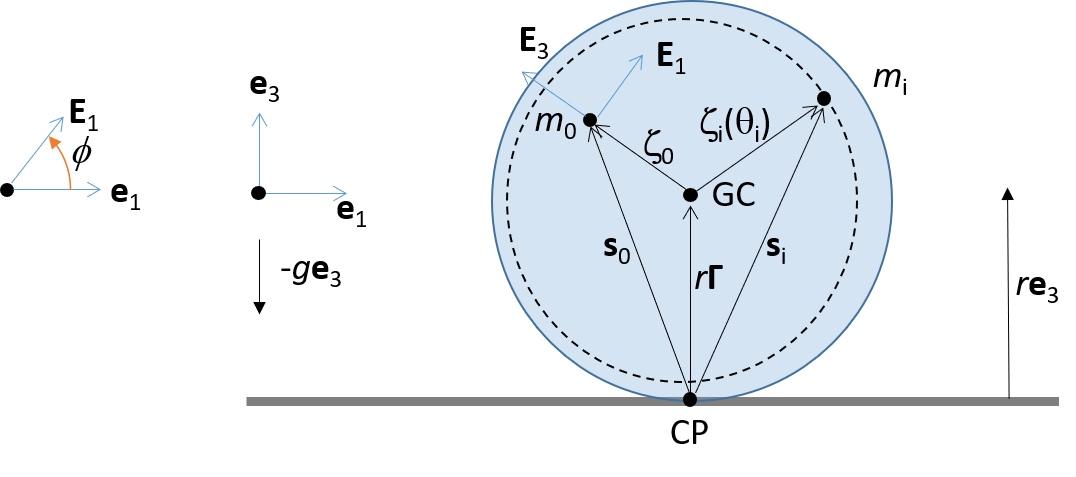}
	\caption{A disk of radius $r$ and mass $m_0$ rolls without slipping in the $\mathbf{e}_1$-$\mathbf{e}_3$ plane. $\mathbf{e}_2$ and $\mathbf{E}_2$ are directed into the page and are omitted from the figure. The disk's center of mass is denoted by $m_0$. The disk's motion is actuated by $n$ point masses, each of mass $m_i$, $1 \le i \le n$, that move along rails fixed inside the disk. The point mass depicted here by $m_i$ moves along a circular hoop in the disk that is not centered on the disk's geometric center (GC). The disk's orientation is determined by $\phi$, the angle measured counterclockwise from $\mathbf{e}_1$ to $\mathbf{E}_1$.  }
	\label{fig:detailed_rolling_disk}
\end{figure}

Let us now  consider the special case when the motion of the  rolling ball is purely planar, which  is the case of a rolling disk. In order to perform this two-dimensional reduction,  suppose that the ball's inertia is such that one of the ball's principal axes, say the one labeled $\mathbf{E}_2$, is orthogonal to the plane containing the GC and CM. Also assume that all the  point masses move along 1-d  rails which lie in the plane containing the GC and CM. Moreover, suppose that the ball is oriented initially so that the plane containing the GC and CM coincides with the $\mathbf{e}_1$-$\mathbf{e}_3$ plane and that the external force $\mathbf{F}_\mathrm{e}$ acts in the $\mathbf{e}_1$-$\mathbf{e}_3$ plane. Then for all time, the ball will remain oriented so that the plane containing the GC and CM coincides with the $\mathbf{e}_1$-$\mathbf{e}_3$ plane and the ball will only move in the $\mathbf{e}_1$-$\mathbf{e}_3$ plane, with the ball's rotation axis always parallel to $\mathbf{e}_2$. Note that the dynamics of this system are equivalent to that of the Chaplygin disk \cite{Ho2011_pII}, equipped with point masses, rolling in the $\mathbf{e}_1$-$\mathbf{e}_3$ plane, and where the Chaplygin disk (minus the  point masses) has polar moment of inertia $d_2$. This particular ball with this special inertia, orientation,  and placement of the rails and point masses, may be referred to as the disk or the rolling disk. Figure~\ref{fig:detailed_rolling_disk} depicts the rolling disk.

Let $\phi$ denote the angle between $\mathbf{e}_1$ and $\mathbf{E}_1$, measured counterclockwise from $\mathbf{e}_1$ to $\mathbf{E}_1$. Thus, if $\dot \phi > 0$, the disk rolls in the $-\mathbf{e}_1$ direction and $\bOm$ has the same direction as $-\mathbf{e}_2$, and  if $\dot \phi < 0$, the disk rolls in the $\mathbf{e}_1$ direction and $\bOm$ has the same direction as $\mathbf{e}_2$.

For the rolling disk with 1-d parameterizations of the point mass trajectories, \eqref{eq_Om_1d} becomes \cite{Putkaradze2018dynamicsP}
\begin{equation} \label{eq_ddphi_disk}
\ddot \phi = \frac{ \sum_{i=0}^n m_i K_i -r F_{\mathrm{e},1}}{d_2+\sum_{i=0}^n m_i \left[\left( r \sin \phi + \zeta_{i,1} \right)^2+\left( r \cos \phi+ \zeta_{i,3} \right)^2 \right]},
\end{equation}
where 
\begin{equation} \label{eq_K_i}
\begin{split}
K_i &\equiv \left(g+ r {\dot \phi}^2 \right) \left(\zeta_{i,3} \sin \phi - \zeta_{i,1} \cos \phi  \right)+
\left(r \cos \phi + \zeta_{i,3} \right) \left(- 2 \dot \phi {\dot \theta}_i \zeta_{i,3}^{\prime} + {\dot \theta}_i^2 \zeta_{i,1}^{\dprime} + {\ddot \theta}_i \zeta_{i,1}^{\prime} \right)\\
&\hphantom{\equiv}   - \left(r \sin \phi + \zeta_{i,1} \right) \left( 2 \dot \phi {\dot \theta}_i \zeta_{i,1}^{\prime}+ {\dot \theta}_i^2 \zeta_{i,3}^{\dprime} + {\ddot \theta}_i \zeta_{i,3}^{\prime} \right), 
\end{split}
\end{equation}
\eqref{eq_normal_1d} becomes
\begin{equation} \label{eq_normal_disk}
\begin{split}
N&=Mg+\sum_{i=0}^n m_i \Big[
 \left(-\ddot \phi \zeta_{i,3} - {\dot \phi}^2 \zeta_{i,1} - 2 \dot \phi {\dot \theta}_i \zeta_{i,3}^{\prime} + {\dot \theta}_i^2 \zeta_{i,1}^{\dprime} + {\ddot \theta}_i \zeta_{i,1}^{\prime} \right) \sin \phi \\
&\hphantom{=Mg+\sum_{i=0}^n m_i \Big[}   + \left(\ddot \phi \zeta_{i,1} - {\dot \phi}^2 \zeta_{i,3}+ 2 \dot \phi {\dot \theta}_i \zeta_{i,1}^{\prime}+ {\dot \theta}_i^2 \zeta_{i,3}^{\dprime} + {\ddot \theta}_i \zeta_{i,3}^{\prime} \right) \cos \phi \Big]-F_{\mathrm{e},3},
\end{split}
\end{equation}
and \eqref{eq_static_friction_1d} becomes
\begin{equation} \label{eq_static_friction_disk}
\begin{split}
-f_\mathrm{s} \fricudir &= -\Big\{ Mr \ddot \phi + \sum_{i=0}^n m_i \Big[\left( \ddot \phi \zeta_{i,3} + {\dot \phi}^2 \zeta_{i,1} + 2 \dot \phi {\dot \theta}_i \zeta_{i,3}^{\prime} - {\dot \theta}_i^2 \zeta_{i,1}^{\dprime} - {\ddot \theta}_i \zeta_{i,1}^{\prime} \right) \cos \phi \\
&\hphantom{= -\Big\{ Mr \ddot \phi + \sum_{i=0}^n m_i \Big[} +\left(\ddot \phi \zeta_{i,1} - {\dot \phi}^2 \zeta_{i,3}+ 2 \dot \phi {\dot \theta}_i \zeta_{i,1}^{\prime}+ {\dot \theta}_i^2 \zeta_{i,3}^{\dprime} + {\ddot \theta}_i \zeta_{i,3}^{\prime} \right) \sin \phi \Big] + F_{\mathrm{e},1} \Big\} \mathbf{e}_1.
\end{split}
\end{equation}
Appendix~\ref{app_disk} provides detailed calculations justifying how \eqref{eq_normal_disk} and \eqref{eq_static_friction_disk} follow from \eqref{eq_normal_1d} and \eqref{eq_static_friction_1d}, respectively. The $\mathbf{e}_1$-component of $-f_\mathrm{s} \fricudir$ is denoted by $\pi f_\mathrm{s}$.

\paragraph{Rolling Disk with Static Internal Structure} \revision{R1Q8}{ By setting the number of point masses $n$ to 0, the dynamics and contact point forces for the rolling disk with static internal structure are readily obtained.} Letting $n=0$, \eqref{eq_ddphi_disk} simplifies to 
\begin{equation} \label{eq_ddphi_sdisk}
\ddot \phi = \frac{ m_0 \left(g+ r {\dot \phi}^2 \right) \left(\zeta_{0,3} \sin \phi - \zeta_{0,1} \cos \phi  \right)-r F_{\mathrm{e},1}}{d_2+ m_0 \left[\left( r \sin \phi + \zeta_{0,1} \right)^2+\left( r \cos \phi+ \zeta_{0,3} \right)^2 \right]},
\end{equation}
\eqref{eq_normal_disk} simplifies to
\begin{equation} \label{eq_normal_sdisk}
N= m_0 \left[g-
\left(\ddot \phi \zeta_{0,3} + {\dot \phi}^2 \zeta_{0,1}  \right) \sin \phi   + \left(\ddot \phi \zeta_{0,1} - {\dot \phi}^2 \zeta_{0,3} \right) \cos \phi \right]-F_{\mathrm{e},3},
\end{equation}
and \eqref{eq_static_friction_disk} simplifies to
\begin{equation} \label{eq_static_friction_sdisk}
-f_\mathrm{s} \fricudir = -\left\{ m_0 \left[r \ddot \phi +
\left(\ddot \phi \zeta_{0,3} + {\dot \phi}^2 \zeta_{0,1}  \right) \cos \phi   + \left(\ddot \phi \zeta_{0,1} - {\dot \phi}^2 \zeta_{0,3} \right) \sin \phi \right]+F_{\mathrm{e},1} \right\} \mathbf{e}_1.
\end{equation}

\section{Numerical Simulations of the Dynamics of the Rolling Disk}
\label{sec_disk_numerical}
To write the equations of motion for the rolling disk in the standard ODE form, the state of the system is defined as 
\begin{equation}
\bx \equiv \begin{bmatrix} \btheta \\ \dot \btheta \\ \phi \\ \dot \phi  \end{bmatrix},
\end{equation}
where $\btheta, \dot \btheta \in \mathbb{R}^n$ and $\phi, \dot \phi  \in \mathbb{R}$.
The ODE formulation of the rolling disk's system dynamics defined for $a \le t \le b$ is
\begin{equation} \label{eq_rdisk_dynamics}
\dot {\bx} = \begin{bmatrix} \dot \btheta \\ \ddot \btheta \\ \dot \phi \\ \ddot \phi  \end{bmatrix}  = \mathbf{f}\left(t,\bx,\bu\right) \equiv \begin{bmatrix} \dot \btheta \\ \bu  \\ \dot \phi \\ \kappa\left(t,\bx,\bu\right)  \end{bmatrix},
\end{equation}
where $\bu \colon \mathbb{R} \to \mathbb{R}^n$ is a prescribed function of $t$ such that $\bu(t)=\ddot \btheta(t) \in \mathbb{R}^n$ and $\kappa\left(t,\bx,\bu \right)$ is given by the right-hand side of the formula for $\ddot \phi$ in \eqref{eq_ddphi_disk}. In order to simulate the rolling disk's dynamics, \eqref{eq_rdisk_dynamics} must be integrated with  prescribed initial conditions at time $t=a$:
\begin{equation} \label{eq_disk_initial_conds}
\bx\left(a\right) = \begin{bmatrix} \btheta(a) \\ \dot \btheta(a)  \\ \phi(a) \\ \dot \phi(a)  \end{bmatrix} = \begin{bmatrix} \btheta_a \\ {\dot \btheta}_a \\ \phi_a \\  \unaryminus \frac{{\dot z}_a}{r}  \end{bmatrix} \equiv \bx_a.
\end{equation}
\eqref{eq_rdisk_dynamics} and \eqref{eq_disk_initial_conds} constitute an ODE IVP.
For   the ODE systems considered here, one can  choose $a=0$ without loss of generality;  however, we shall let $a$ be  arbitrary to keep our discussion general and consistent with the notation used in the literature on the numerical solution of boundary value problems \cite{ascher1994numerical}. Given $\phi$, the spatial $\mathbf{e}_1$-component $z$ of the disk's GC and CP is $z = z_a-r \left(\phi-\phi_a \right)$, where $z_a$ is the spatial $\mathbf{e}_1$-component of the disk's GC and CP at time $t=a$ and $\phi_a$ is the disk's angle at time $t=a$.

\begin{figure}[h] 
	\centering
	\includegraphics[scale=.7]{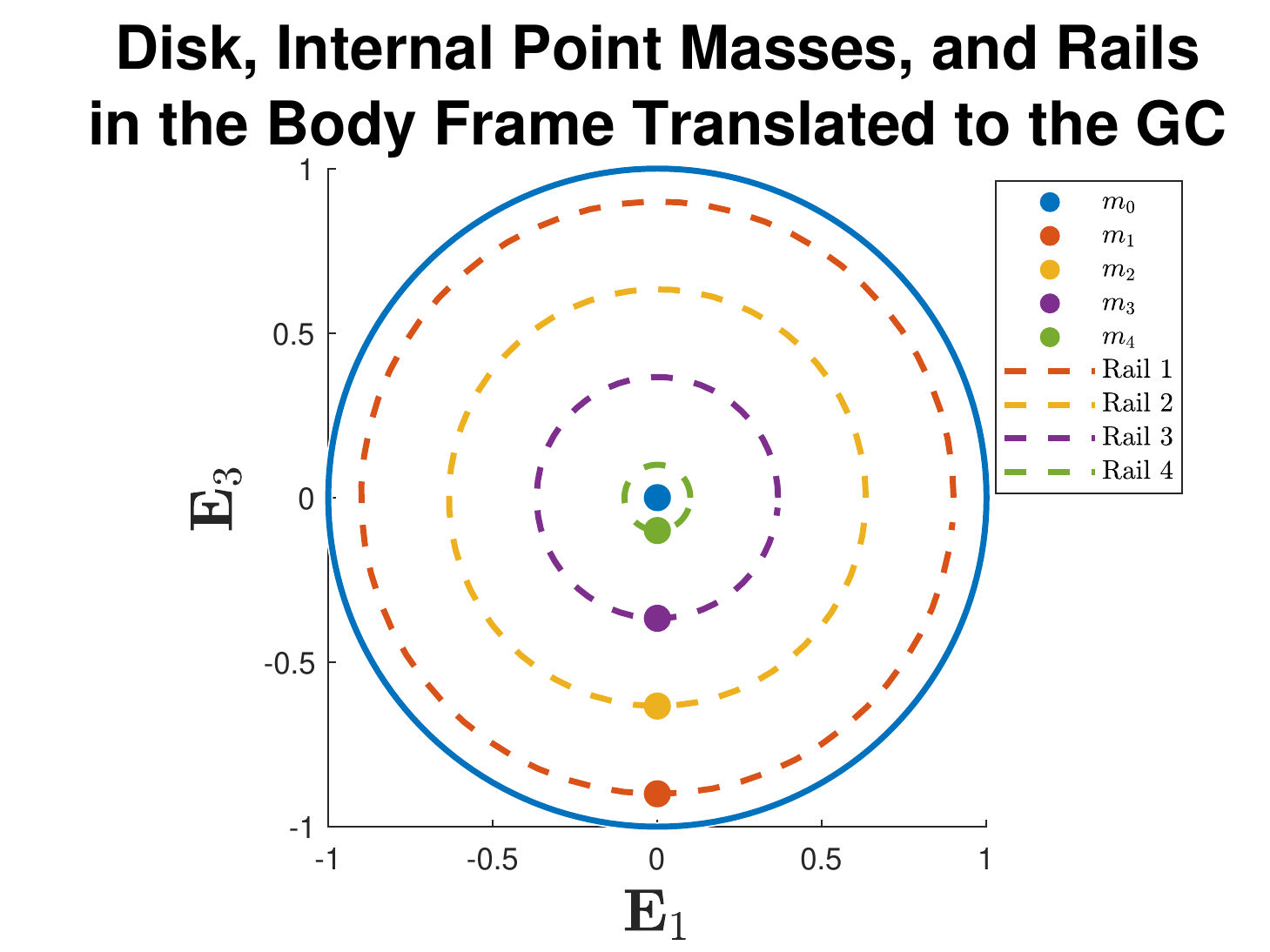}
	\caption{A disk of radius $r=1$ actuated by $4$ internal point masses, $m_1$, $m_2$, $m_3$, and $m_4$, each on its own circular rail of radius $r_1=.9$, $r_2=.6\overline{3}$, $r_3=.3\overline{6}$, and $r_4=.1$, respectively. The location of the disk's CM coincides with the GC and is denoted by $m_0$. $m_0=m_1=m_2=m_3=m_4=1$ and $g=1$. The configuration at the initial time $t=0$ is shown.}
	\label{fig_disk_masses_rails_bf_gc}
\end{figure}

Consider a rolling disk of mass $m_0=1$, radius $r=1$, polar moment of inertia $d_2=1$, and with the CM coinciding with the GC (i.e. $\bzeta_0=\mathbf{0}$). The disk contains $n=4$ internal point masses, each of mass $1$ so that $m_1=m_2=m_3=m_4=1$ and each located on its own concentric circle centered on the GC of radius $r_1=.9$, $r_2=.6\overline{3}$, $r_3=.3\overline{6}$, and $r_4=.1$, respectively, as shown in Figure~\ref{fig_disk_masses_rails_bf_gc}. For $1 \le i \le n$, the position of $m_i$ in the body frame centered on the GC is:
\begin{equation}
\bzeta_i\left(\theta_i\right) = r_i \begin{bmatrix} \cos \theta_i \\ 0 \\ \sin \theta_i  \end{bmatrix}.
\end{equation}
The disk's total system mass is $M=5$, and gravity is rescaled to be $g=1$. There is no external force acting on the disk's GC so that $F_{\mathrm{e},1} = 0$ in the right-hand side of \eqref{eq_rdisk_dynamics}. This disk's dynamics are simulated with initial time $a=0$ and final time $b=20$, so that the simulation time interval is $\left[0,20\right]$. The parameterized acceleration of each internal point mass is a continuous approximation of a short duration unit amplitude step function:
\rem{\todo{VP: Are you trying to say that the acceleration is a linear function on an interval and zero otherwise? \\ SMR: The acceleration is constant at 1 for a short time, then decreases linearly from 1 to 0 for a short time, and then stays constant at 0 for the rest of time. $u_i = \ddot \theta_i$ is constructed to be continuous (instead of discontinuous) so that $\dot \btheta$ and $\btheta$ are differentiable, by the Fundamental Theorem of Calculus. The derivation of the equations of motion assumed that $\btheta$ and $\dot  \btheta$ are differentiable. Should I add in a plot of $u_i$? I didn't make the plot in the interest of keeping the paper short. The coefficient $\left(\unaryminus 1\right)^i$ in $u_i$ makes the simulation a little more interesting so that odd masses rotate clockwise and even masses rotate counterclockwise in the disk's body frame. \\ 
		VP: Yes, please put a plot of $u_i(t)$ to explain. \\ SMR: I added in a plot of $|u_i(t)|$.  }} 
\begin{equation} \label{eq_prescribed_u_disk}
u_i(t) = \ddot \theta_i(t) = \left(\unaryminus 1\right)^i \left\{ \begin{array}{ll} 1, & 0 \le t \le .1, \\ \unaryminus 10t+2, & .1 \le t \le .2, \\ 0, & .2 \le t \le 20, \end{array} \right. \quad \mbox{for} \quad 1 \le i \le n.
\end{equation}
\begin{figure}[h] 
	\centering
	\includegraphics[scale=.5]{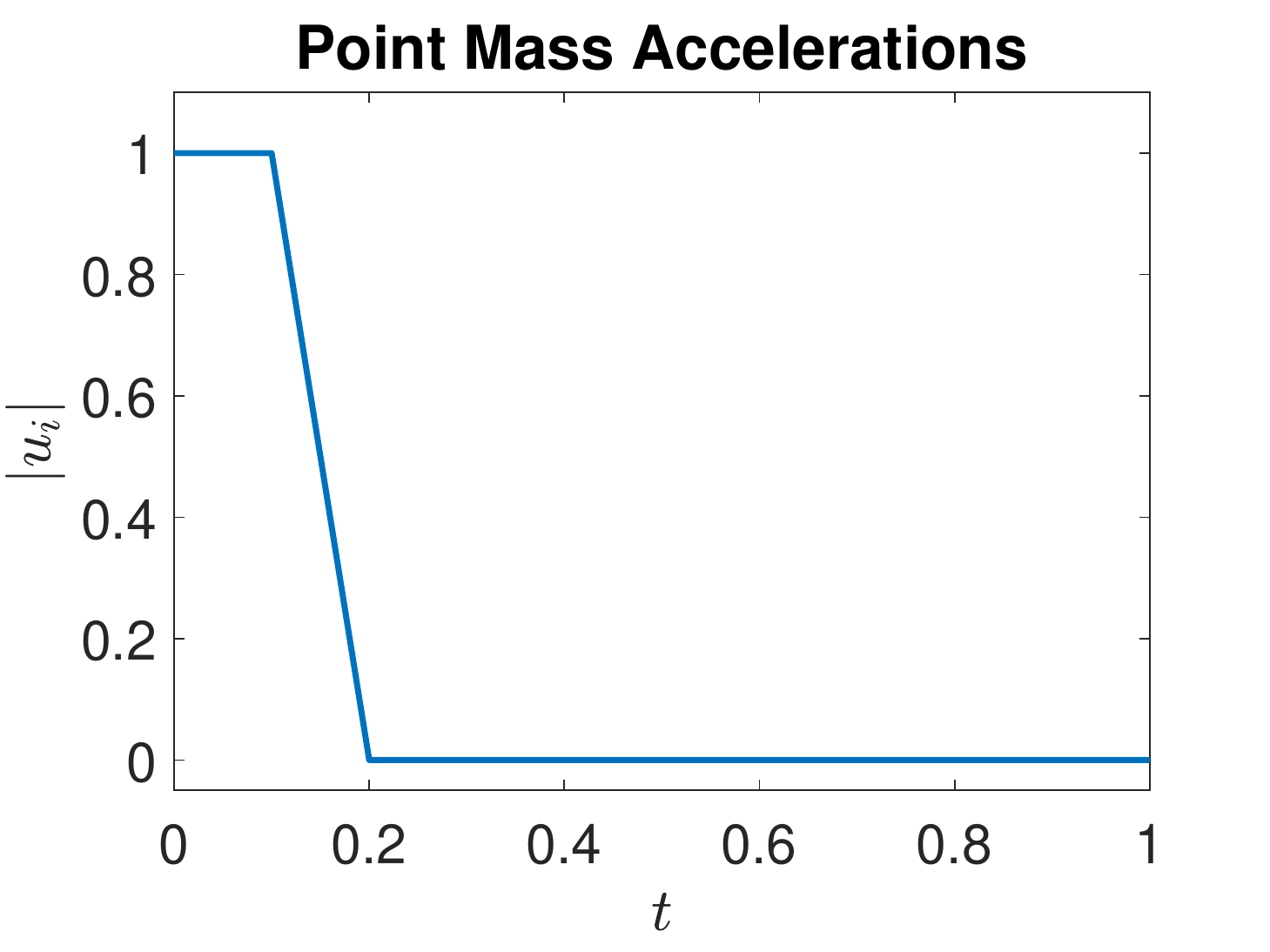}
	\caption{The magnitude of the parameterized acceleration, $u_i(t)=\ddot \theta_i(t)$, of each point mass, $1\le i \le n$.}
	\label{fig_abs_ui}
\end{figure} 
The  magnitudes of the functions $u_i(t)$ are illustrated in Figure~\ref{fig_abs_ui}. For each $i$, the magnitude of the parameterized acceleration $u_i$ is chosen to be $1$ for the short time interval $0\le t \le 0.1$, then decreases linearly from $1$ to $0$ for the short time interval $0.1 \le t \le 0.2$, and finally stays constant at $0$ for the rest of time. \revision{R1Q6}{These motions can be realized by finite, continuous forces and torques applied by the driving motors, as long as it can be assumed that the motion of the masses can be prescribed without the need to solve additional differential equations for the masses. See the discussion concerning Figure~\ref{fig:stepper_motor} in Section~\ref{sec_3d_traj} and also the discussion of the same topic related to the motion of the rolling ball after \eqref{eq_prescribed_u_ball} below. } The parameterized accelerations $u_i = \ddot \theta_i$ are constructed to be continuous (instead of discontinuous) so that $\dot \btheta$ and $\btheta$ are differentiable. We have used these parameterized accelerations since the derivation of the equations of motion \eqref{eq_Om_1d} and \eqref{eq_ddphi_disk} assumed that $\btheta$ and $\dot  \btheta$ are differentiable.

The rolling disk's initial conditions are selected so that the disk starts at rest at the origin. Table~\ref{table_disk_ICs} shows parameter values used in the rolling disk's initial conditions \eqref{eq_disk_initial_conds}. Since the initial orientation of the disk is $\phi_a=0$ and since the initial configurations of the internal point masses are given by $\btheta_a=\begin{bmatrix} \unaryminus \frac{\pi}{2} & \unaryminus \frac{\pi}{2} & \unaryminus \frac{\pi}{2} & \unaryminus \frac{\pi}{2} \end{bmatrix}^\mathsf{T}$, all the internal point masses are initially located directly below the GC, so that the disk's total system CM is initially located below the GC. To ensure that the disk is initially at rest, ${\dot \btheta}_a = \begin{bmatrix} 0 & 0 & 0 & 0 \end{bmatrix}^\mathsf{T}$ and $\dot \phi_a = \unaryminus \frac{{\dot z}_a}{r} = 0$. To ensure that the disk's GC is initially located at the origin, $z_a = 0$. In summary, the rolling disk's initial conditions are
\begin{equation} \label{eq_disk_initial_conds_spec}
\bx_a = \begin{bmatrix} \unaryminus \frac{\pi}{2} & \unaryminus \frac{\pi}{2} & \unaryminus \frac{\pi}{2} & \unaryminus \frac{\pi}{2} & 0 & 0 & 0 & 0 & 0 & 0  \end{bmatrix}^\mathsf{T} .
\end{equation}

\begin{table}[h!]
	\centering 
	{ 
		\setlength{\extrarowheight}{1.5pt}
		\begin{tabular}{| c | c |} 
			\hline
			\textbf{Parameter} & \textbf{Value} \\ 
			\hline\hline 
			$\btheta_a$ & $\begin{bmatrix} \unaryminus \frac{\pi}{2} & \unaryminus \frac{\pi}{2} & \unaryminus \frac{\pi}{2} & \unaryminus \frac{\pi}{2} \end{bmatrix}^\mathsf{T}$  \\  
			\hline
			$\dot \btheta_a$ & $\begin{bmatrix} 0 & 0 & 0 & 0 \end{bmatrix}^\mathsf{T}$ \\ 
			\hline
			$\phi_a$ & $0$ \\
			\hline
			$\dot \phi_a$ & $0$ \\  
			\hline
			$z_a$ & $0$ \\
			\hline
			$\dot z_a$ & $0$ \\ 
			\hline
		\end{tabular} 
	}
	\caption{Initial condition parameter values for the rolling disk.}
	\label{table_disk_ICs}
\end{table}

\begin{figure}[!ht] 
	\centering
	\subfloat[Trajectories of the disk's internal point masses and of the total system center of mass in the body frame translated to the GC.]{\includegraphics[scale=.5]{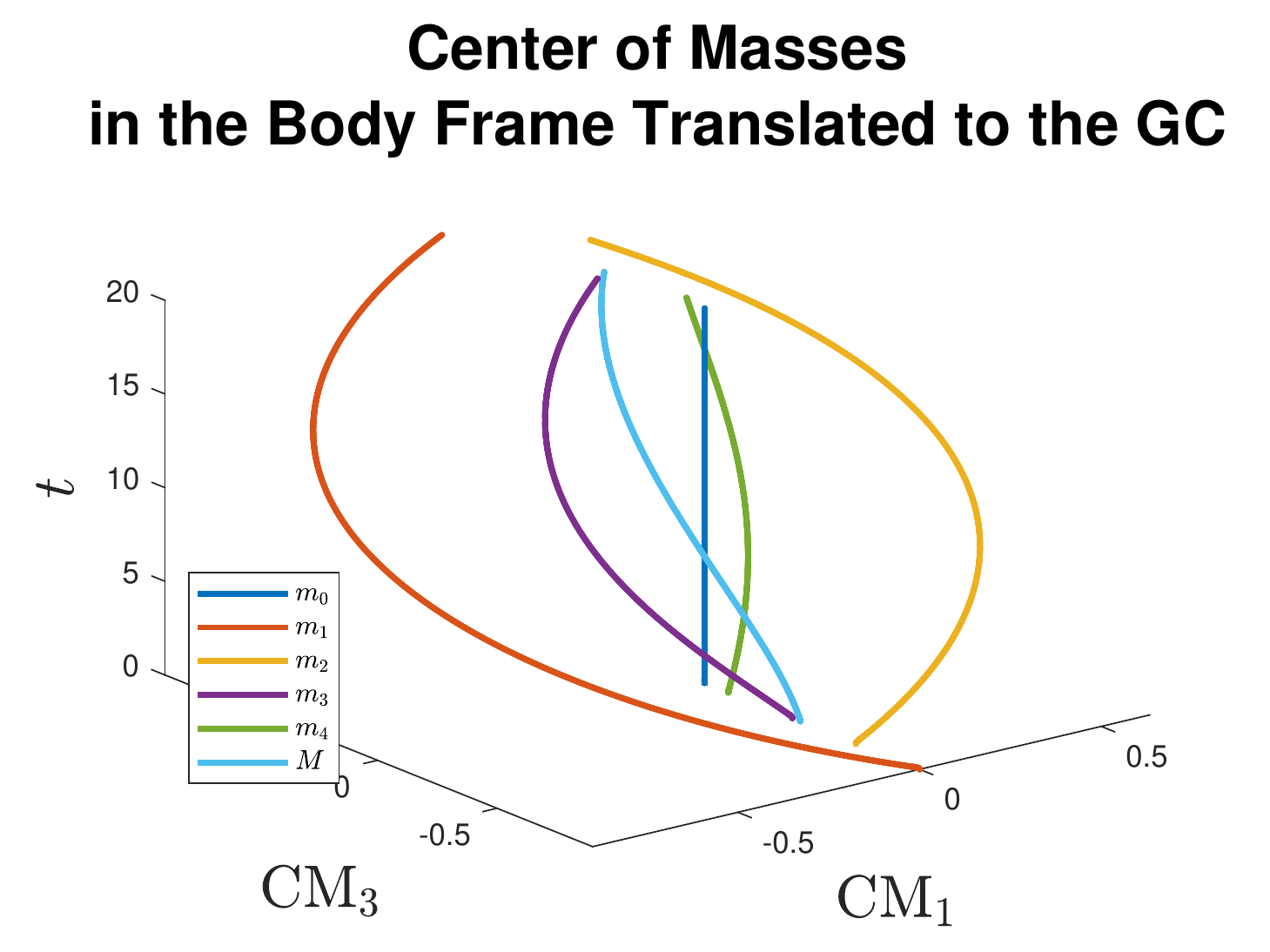}\label{fig_disk_cm_time_bf_gc}}
	\hspace{5mm}
	\subfloat[Trajectories of the disk's internal point masses and of the total system center of mass in the spatial frame translated to the GC.]{\includegraphics[scale=.5]{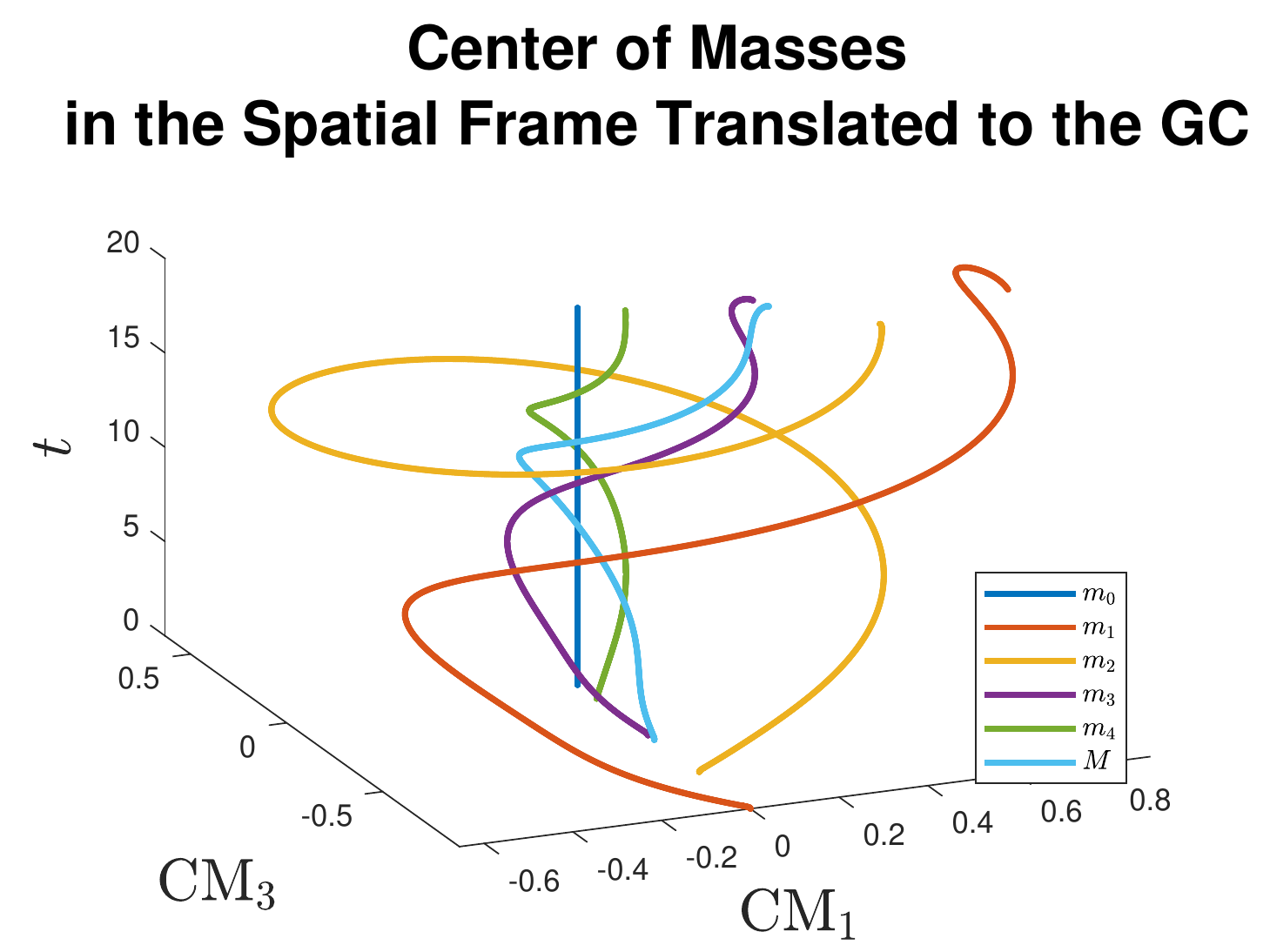}\label{fig_disk_cm_time_spatial_gc}}
	\\
	\subfloat[Evolution of the time derivative of the disk's rotation angle.]{\includegraphics[scale=.5]{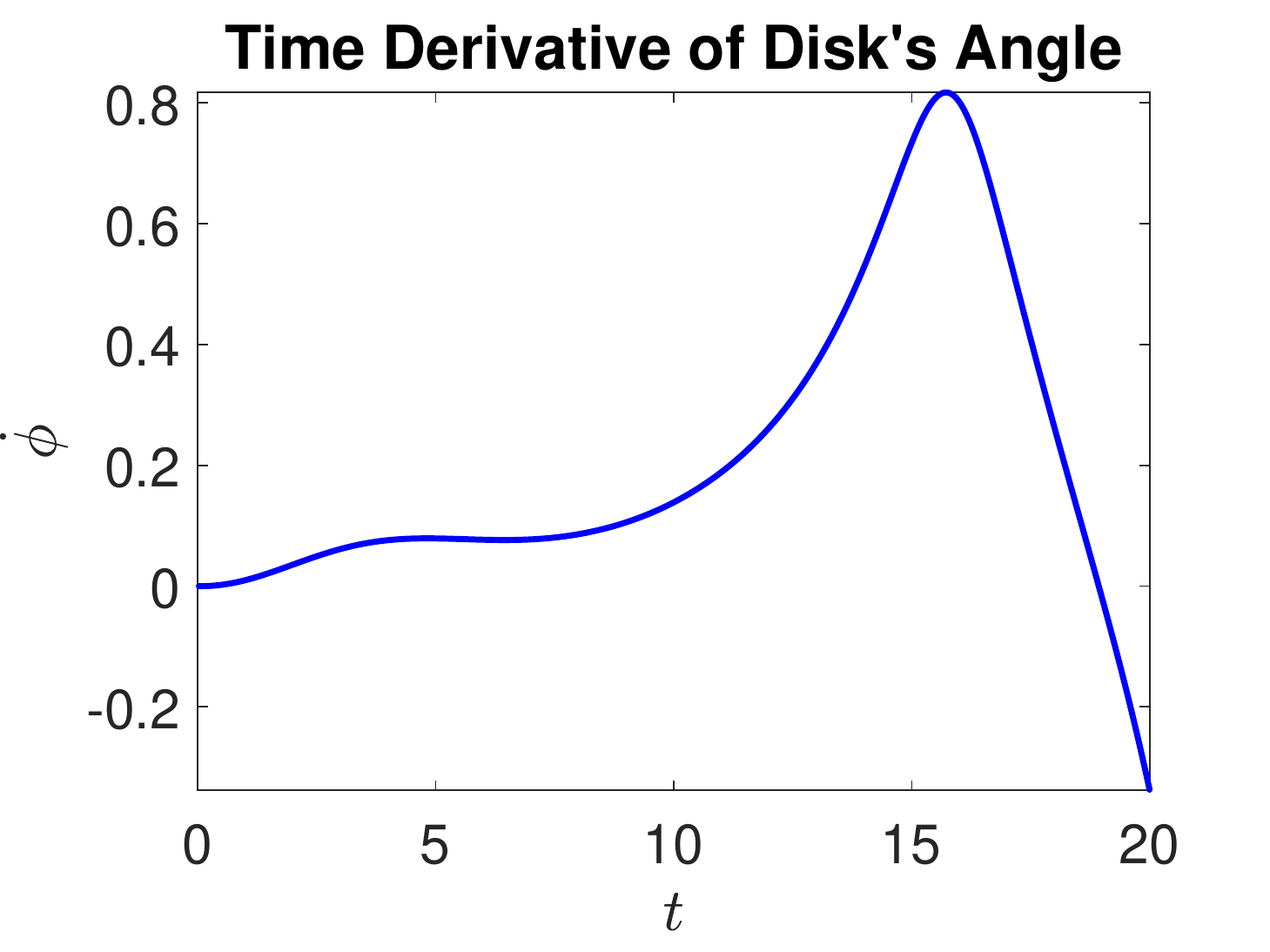}\label{fig_disk_dot_angle}}
	\hspace{5mm}
	\subfloat[Trajectory of the disk's GC and CP.]{\includegraphics[scale=.5]{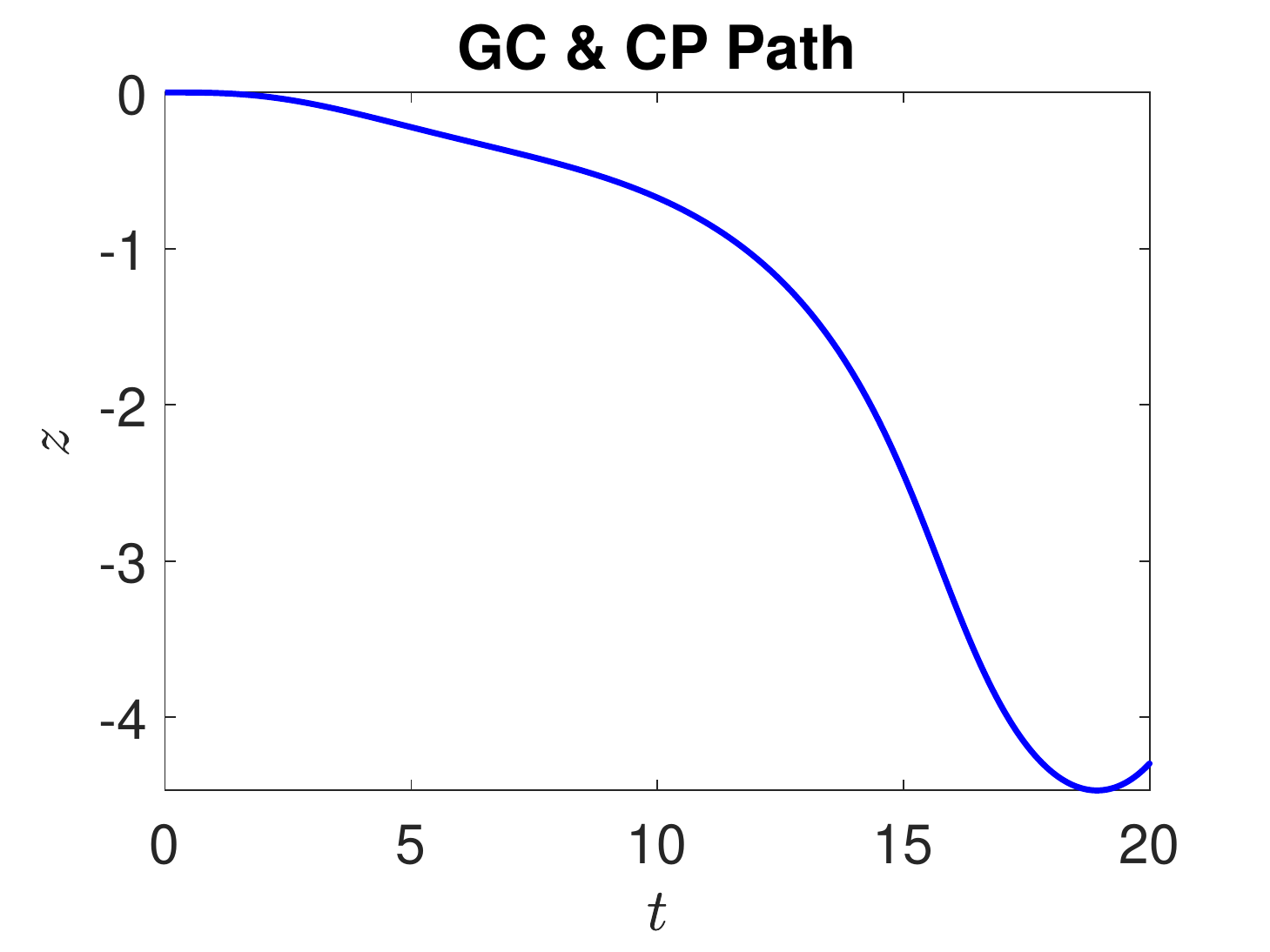}\label{fig_disk_gc}}
	\caption{Dynamics of the rolling disk shown in Figure~\ref{fig_disk_masses_rails_bf_gc} obtained by numerically integrating the ODE IVP \eqref{eq_rdisk_dynamics}, \eqref{eq_disk_initial_conds_spec} with \mcode{ode45} over the time interval $\left[0,20\right]$. The parameterized accelerations of the internal point masses are given in \eqref{eq_prescribed_u_disk}.}
	\label{fig_disk_ode_sims}
\end{figure}

\begin{figure}[!ht] 
	\centering
	\subfloat[The magnitude of the disk's normal force is always positive so that the disk rolls without slipping if the coefficient of static friction exceeds $.2951$.]{\includegraphics[scale=.5]{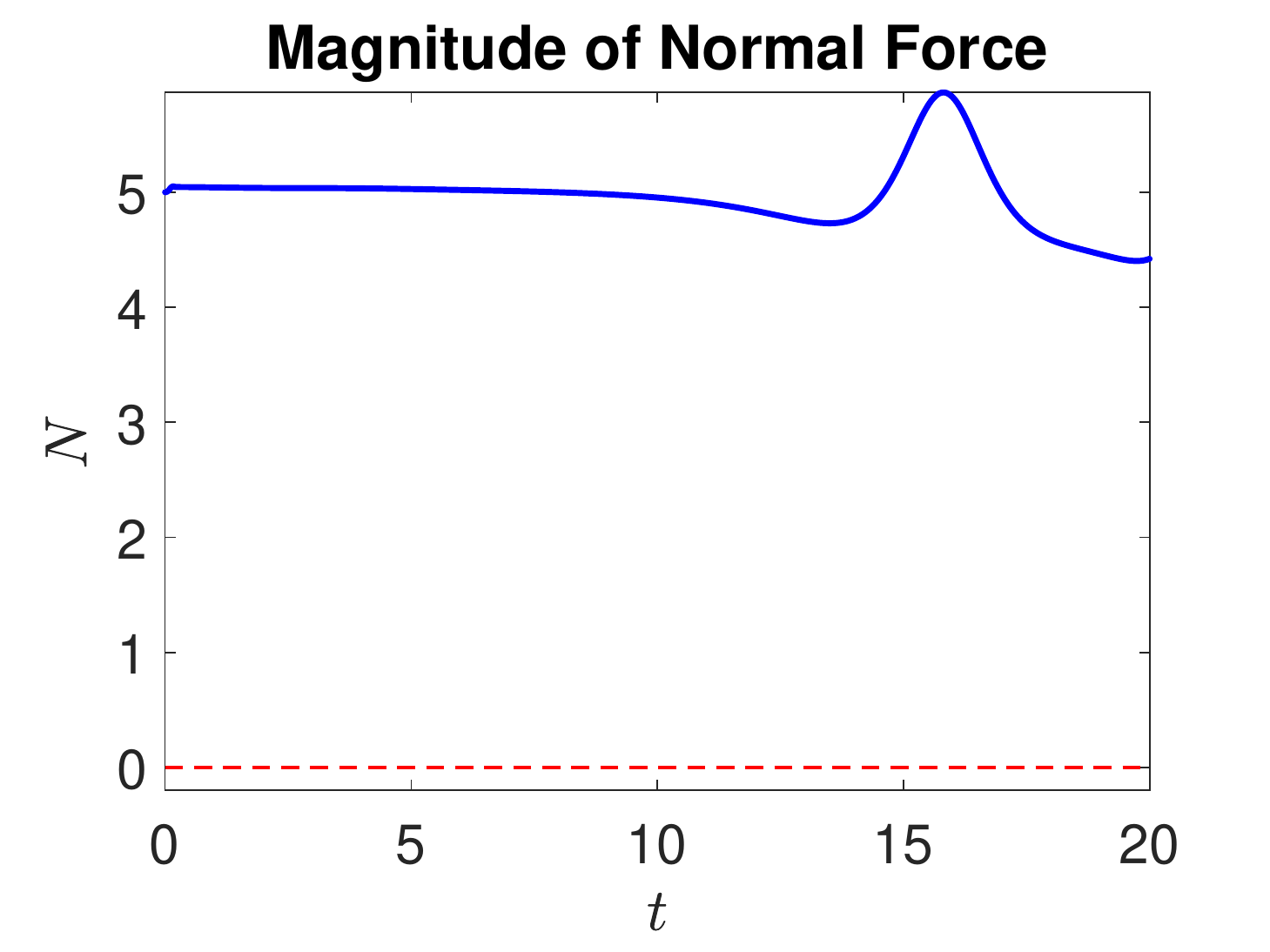}\label{fig_disk_normal}}
	\hspace{5mm}
	\subfloat[$\mathbf{e}_1$-component of the disk's static friction.]{\includegraphics[scale=.5]{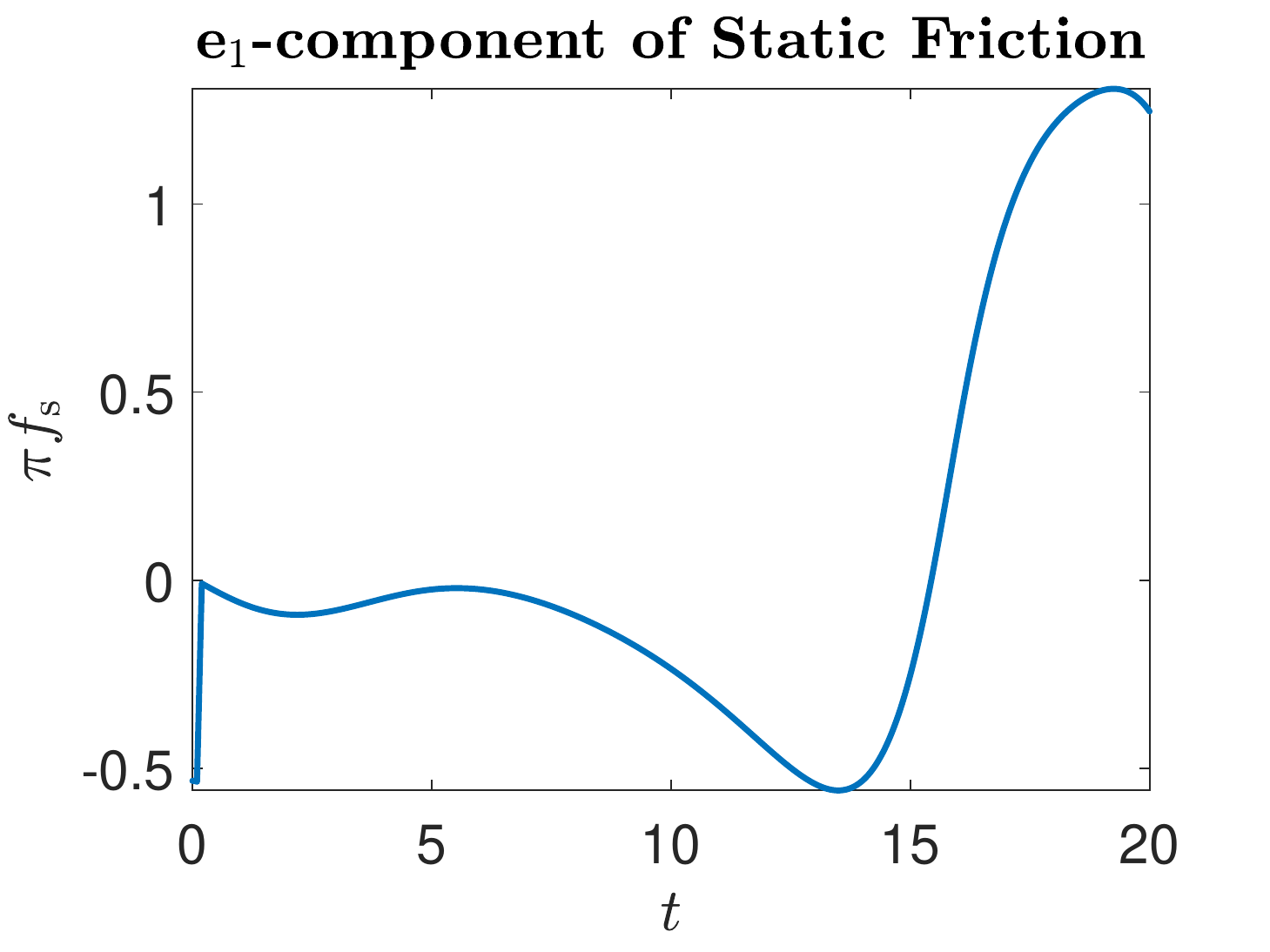}\label{fig_disk_sf}}
	\\
	\subfloat[Ratio of the magnitude of the static friction to the magnitude of the normal force. The minimum coefficient of static friction for the disk is $.2951$.]{\includegraphics[scale=.5]{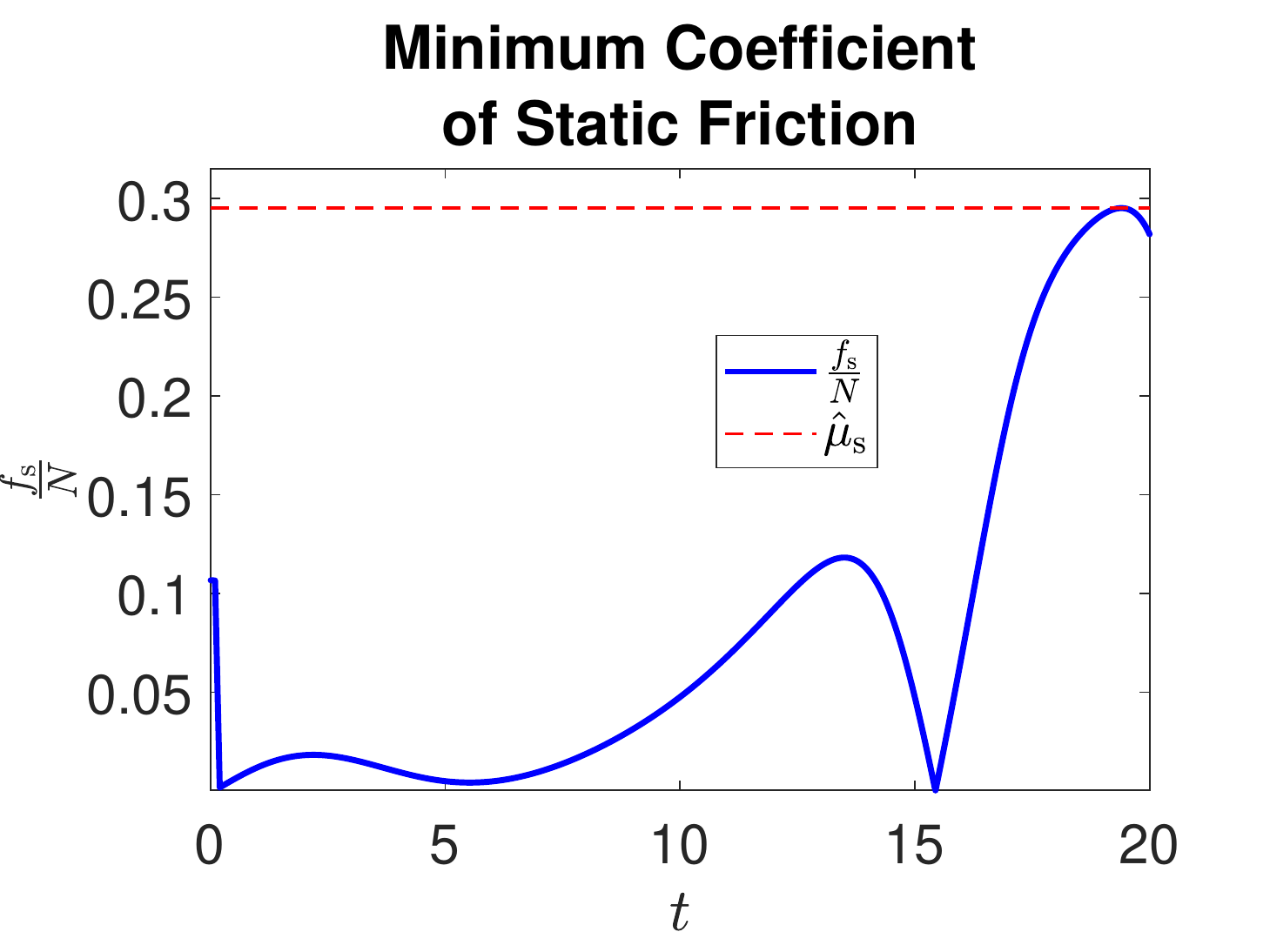}\label{fig_disk_coeff_sf}}
	\caption{Contact point forces acting on the rolling disk shown in Figure~\ref{fig_disk_masses_rails_bf_gc} obtained by numerically integrating the ODE IVP \eqref{eq_rdisk_dynamics}, \eqref{eq_disk_initial_conds_spec} with \mcode{ode45} over the time interval $\left[0,20\right]$. The parameterized accelerations of the internal point masses are given in \eqref{eq_prescribed_u_disk}. Since the magnitude of the disk's normal force is always positive, the disk rolls without slipping if the coefficient of static friction exceeds $.2951$.}
	\label{fig_disk_ode_sims_cp_forces}
\end{figure}

The dynamics of this rolling disk are simulated by numerically integrating the ODE IVP \eqref{eq_rdisk_dynamics}, \eqref{eq_disk_initial_conds_spec} via \mcode{MATLAB} R2017b and Fortran  ODE-integration routines.  For ODE integrators, we have used the \mcode{MATLAB} R2017b routines \mcode{ode45}, \mcode{ode113}, \mcode{ode15s}, \mcode{ode23t}, and \mcode{ode23tb} and a \mcode{MATLAB} \mcode{MEX} wrapper of the Fortran routine radau5 \cite{hairer1996solving}, using the default input options except for the absolute and relative error tolerances and the Jacobian. The absolute and relative error tolerances supplied to the numerical integrators are both set to $1\mathrm{e}{-12}$. The Jacobian of $\mathbf{f}$ with respect to the state $\bx$, obtained via complex-step differentiation \cite{squire1998using,martins2001connection,martins2003complex}, is supplied to \mcode{ode15s}, \mcode{ode23t}, \mcode{ode23tb}, and radau5. Since excellent agreement was observed between all the numerical integrators, only the results obtained by numerically integrating the ODE IVP \eqref{eq_rdisk_dynamics}, \eqref{eq_disk_initial_conds_spec} with \mcode{ode45} are shown in Figures~\ref{fig_disk_ode_sims} and \ref{fig_disk_ode_sims_cp_forces}. We shall also note that while all the numerical integrators yielded identical results, \mcode{ode113} completed the numerical integration in the shortest time. Figure~\ref{fig_disk_normal} shows that the magnitude of the disk's normal force is always positive and Figure~\ref{fig_disk_coeff_sf} shows that the minimum coefficient of static friction required for the disk to roll without slipping is $\hat \mu_\mathrm{s}=.2951$. The reader is referred to \cite{totten2017asm} for listings of the coefficient of static friction for pairs of materials to see which materials could be used to make this particular disk roll without slipping on the surface.  For example, if the disk's shell were made from aluminum, then it could roll without slipping on an aluminum ($\mu_\mathrm{s}=.42$), steel ($\mu_\mathrm{s}=.35$), titanium ($\mu_\mathrm{s}=.34$), or nickel ($\mu_\mathrm{s}=.33$) surface, but not on a copper ($\mu_\mathrm{s}=.28$), chromium ($\mu_\mathrm{s}=.27$), glass ($\mu_\mathrm{s}=.17$), or graphite ($\mu_\mathrm{s}=.16$) surface.

\section{Numerical Simulations of the Dynamics of the Rolling Ball}
\label{sec_ball_numerical}
To write the equations of motion for the rolling ball in the standard ordinary differential/algebraic equation (ODE/DAE) form, the state of the system is defined as
\begin{equation} \label{eq_ball_sv}
\bx \equiv \begin{bmatrix} \btheta \\ \dot \btheta \\ \mathfrak{q} \\ \bOm \\ \bz  \end{bmatrix},
\end{equation}
where $\btheta, \, \dot \btheta \in \mathbb{R}^n$ encode the positions and velocities of the moving masses, the versor $\mathfrak{q} \in \mathscr{S} \cong \mathbb{S}^3 \subset \mathbb{R}^4$ encodes the orientation of the rolling ball, $\bOm \in \mathbb{R}^3$ is the body angular velocity, and $\bz \in \mathbb{R}^2$ denotes the spatial $\mathbf{e}_1$- and $\mathbf{e}_2$-components of the GC and CP.  \revision{R1Q7}{Appendix D of \cite{Putkaradze2018dynamicsP} provides a brief review of quaternions and versors. Recall from \cite{Putkaradze2018dynamicsP} that given a column vector $\bv \in \mathbb{R}^3$, $\bv^\sharp$ is the quaternion 
\begin{equation}
\bv^\sharp = \begin{bmatrix}0 \\ \bv \end{bmatrix},
\end{equation}
and given a quaternion $\mathfrak{p} \in \mathbb{H}$, $\mathfrak{p}^\flat \in \mathbb{R}^3$ is the column vector such that 
\begin{equation}
\mathfrak{p} = \begin{bmatrix} p_0 \\ \mathfrak{p}^\flat \end{bmatrix}.
\end{equation}
Using a versor to parameterize the ball's orientation implies that the state vector \eqref{eq_ball_sv} consists of $2n+9$ components, whereas the state vector would be comprised of only $2n+8$ components if Euler angles were used instead. While the versor is less efficient than Euler angles at parameterizing the ball's orientation, the versor parameterization, which is a mapping from the unit 3-sphere $\mathbb{S}^3$ to $SO(3)$, provides a double covering of $SO(3)$ and therefore gives a local homeomorphism about each point in $SO(3)$ \cite{schroder1993transferring,stuelpnagel1964parametrization}. In contrast, the Euler angle parameterization, which is a mapping from the 3-torus $\mathbb{T}^3$ to $SO(3)$, is not a covering map of $SO(3)$ and therefore does not give a local homeomorphism about each point in $SO(3)$, which causes gimbal lock at those points where the parameterization is not a local homeomorphism \cite{schroder1993transferring,stuelpnagel1964parametrization}.}
ODE and DAE formulations of the rolling ball's system dynamics defined for $a \le t \le b$ are
\begin{equation} \label{rolling_ball_ode_f}
\dot {\bx} = \begin{bmatrix} \dot \btheta \\ \ddot \btheta \\ \dot {\mathfrak{q}} \\ \dot \bOm \\ \dot \bz  \end{bmatrix}  = \mathbf{f}\left(t,\bx,\bu\right) \equiv \begin{bmatrix} \dot \btheta \\ \bu  \\ \frac{1}{2} \mathfrak{q} \bOm^\sharp \\ \bkappa\left(t,\bx,\bu \right) \\ \left( \left[\mathfrak{q} \bOm^\sharp \mathfrak{q}^{-1} \right]^\flat \times r \mathbf{e}_3  \right)_{12}  \end{bmatrix}
\end{equation}
and
\begin{equation} \label{rolling_ball_dae_g}
\mathcal{M} \dot {\bx} = \begin{bmatrix} \dot \btheta \\ \ddot \btheta \\ 0 \\ {\dot {\mathfrak{q}}}^\flat \\ \dot \bOm \\ \dot \bz  \end{bmatrix}  = \mathbf{g}\left(t,\bx,\bu\right) \equiv \begin{bmatrix} \dot \btheta \\ \bu  \\ \left| \mathfrak{q} \right|^2-1 \\ \left[\frac{1}{2} \mathfrak{q} \bOm^\sharp\right]^\flat \\ \bkappa\left(t,\bx,\bu \right) \\ \left( \left[\mathfrak{q} \bOm^\sharp \mathfrak{q}^{-1} \right]^\flat \times r \mathbf{e}_3  \right)_{12}  \end{bmatrix},
\end{equation}
respectively, where $\bu \colon \mathbb{R} \to \mathbb{R}^n$ is a prescribed function of $t$ such that $\bu(t)=\ddot \btheta(t) \in \mathbb{R}^n$, $\bkappa\left(t,\bx,\bu \right)$ is given by the right-hand side of the formula for $\dot \bOm$ in \eqref{eq_Om_1d}, and
\begin{equation}
\mathcal{M} \equiv \diag \left( \begin{bmatrix} \mathbf{1}_{1 \times 2n} & 0 & \mathbf{1}_{1 \times 8} \end{bmatrix} \right)
\end{equation}
is a diagonal DAE mass matrix. Observe that \eqref{rolling_ball_dae_g} is a semi-explicit DAE of index 1, \revision{R1Q7}{since differentiation of the algebraic constraint $\left| \mathfrak{q} \right|^2-1$ with respect to time followed by using ${\dot {\mathfrak{q}}}^\flat = \left[\frac{1}{2} \mathfrak{q} \bOm^\sharp\right]^\flat$ and algebraic manipulation yield the $\left(2n+1\right)^{\mathrm{st}}$ equation in \eqref{rolling_ball_ode_f}, $\dot q_0 = -\frac{1}{2} {\mathfrak{q}}^\flat \cdot \bOm$}. The reader is referred to \cite{Putkaradze2018dynamicsP} for details on the most efficient way to compute $\bGamma \equiv \Lambda^{-1} \mathbf{e}_3=\left[\mathfrak{q}^{-1} \mathbf{e}_3^\sharp \mathfrak{q} \right]^\flat$, $\tilde \bGamma \equiv \Lambda^{-1} \mathbf{F}_\mathrm{e} = \left[\mathfrak{q}^{-1} \mathbf{F}_\mathrm{e}^\sharp \mathfrak{q} \right]^\flat$, and $\bomega \equiv \left[ \dot{\Lambda} \Lambda^{-1} \right]^\vee =  \Lambda \bOm=\left[\mathfrak{q} \bOm^\sharp \mathfrak{q}^{-1} \right]^\flat$, which appear on the right-hand sides of \eqref{rolling_ball_ode_f} and \eqref{rolling_ball_dae_g}.

In order to simulate the rolling ball's dynamics, \eqref{rolling_ball_ode_f} or \eqref{rolling_ball_dae_g} must be integrated with prescribed initial conditions at time $t=a$:
\begin{equation} \label{eq_ball_initial_conds}
\bx\left(a\right) = \begin{bmatrix} \btheta(a) \\ \dot \btheta(a) \\ \mathfrak{q}(a) \\ \bOm(a) \\ \bz(a) \end{bmatrix} = \begin{bmatrix}  \btheta_a \\  {\dot \btheta}_a \\ \mathfrak{q}_a \\ \bOm_a \\ \bz_a \end{bmatrix} \equiv \bx_a.
\end{equation}
\eqref{rolling_ball_ode_f} and \eqref{eq_ball_initial_conds} constitute an ODE IVP, while \eqref{rolling_ball_dae_g} and \eqref{eq_ball_initial_conds} constitute a DAE IVP.

\begin{figure}[h] 
	\centering
	\includegraphics[scale=.7]{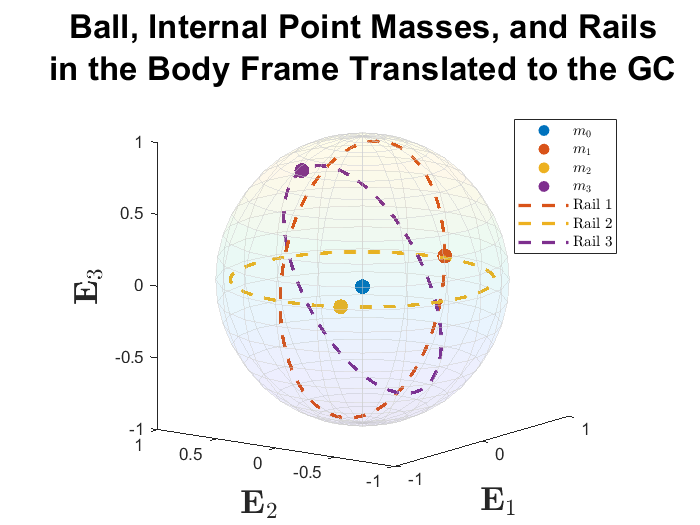}
	\caption{A ball of radius $r=1$ actuated by $3$ internal point masses, $m_1$, $m_2$, and $m_3$, each on its own circular rail of radius $r_1=.95$, $r_2=.9$, and $r_3=.85$, respectively. The location of the ball's CM is shifted slightly away from the GC and is denoted by $m_0$. $m_0=m_1=m_2=m_3=1$ and $g=1$. The configuration at the initial time $t=0$ is shown.}
	\label{fig_ball_masses_rail_bf_gc}
\end{figure}

In the simulations, we consider a rolling ball of mass $m_0=1$, radius $r=1$, principal moments of inertia $d_1=.9$, $d_2=1$, and $d_3=1.1$, and with the CM shifted slightly away from the GC at $\bzeta_0=\begin{bmatrix} 0 & 0 & -.05 \end{bmatrix}^\mathsf{T}$. The ball contains $n=3$ internal point masses, each of mass $1$ so that $m_1=m_2=m_3=1$ and each located on its own circular rail centered on the GC of radius $r_1=.95$, $r_2=.9$, and $r_3=.85$, respectively, oriented as shown in Figure~\ref{fig_ball_masses_rail_bf_gc}. 
\rem{\todo{VP: The example is quite simple, CM is in GC and inertia matrix is proportional to unity, which makes a lot of terms cancel. Of course, the total moment of inertia and CM is no longer trivial. Was there a particular reason you chose that simple case for the ball without masses? Some terms in equations will cancel, although it is still non-trivial. \\ SMR: I re-simulated the ball using a CM slightly shifted from the GC and using unequal principal moments of inertia. Are these new CM and principal moment values and the associated simulation plots ok? Do you also want me to re-simulate the disk using an off-center CM? I am a bit hesitant to re-simulate the controlled ball using these new values for the next paper on control, since the simulations will be tedious to regenerate for new ball parameters due to having to find new predictor-corrector tangent step lengths. \\ 
		VP: OK, fair enough. Hope the Referees won't bug us about it.  } }
For $1 \le i \le n$, the position of $m_i$ in the body frame centered on the GC is:
\begin{equation}
\bzeta_i\left(\theta_i\right) = r_i \mathcal{B}_i \left(\boldsymbol{\varsigma} \left(\mathbf{v}_i\right) \right) \begin{bmatrix} \cos \theta_i \\ 0 \\ \sin \theta_i  \end{bmatrix},
\end{equation}
where $\mathcal{B}_i \left(\mathbf{n} \right) \in SO(3)$ is a rotation matrix whose columns are the right-handed orthonormal basis constructed from the unit vector $\mathbf{n} \in \mathbb{R}^3$ based on the algorithm given in Section 4 and Listing 2 of \cite{frisvad2012building}, $\boldsymbol{\varsigma} \colon \mathbb{R}^3 \to \mathbb{R}^3$ maps spherical coordinates to Cartesian coordinates:
\begin{equation}
\boldsymbol{\varsigma}\left(\begin{bmatrix} \phi \\ \theta \\ \rho \end{bmatrix} \right) = \begin{bmatrix} \rho \cos \theta \cos \phi  \\ \rho \cos \theta \sin \phi \\ \rho \sin \theta \end{bmatrix},
\end{equation}
and
\begin{equation}
\mathbf{v}_1 = \begin{bmatrix} 0 & 0 & 1 \end{bmatrix}^\mathsf{T}, \quad \mathbf{v}_2 = \begin{bmatrix} \frac{\pi}{2} & 0 & 1 \end{bmatrix}^\mathsf{T}, \quad \mathrm{and} \quad \mathbf{v}_3 = \begin{bmatrix} \frac{\pi}{4} & \frac{\pi}{4} & 1 \end{bmatrix}^\mathsf{T}
\end{equation}
are spherical coordinates of unit vectors in $\mathbb{R}^3$. The total mass of the ball's  system  is $M=4$, and  gravity is rescaled to be $g=1$. There is no external force acting on the ball's GC so that $\mathbf{F}_\mathrm{e} = \tilde \bGamma  = \mathbf{0}$ in the right-hand sides of \eqref{rolling_ball_ode_f} and \eqref{rolling_ball_dae_g}. This ball's dynamics are simulated with initial time $a=0$ and final time $b=20$, so that the simulation time interval is $\left[0,20\right]$. The parameterized acceleration of each internal point mass is a continuous approximation of a short duration unit amplitude step function:
\begin{equation} \label{eq_prescribed_u_ball}
u_i(t) = \ddot \theta_i(t) = \left\{ \begin{array}{ll} 1, & 0 \le t \le .1, \\ \unaryminus 10t+2, & .1 \le t \le .2, \\ 0, & .2 \le t \le 20, \end{array} \right. \quad \mbox{for} \quad 1 \le i \le n.
\end{equation}
A plot of the magnitude of \eqref{eq_prescribed_u_ball} is depicted in Figure~\ref{fig_abs_ui}. \revision{R1Q6}{Physically, these motions of the internal masses are realized by applying finite forces and torques during the initial time interval $0\leq t \leq .1$, ramping these forces/torques to other values  during the time interval $.1 \leq t \leq .2$, and maintaining a uniform angular speed of the masses for all later times.} \revisionS{RR1Q1 \\ RR1Q2}{ If electric motors are used to actuate the masses, the actuation dynamics are coupled with the ball's dynamics \cite{balandin2013motion,ivanova2013dynamics,ivanovacontrolled} and \eqref{eq_prescribed_u_ball} is not physically realizable. In this work, we assume that the masses are actuated by stepper motors so that \eqref{eq_prescribed_u_ball} is realizable, as discussed in Section~\ref{sec_3d_traj}. }

The rolling ball's initial conditions are selected so that the ball starts at rest at the origin. Table~\ref{table_ball_ICs} shows parameter values used in the rolling ball's initial conditions \eqref{eq_ball_initial_conds}.   The initial orientation matrix is selected to be the identity matrix so that $\mathfrak{q}_a = \begin{bmatrix} 1 & 0 & 0 & 0 \end{bmatrix}^\mathsf{T}$ and the initial configurations of the internal point masses are given by $\btheta_a = \begin{bmatrix} 0 & 2.0369 & 0.7044 \end{bmatrix}^\mathsf{T}$, so that the ball's total system center of mass is initially located above the GC. These particular initial configurations of the point masses were obtained by solving a system of algebraic equations for mass positions  based on the requirement that the ball's  total system center of mass be  directly above  or below the GC.  
To ensure that the ball is initially at rest, ${\dot \btheta}_a = \begin{bmatrix} 0 & 0 & 0 \end{bmatrix}^\mathsf{T}$ and $\bOm_a = \begin{bmatrix} 0 & 0 & 0 \end{bmatrix}^\mathsf{T}$. To ensure that the ball's GC is initially located at the origin, $\bz_a = \begin{bmatrix} 0 & 0 \end{bmatrix}^\mathsf{T}$. In summary, the rolling ball's initial conditions are
\begin{equation} \label{eq_ball_initial_conds_spec}
\bx_a = \begin{bmatrix} 0 & 2.0369 & 0.7044 & 0 & 0 & 0 & 1 & 0 & 0 & 0 & 0 & 0 & 0 & 0 & 0 \end{bmatrix}^\mathsf{T}.
\end{equation}

\begin{table}[h!]
	\centering 
	{ 
		\setlength{\extrarowheight}{1.5pt}
		\begin{tabular}{| c | c |} 
			\hline
			\textbf{Parameter} & \textbf{Value} \\ 
			\hline\hline 
			$\btheta_a$ & $\begin{bmatrix}  0 & 2.0369 & .7044 \end{bmatrix}^\mathsf{T}$  \\  
			\hline
			$\dot \btheta_a$ & $\begin{bmatrix} 0 & 0 & 0 \end{bmatrix}^\mathsf{T}$ \\ 
			\hline
			$\mathfrak{q}_a$ & $\begin{bmatrix} 1 &  0 & 0 & 0 \end{bmatrix}^\mathsf{T}$ \\
			\hline
			$\bOm_a$ & $\begin{bmatrix}  0 & 0 & 0 \end{bmatrix}^\mathsf{T}$ \\ 
			\hline
			$\bz_a$ & $\begin{bmatrix}  0 & 0 \end{bmatrix}^\mathsf{T}$ \\ 
			\hline
		\end{tabular} 
	}
	\caption{Initial condition parameter values for the rolling ball.}
	\label{table_ball_ICs}
\end{table}

\begin{figure}[!ht] 
	\centering
	\subfloat[Trajectories of the ball's internal point masses and of the total system center of mass in the body frame translated to the GC.]{\includegraphics[scale=.5]{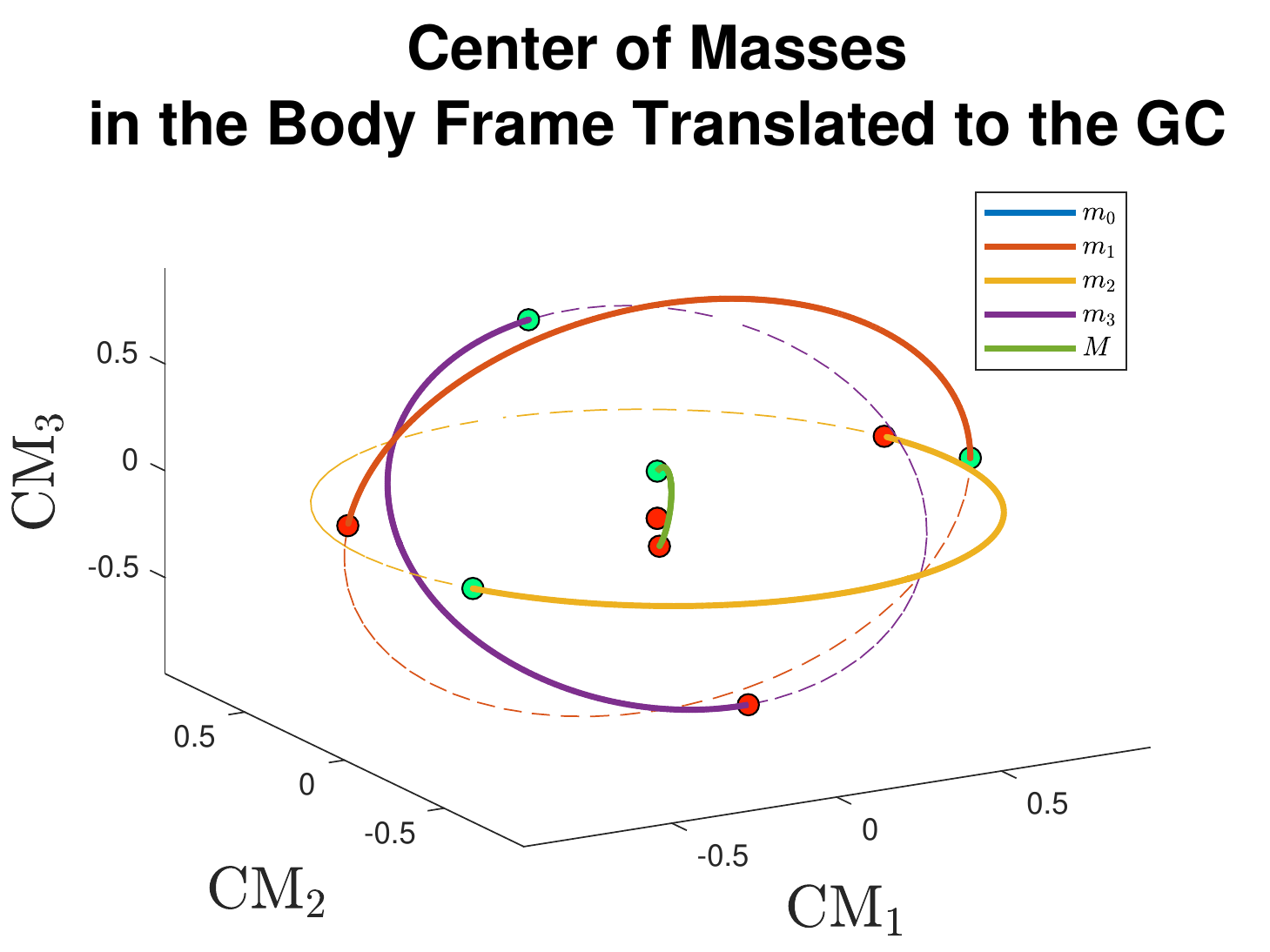}\label{fig_ball_cm_bf_gc}}
	\hspace{5mm}
	\subfloat[Trajectories of the ball's internal point masses and of the total system center of mass in the spatial frame translated to the GC.]{\includegraphics[scale=.5]{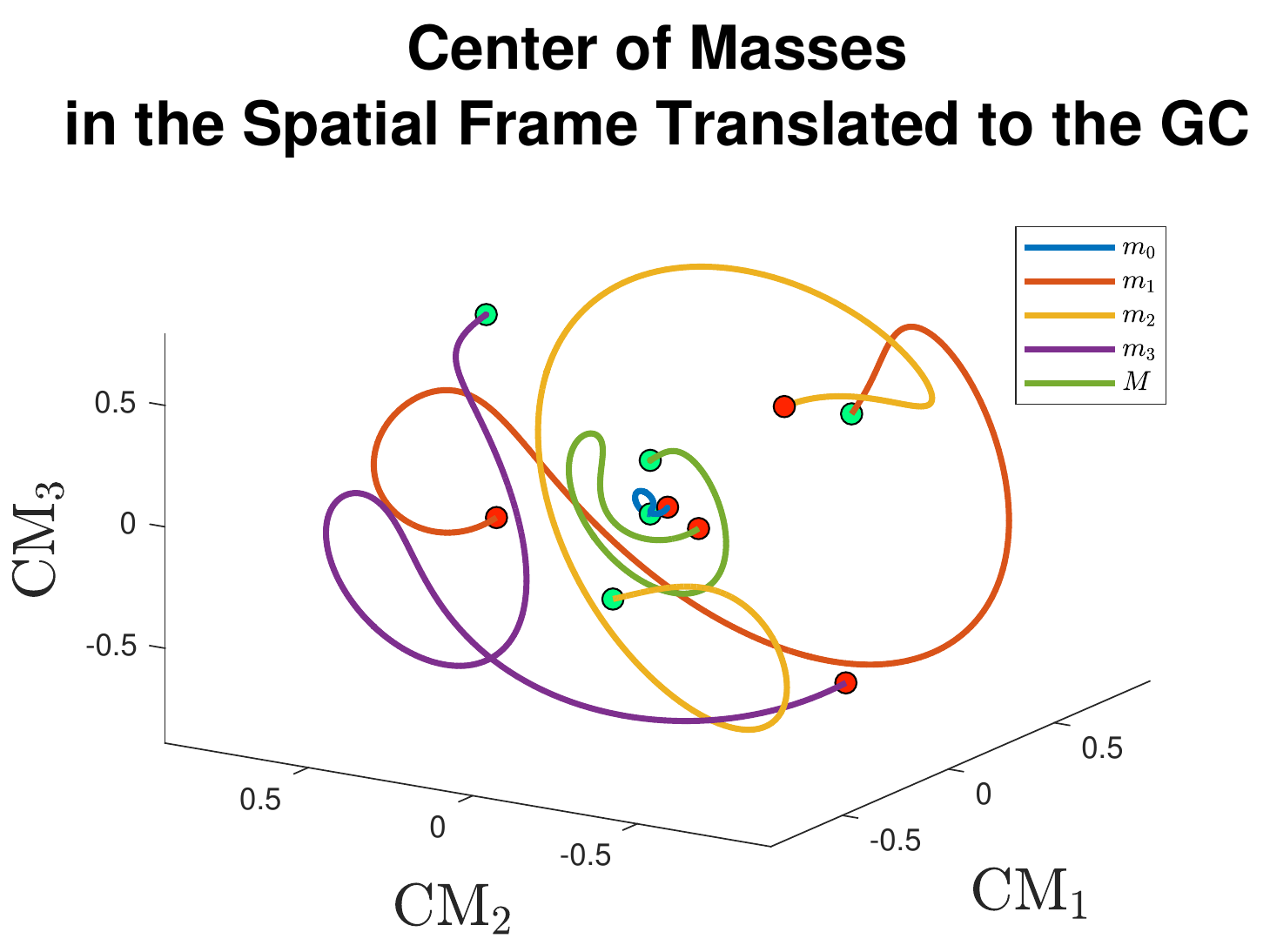}\label{fig_ball_cm_spatial_gc}}
	\\
	\subfloat[Evolution of the ball's body angular velocity.]{\includegraphics[scale=.5]{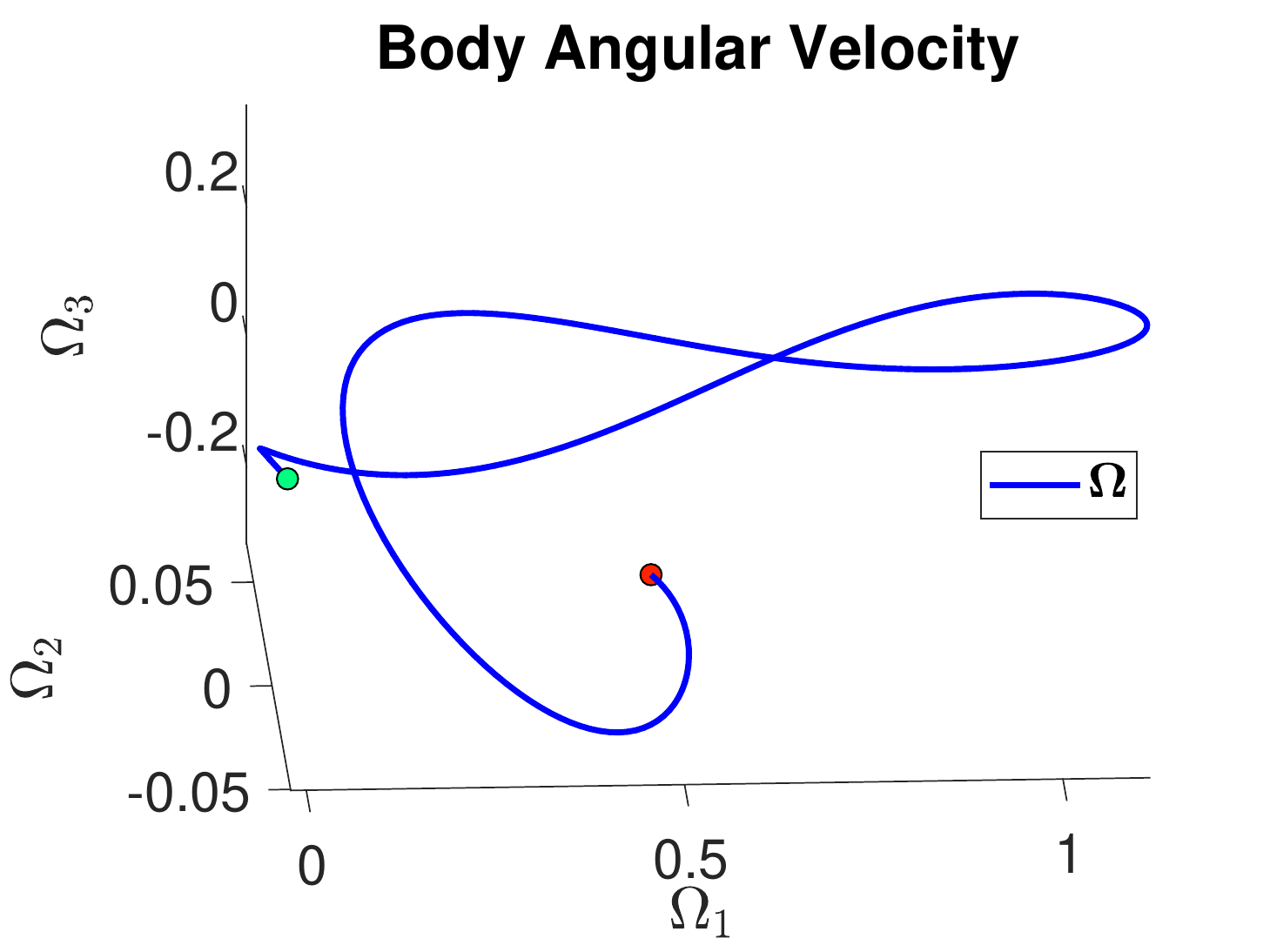}\label{fig_ball_Omega}}
	\hspace{5mm}
	\subfloat[Trajectory of the ball's GC and CP.]{\includegraphics[scale=.5]{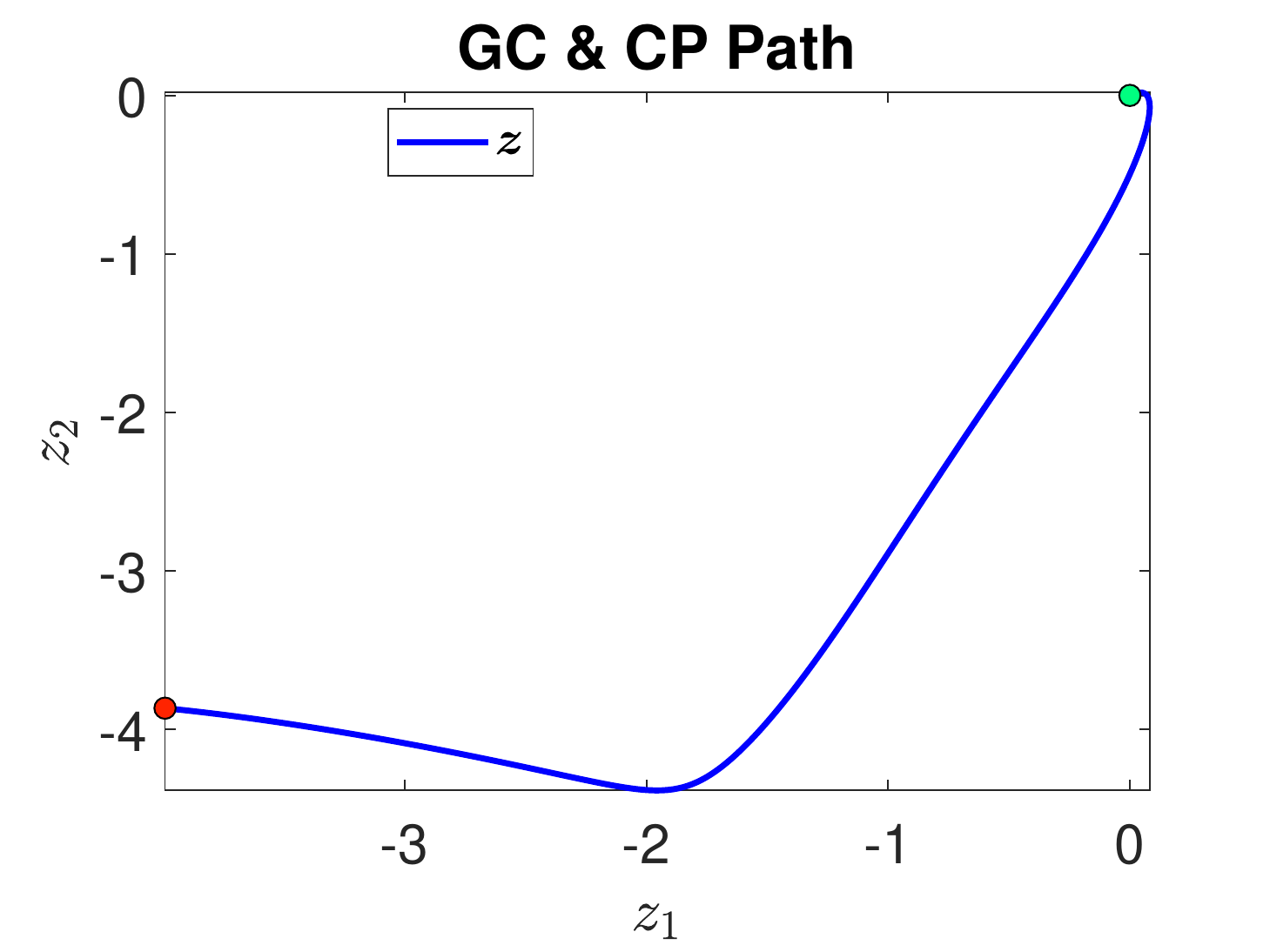}\label{fig_ball_gc_path}}
	\caption{Dynamics of the rolling ball shown in Figure~\ref{fig_ball_masses_rail_bf_gc} obtained by numerically integrating the DAE IVP \eqref{rolling_ball_dae_g}, \eqref{eq_ball_initial_conds_spec} with radau5 over the time interval $\left[0,20\right]$. The parameterized accelerations of the internal point masses are given in \eqref{eq_prescribed_u_ball}.}
	\label{fig_ball_dae_sims}
\end{figure}

\begin{figure}[!ht] 
	\centering
	\subfloat[The magnitude of the ball's normal force is always positive so that the ball rolls without slipping if the coefficient of static friction exceeds $.19$.]{\includegraphics[scale=.5]{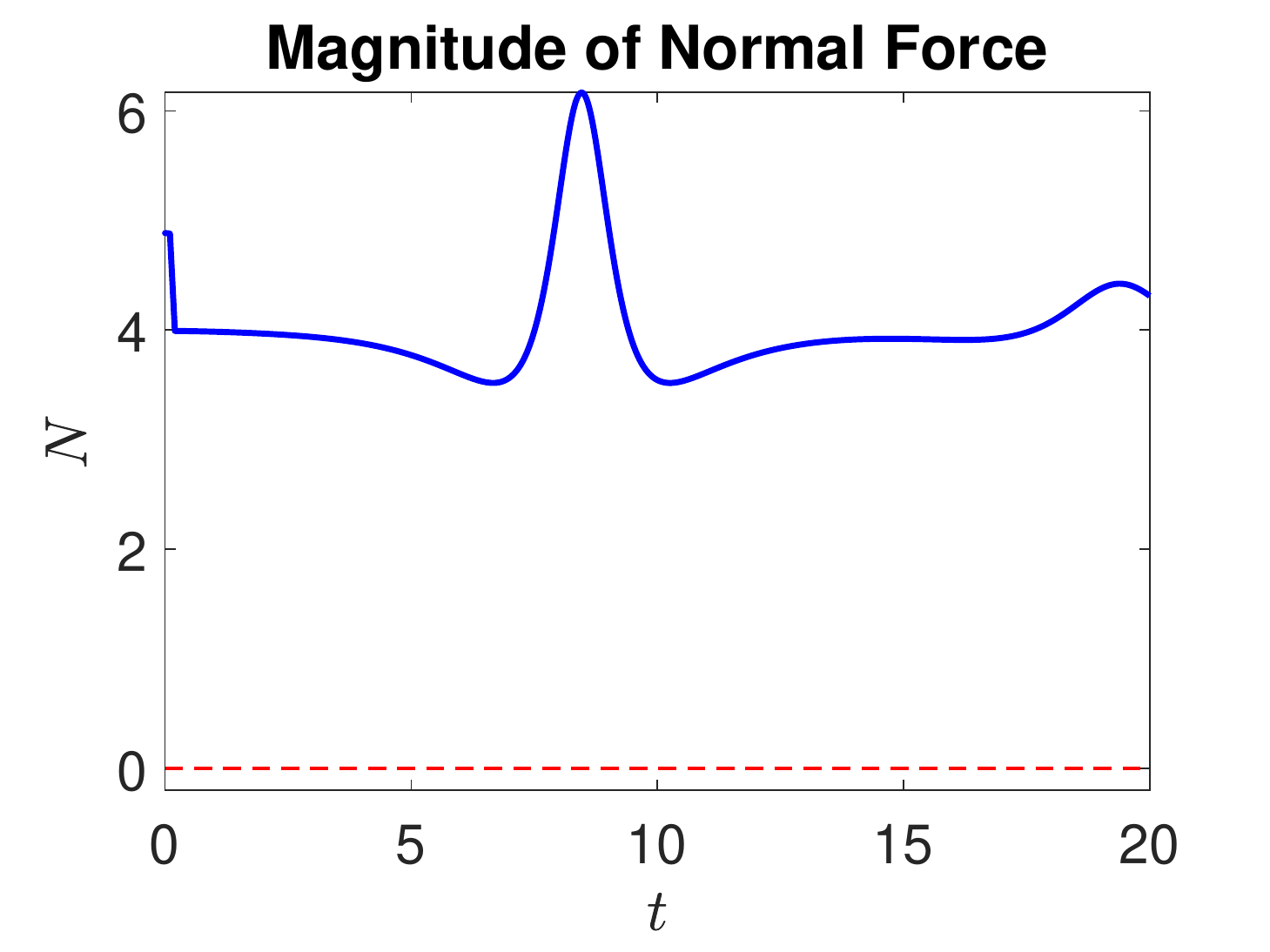}\label{fig_ball_normal}}
	\hspace{5mm}
	\subfloat[Magnitude of the ball's static friction.]{\includegraphics[scale=.5]{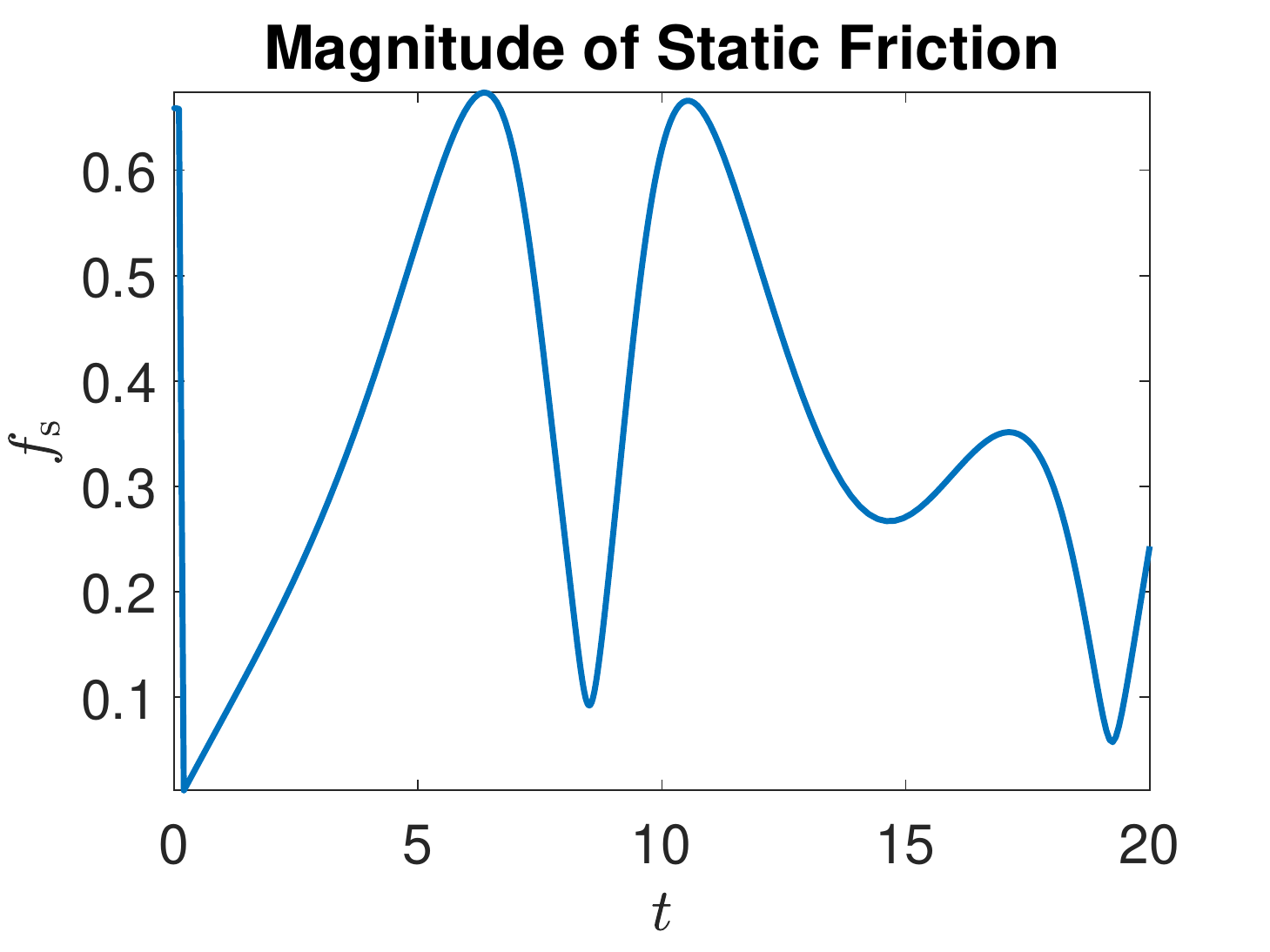}\label{fig_ball_mag_sf}}
	\\
	\subfloat[Ratio of the magnitude of the static friction to the magnitude of the normal force. The minimum coefficient of static friction for the ball is $.19$.]{\includegraphics[scale=.5]{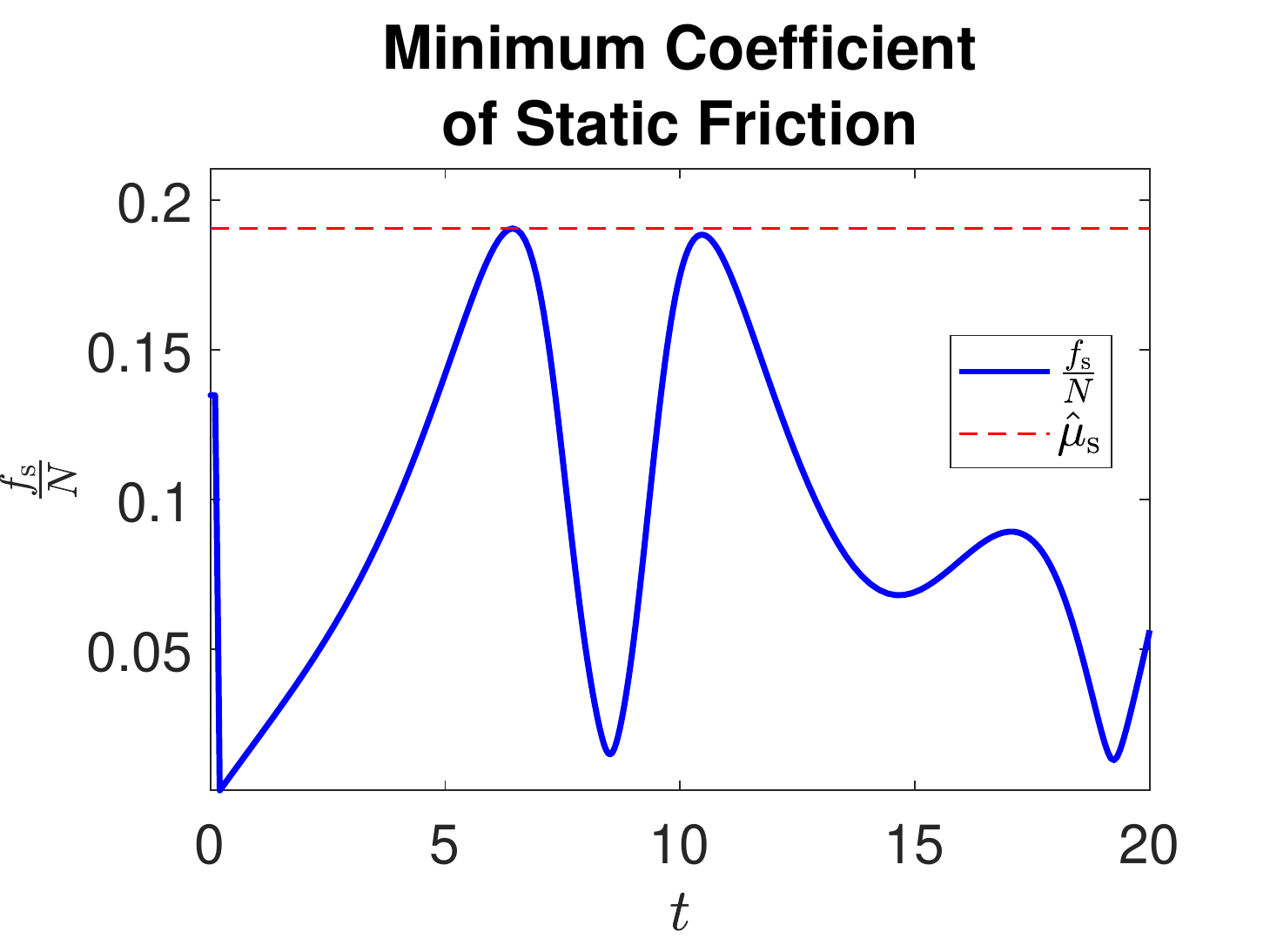}\label{fig_ball_coeff_sf}}
	\hspace{5mm}
	\subfloat[The static friction is plotted atop the trajectory of the ball's GC and CP.]{\includegraphics[scale=.5]{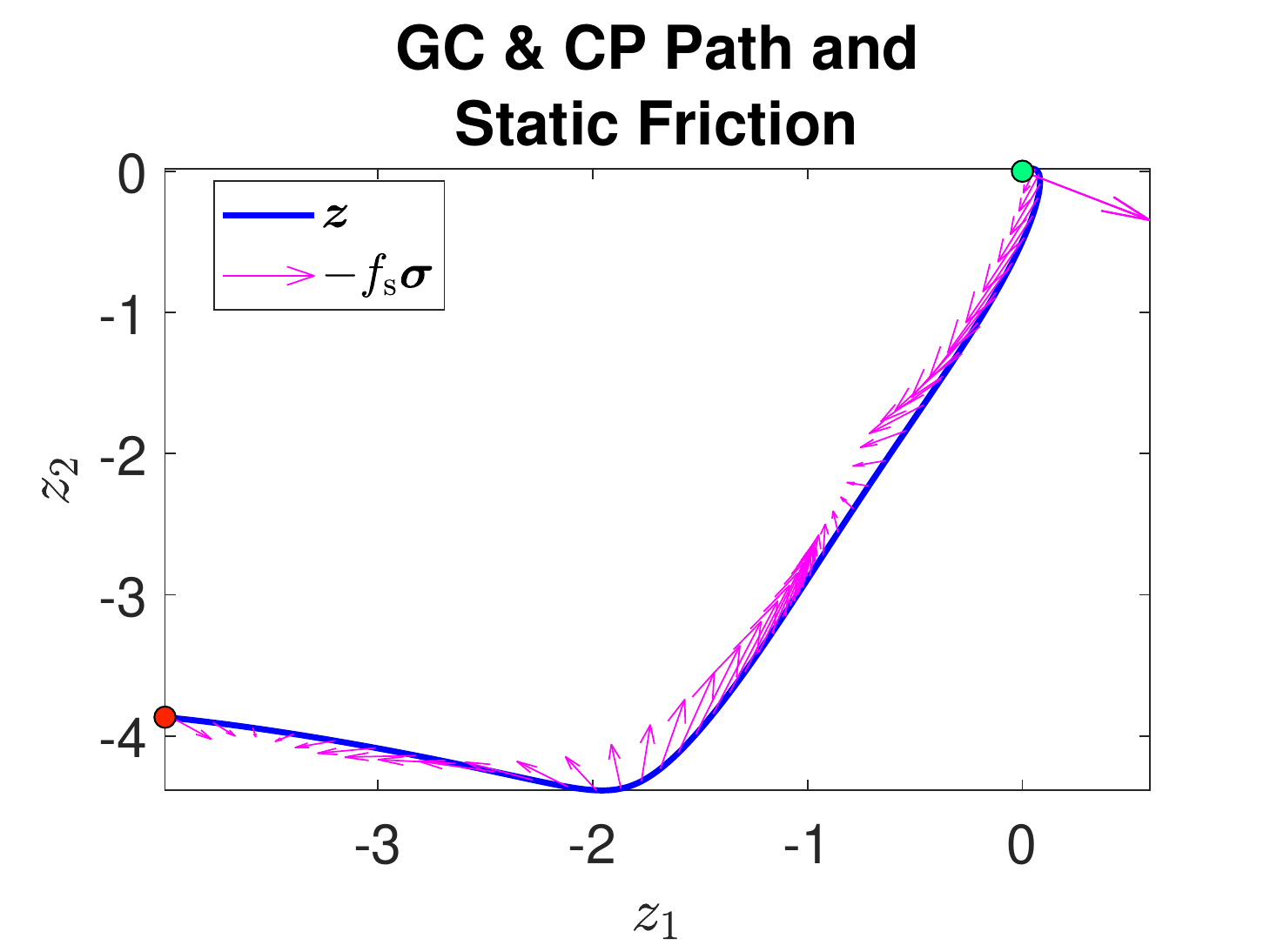}\label{fig_ball_sf_sparse}}
	\caption{Contact point forces acting on the rolling ball shown in Figure~\ref{fig_ball_masses_rail_bf_gc} obtained by numerically integrating the DAE IVP \eqref{rolling_ball_dae_g}, \eqref{eq_ball_initial_conds_spec} with radau5 over the time interval $\left[0,20\right]$. The parameterized accelerations of the internal point masses are given in \eqref{eq_prescribed_u_ball}. Since the magnitude of the ball's normal force is always positive, the ball rolls without slipping if the coefficient of static friction exceeds $.19$.}
	\label{fig_ball_dae_sims_cp_forces}
\end{figure}

The dynamics of this rolling ball are simulated by numerically integrating the ODE IVP \eqref{rolling_ball_ode_f}, \eqref{eq_ball_initial_conds_spec} or the DAE IVP \eqref{rolling_ball_dae_g}, \eqref{eq_ball_initial_conds_spec}. The ODE IVP \eqref{rolling_ball_ode_f}, \eqref{eq_ball_initial_conds_spec} is numerically integrated via the \mcode{MATLAB} R2017b routines \mcode{ode45}, \mcode{ode113}, \mcode{ode15s}, \mcode{ode23t}, and \mcode{ode23tb} and a \mcode{MATLAB} \mcode{MEX} wrapper of the Fortran routine radau5 \cite{hairer1996solving}, while the DAE IVP \eqref{rolling_ball_dae_g}, \eqref{eq_ball_initial_conds_spec} is numerically integrated via the \mcode{MATLAB} R2017b routines \mcode{ode15s} and \mcode{ode23t} and a \mcode{MATLAB} \mcode{MEX} wrapper of the Fortran routine radau5. Except for the absolute and relative error tolerances and the Jacobian, all the numerical integrators are used with the default input options. The absolute and relative error tolerances supplied to the numerical integrators are both set to $1\mathrm{e}{-10}$. Jacobions of $\mathbf{f}$ and $\mathbf{g}$ with respect to the state $\bx$, obtained via complex-step differentiation \cite{squire1998using,martins2001connection,martins2003complex}, are supplied to \mcode{ode15s}, \mcode{ode23t}, \mcode{ode23tb}, and radau5, depending on whether the ODE or DAE IVP is numerically integrated. Since excellent agreement was observed between all the numerical integrators, only the results obtained by numerically integrating the DAE IVP \eqref{rolling_ball_dae_g}, \eqref{eq_ball_initial_conds_spec} with radau5 are shown in Figures~\ref{fig_ball_dae_sims} and \ref{fig_ball_dae_sims_cp_forces}. As was the case for the rolling disk, \mcode{ode113} completed the numerical integration of the rolling ball's equations of motion in the shortest time. Figure~\ref{fig_ball_normal} shows that the magnitude of the ball's normal force is always positive and Figure~\ref{fig_ball_coeff_sf} shows that the minimum coefficient of static friction required for the ball to roll without slipping is $\hat \mu_\mathrm{s}=.19$. The reader is referred to \cite{totten2017asm} for listings of the coefficient of static friction for pairs of materials to see which materials could be used to make this particular ball roll without slipping on the surface. Similarly to the example of the rolling disk,  if the ball's shell were made from aluminum, then it could roll without slipping on an aluminum ($\mu_\mathrm{s}=.42$), steel ($\mu_\mathrm{s}=.35$), titanium ($\mu_\mathrm{s}=.34$), nickel ($\mu_\mathrm{s}=.33$), copper ($\mu_\mathrm{s}=.28$), or chromium ($\mu_\mathrm{s}=.27$) surface, but not on a glass ($\mu_\mathrm{s}=.17$) or graphite ($\mu_\mathrm{s}=.16$) surface.

\begin{figure}[!ht] 
	\centering
	\subfloat[The ball detaches at $t=3.7358$ when the magnitude of the ball's normal force vanishes.]{\includegraphics[scale=.5]{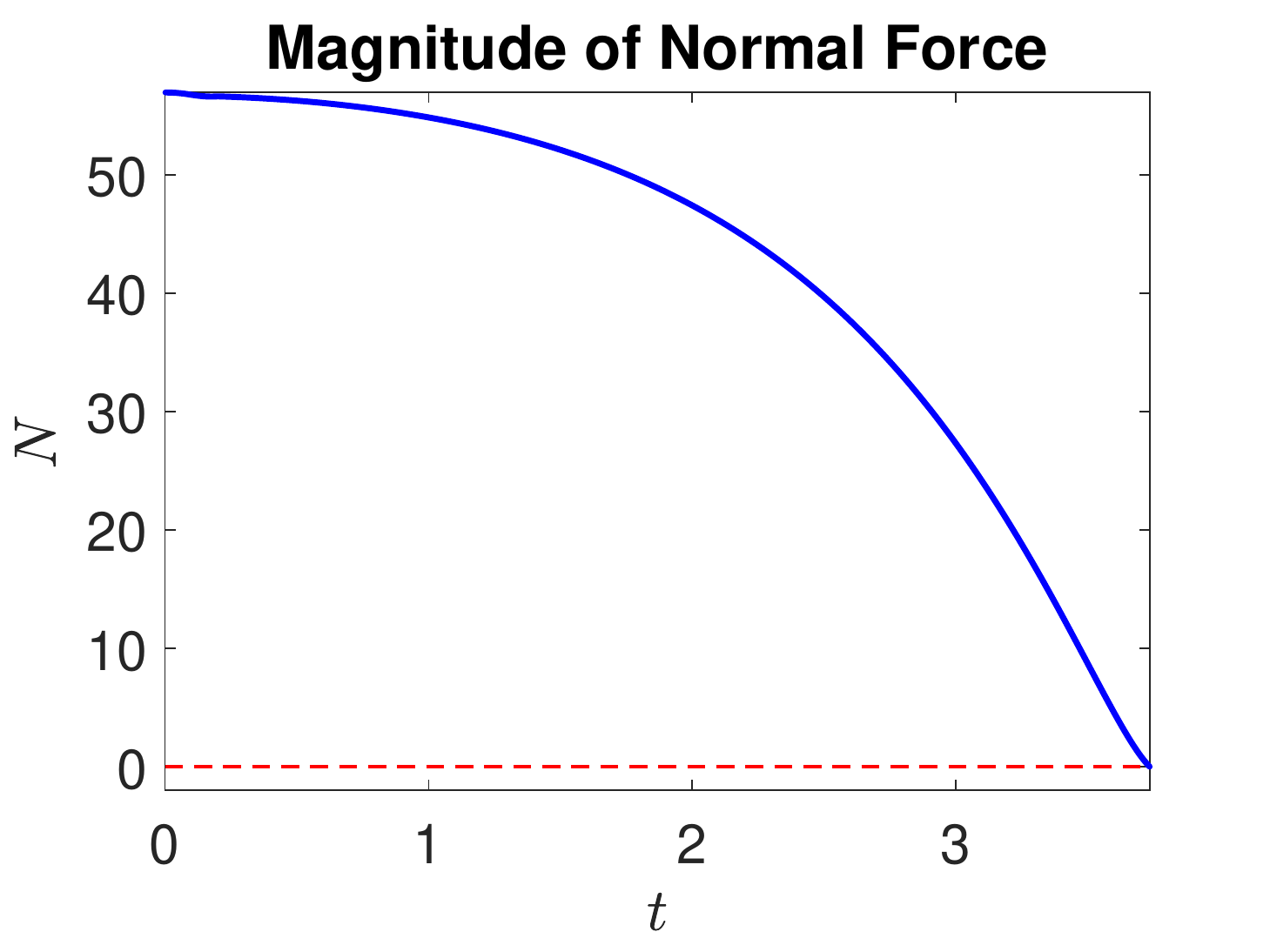}\label{fig_dball_normal}}
	\hspace{5mm}
	\subfloat[Magnitude of the ball's static friction.]{\includegraphics[scale=.5]{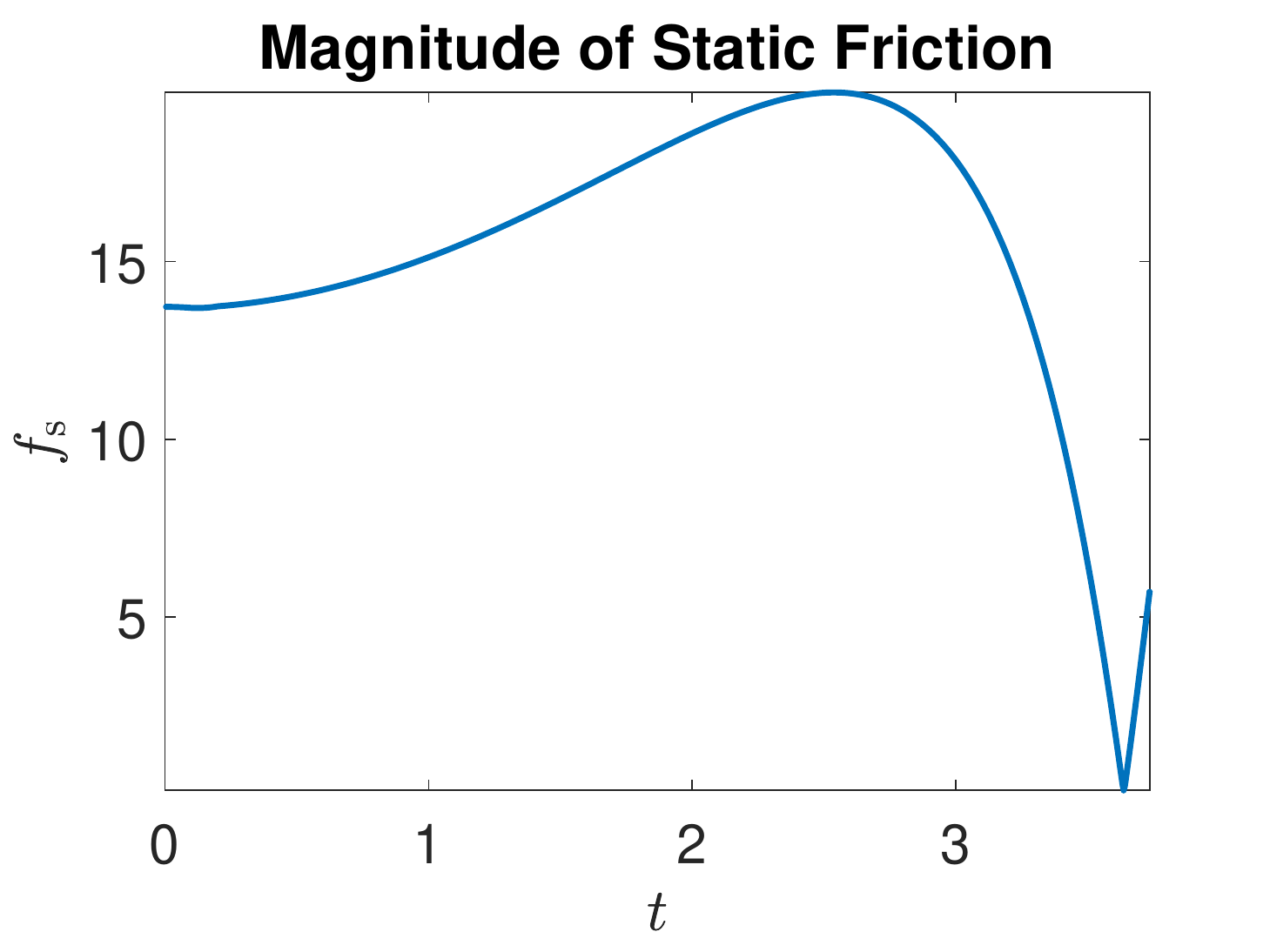}\label{fig_dball_mag_sf}}
	\\
	\subfloat[Evolution of the ball's body angular velocity.]{\includegraphics[scale=.5]{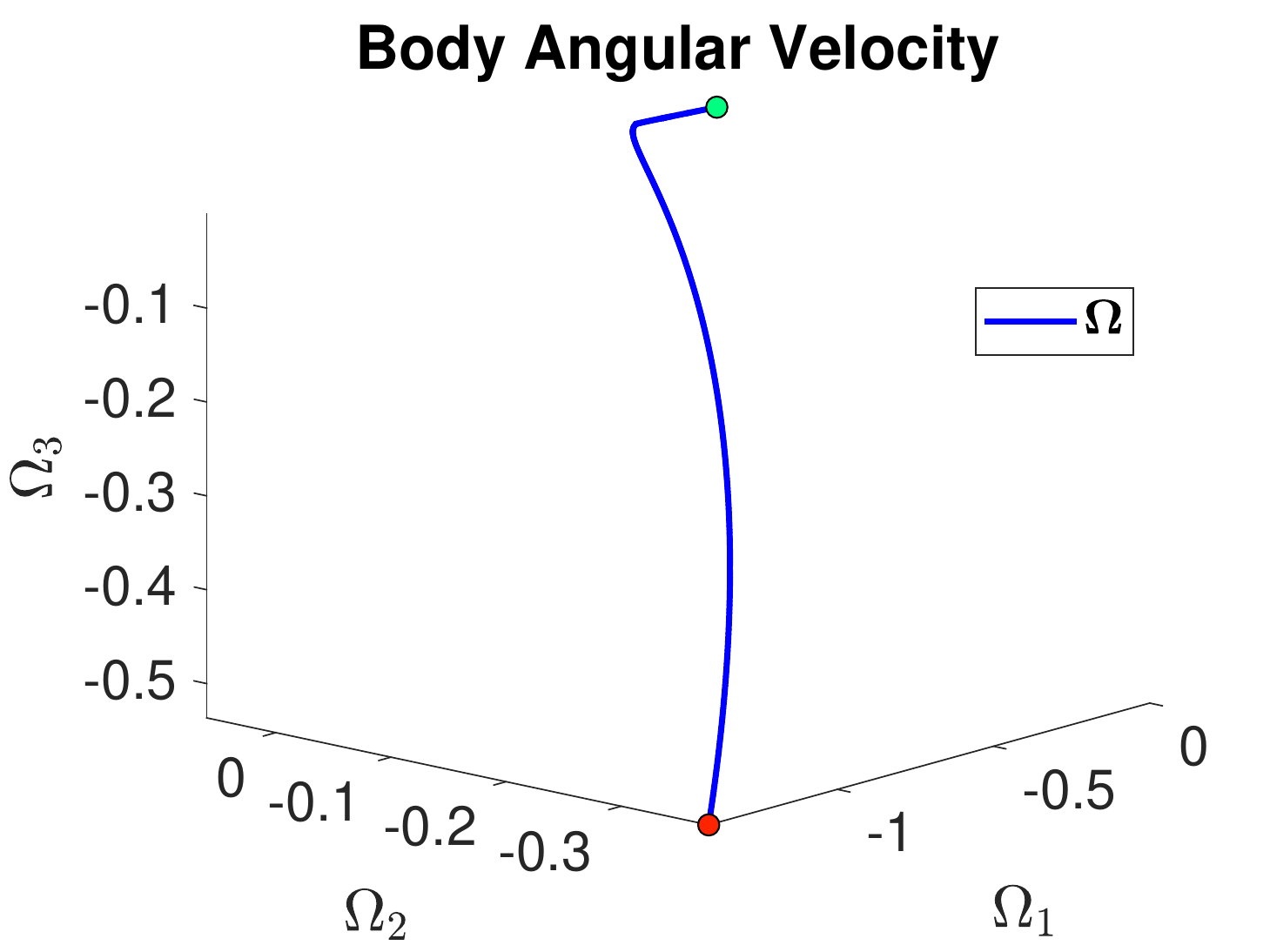}\label{fig_dball_Omega}}
	\hspace{5mm}
	\subfloat[The static friction is plotted atop the trajectory of the ball's GC and CP.]{\includegraphics[scale=.5]{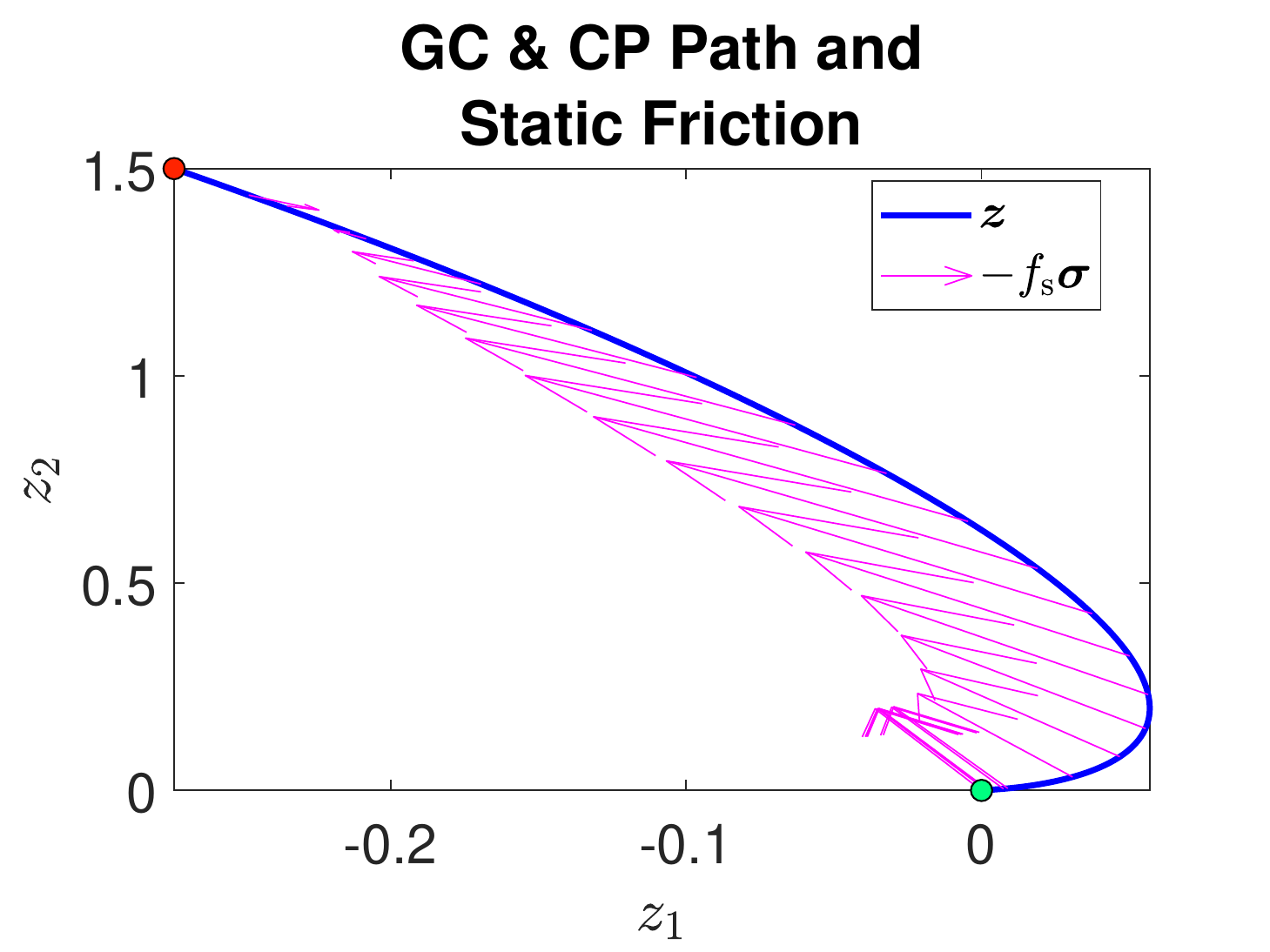}\label{fig_dball_sf_sparse}}
	\caption{Dynamics and contact point forces for the rolling ball shown in Figure~\ref{fig_ball_masses_rail_bf_gc} obtained by numerically integrating the ODE IVP \eqref{rolling_ball_ode_f}, \eqref{eq_ball_initial_conds_spec} with \mcode{ode45} over the time interval $\left[0,3.7358\right]$. The parameterized accelerations of the internal point masses are given in \eqref{eq_prescribed_u_ball}. Detachment occurs at the final time when $N=0$.}
	\label{fig_ball_ode_sims_detach}
\end{figure}

\revision{R1Q9 \\ R2Q2}{\paragraph{Detachment} There are three ways to numerically simulate detachment of the ball from the horizontal surface:
\begin{enumerate}[1)]
	\item Assume perfect friction (i.e. $\mu_\mathrm{s} = \infty$), which is not physically possible.
	\item Assume that $\mu_\mathrm{s}$ is finite and be able to model slipping, which we do not know how to do at this time.
	\item Assume that $\mu_\mathrm{s}$ is finite and construct an example for which $N=f_\mathrm{s}=0$ at the detachment time and for which the no-slip condition $f_\mathrm{s} \le \mu_\mathrm{s} N$ is satisfied prior to the detachment time.
\end{enumerate} 
\revisionS{RR2Q0}{Figure~\ref{fig_ball_ode_sims_detach} illustrates the dynamics and contact point forces of a ball that detaches under the assumption of perfect friction (i.e. $\mu_\mathrm{s} = \infty$), where $N=0$ and $f_\mathrm{s} > 0$ at the detachment time $t=3.7358$. This example is obtained by simulating the same ball as that depicted in Figure~\ref{fig_ball_masses_rail_bf_gc}, with the same initial conditions as shown in Table~\ref{table_ball_ICs}, the same mass excitations \eqref{eq_prescribed_u_ball}, and the same physical parameters as described at the beginning of this section,  except that the masses have been modified so that $m_0=m_1=m_2=.1$ and $m_3=60$. The ODE IVP \eqref{rolling_ball_ode_f}, \eqref{eq_ball_initial_conds_spec} is numerically integrated with \mcode{ode45} using the same settings as before, except that \mcode{MATLAB} ODE event location is used to stop the numerical integration when $N=0$.}  However, this example is unphysical since  $\mu_\mathrm{s}$ must be finite in reality. In reality, such an example of perfect friction detachment would slip just prior to detachment as $N \searrow 0$, since $\mu_\mathrm{s}$ must be finite in reality. We believe that the third option is quite exceptional in practice, and we believe it would be difficult to construct such an example.}

\section{Conclusions} \label{sec_conclusions}
Newton's laws were used to derive the equations of motion, normal force, and static friction for several cases of a ball, actuated by internal point masses, that rolls without slipping on a horizontal surface. This derivation of the equations of motion via Newton's laws validates a previous derivation via Lagrange-d'Alembert's principle in \cite{Putkaradze2018dynamicsP}. The dynamics of a rolling disk and ball actuated by internal point masses were simulated and the formulas for the normal force and static friction were exploited to calculate the minimum coefficient of static friction required to prevent slipping.

One  may  observe that the main results of the paper, equations \eqref{eq_normal_1d} for  the magnitude of the normal force  and \eqref{eq_static_friction_1d} for the static friction, connect  the contact point forces with the dynamic variables  $\dot \bOm$  computed  by  \eqref{eq_Om_1d}.  Thus, equations \eqref{eq_normal_1d} and \eqref{eq_static_friction_1d},    in concert  with the  no-detachment condition \eqref{eq_no_detachment} and  no-slip condition \eqref{Amontons_law}, form an explicit   performance envelope   within  which the rolling ball, actuated by moving internal  point  masses, must operate in order to avoid detachment and slip. Alternatively,  if detachment is desired, for example, to make the ball climb up stairs or hop over an obstacle, or  if  slip is desired, for example, to realize a  change of orientation without  spatial translation of the geometric center, this performance envelope can be intentionally violated by appropriate accelerations of the masses. These questions,  in part  due to the complexity of  the  transition from slip to no-slip  dynamics and vice versa, should be treated carefully in  future work on the subject. We thus hope that the results presented here will be useful for further study of the dynamics and control of rolling  ball robots.  

\section*{Acknowledgements}
At the Seventh International Conference on Geometry, Dynamics, and Integrable Systems (GDIS) 2018, Phanindra Tallapragada observed that the reaction forces exerted on the ball by the accelerating internal point masses  may cause the ball to detach from the surface, prompting the research reported in this paper. Vakhtang Putkaradze's research was partially supported by an NSERC Discovery Grant and the University of Alberta.  Stuart Rogers' postdoctoral research was supported by Target Corporation and the Institute for Mathematics and its Applications at the University of Minnesota. 

\phantomsection
\addcontentsline{toc}{section}{References}
\printbibliography
\hypertarget{References}{}

\appendix
\section{Rolling Disk Calculations} \label{app_disk}
\revisionS{RR2Q3}{This appendix provides calculations that derive the normal force \eqref{eq_normal_disk} and static friction \eqref{eq_static_friction_disk} acting on the rolling disk. The reader is referred to Sections~\ref{sec_3d_traj}, \ref{sec_1d_traj}, and \ref{sec_disk} for explanations of the notation.} For the rolling disk
\begin{equation}
\Lambda = \begin{bmatrix} \cos \phi & 0 & - \sin \phi \\ 0 & 1 & 0 \\ \sin \phi & 0 & \cos \phi \end{bmatrix},
\end{equation}
\begin{equation}
\bGam \equiv \Lambda^{-1} \mathbf{e}_3 = \Lambda^\mathsf{T} \mathbf{e}_3 = \begin{bmatrix} \sin \phi \\ 0 \\ \cos \phi \end{bmatrix}, \quad \bOm \equiv \left( \Lambda^{-1} \dot \Lambda \right)^\vee = \left( \Lambda^\mathsf{T} \dot \Lambda \right)^\vee = \begin{bmatrix} 0 \\ -1 \\ 0 \end{bmatrix} \dot \phi = -\dot \phi \begin{bmatrix} 0 \\ 1 \\ 0 \end{bmatrix} = - \dot \phi  \mathbf{e}_2,
\end{equation}
\begin{equation}
\bzeta_i = \begin{bmatrix} \zeta_{i,1} \\ 0 \\ \zeta_{i,3} \end{bmatrix}, \quad \bzeta_i^{\prime} = \begin{bmatrix} \zeta_{i,1}^{\prime} \\ 0 \\ \zeta_{i,3}^{\prime} \end{bmatrix}, \quad \bzeta_i^{\dprime} = \begin{bmatrix} \zeta_{i,1}^{\dprime} \\ 0 \\ \zeta_{i,3}^{\dprime} \end{bmatrix}, \quad \mathrm{and} \quad \mathbf{s}_i = r \bGam + \bzeta_i = \begin{bmatrix} r \sin \phi + \zeta_{i,1} \\ 0 \\ r \cos \phi+ \zeta_{i,3} \end{bmatrix}.
\end{equation}
Therefore,
\begin{equation} \label{eq_c1}
\dot \bOm \times \mathbf{s}_i = \left(- \ddot \phi \mathbf{e}_2 \right) \times  \begin{bmatrix} r \sin \phi + \zeta_{i,1} \\ 0 \\ r \cos \phi + \zeta_{i,3} \end{bmatrix} = - \ddot \phi \begin{bmatrix} r \cos \phi + \zeta_{i,3} \\ 0 \\ -r \sin \phi - \zeta_{i,1} \end{bmatrix},
\end{equation}
\begin{equation} \label{eq_c2}
\bOm \times \bzeta_i = \left(- \dot \phi \mathbf{e}_2 \right) \times  \begin{bmatrix} \zeta_{i,1} \\ 0 \\ \zeta_{i,3} \end{bmatrix} = - \dot \phi \begin{bmatrix} \zeta_{i,3} \\ 0 \\ - \zeta_{i,1} \end{bmatrix},
\end{equation}
\begin{equation} \label{eq_c3}
\bOm \times \left( \bOm \times \bzeta_i \right) = \left(- \dot \phi \mathbf{e}_2 \right) \times - \dot \phi \begin{bmatrix} \zeta_{i,3} \\ 0 \\ - \zeta_{i,1} \end{bmatrix} = - {\dot \phi}^2 \begin{bmatrix} \zeta_{i,1} \\ 0 \\ \zeta_{i,3} \end{bmatrix},
\end{equation}
\begin{equation} \label{eq_c4}
2 \dot \theta_i \bOm \times \bzeta_{i,1}^{\prime} = 2 \dot \theta_i \left(- \dot \phi \mathbf{e}_2 \right) \times \begin{bmatrix} \zeta_{i,1}^{\prime} \\ 0 \\ \zeta_{i,3}^{\prime} \end{bmatrix} = -2  \dot \phi \dot \theta_i \begin{bmatrix} \zeta_{i,3}^{\prime} \\ 0 \\ -\zeta_{i,1}^{\prime} \end{bmatrix},
\end{equation}
and
\begin{equation} \label{eq_c5}
{\dot \theta}_i^2 \bzeta_i^{\dprime} +  {\ddot \theta}_i \bzeta_i^{\prime}= \begin{bmatrix} {\dot \theta}_i^2 \zeta_{i,1}^{\dprime}+ {\ddot \theta}_i \zeta_{i,1}^{\prime} \\ 0 \\ {\dot \theta}_i^2 \zeta_{i,3}^{\dprime}+ {\ddot \theta}_i \zeta_{i,3}^{\prime} \end{bmatrix}.
\end{equation}
Combining \eqref{eq_c1}, \eqref{eq_c3}, \eqref{eq_c4}, and \eqref{eq_c5} yields
\begin{equation} \label{eq_c6}
\dot \bOm \times \mathbf{s}_i +  \bOm \times \left(\bOm \times \bzeta_i +2 \dot \theta_i \bzeta_i^{\prime} \right) + \dot \theta_i^2 \bzeta_i^{\dprime} + \ddot \theta_i \bzeta_i^{\prime} = \begin{bmatrix} -r \ddot \phi \cos \phi -\ddot \phi \zeta_{i,3} - {\dot \phi}^2 \zeta_{i,1} - 2 \dot \phi {\dot \theta}_i \zeta_{i,3}^{\prime} + {\dot \theta}_i^2 \zeta_{i,1}^{\dprime} + {\ddot \theta}_i \zeta_{i,1}^{\prime} \\ 0 \\ r \ddot \phi \sin \phi + \ddot \phi \zeta_{i,1} - {\dot \phi}^2 \zeta_{i,3}+ 2 \dot \phi {\dot \theta}_i \zeta_{i,1}^{\prime}+ {\dot \theta}_i^2 \zeta_{i,3}^{\dprime} + {\ddot \theta}_i \zeta_{i,3}^{\prime} \end{bmatrix}.
\end{equation}
Substituting \eqref{eq_c6} into \eqref{eq_normal_1d} and \eqref{eq_static_friction_1d} yields \eqref{eq_normal_disk} and \eqref{eq_static_friction_disk}, respectively.

\section{Detachment Dynamics} \label{app_detachment}
\revisionS{RR2Q3}{This appendix derives the dynamics of the ball and disk when they are detached from the surface. The reader is referred to Sections~\ref{sec_3d_traj}, \ref{sec_1d_traj}, and \ref{sec_disk} for explanations of the notation.} Suppose that the ball is detached from the surface, so that  $\mbox{z}_{\mathrm{GC},3} \ge 0$ and $N=f_\mathrm{s}=0$. Setting $N=f_\mathrm{s}=0$ in \eqref{eq_Newton2}, Newton's laws of linear motion about the ball's CM give
\begin{equation} \label{eq_detach_Newton}
\mathbf{0} = -Mg\mathbf{e}_3+\mathbf{F}_\mathrm{e}-\sum_{i=0}^n m_i {\ddot {\mathbf{z}}}_i. 
\end{equation}
For $0 \le i \le n$,
\begin{equation} \label{eq2_z_i}
\mathbf{z}_i = \mathbf{z}_\mathrm{GC}+\Lambda \bchi_i.
\end{equation}
Therefore,
\begin{equation} \label{eq2_dot_z_i}
{\dot {\mathbf{z}}}_i = {\dot {\mathbf{z}}}_\mathrm{GC} + \Lambda \left[\bOm \times \bchi_i + {\dot \bchi}_i  \right]
\end{equation}
and
\begin{equation} \label{eq2_ddot_z_i}
{\ddot {\mathbf{z}}}_i = {\ddot {\mathbf{z}}}_\mathrm{GC} + \Lambda \left[ \dot \bOm \times \bchi_i+ \bOm \times \left(\bOm \times \bchi_i+2{\dot \bchi}_i \right) + {\ddot \bchi}_i  \right].
\end{equation}
Plugging \eqref{eq2_ddot_z_i} into \eqref{eq_detach_Newton} gives
\begin{equation} \label{eq_detach_Newton1}
\mathbf{0} = -Mg\mathbf{e}_3+\mathbf{F}_\mathrm{e}-\sum_{i=0}^n m_i \left\{ {\ddot {\mathbf{z}}}_\mathrm{GC} + \Lambda \left[ \dot \bOm \times \bchi_i+ \bOm \times \left(\bOm \times \bchi_i+2{\dot \bchi}_i \right) + {\ddot \bchi}_i  \right] \right\}. 
\end{equation}
Solving \eqref{eq_detach_Newton1} for $\ddot {\mathbf{z}}_\mathrm{GC}$ gives
\begin{equation} \label{eq_ddot_z_GC}
\ddot {\mathbf{z}}_\mathrm{GC} = \frac{1}{M}\mathbf{F}_\mathrm{e}-g\mathbf{e}_3-\frac{1}{M}\Lambda \sum_{i=0}^n m_i \left[ \dot \bOm \times \bchi_i+ \bOm \times \left(\bOm \times \bchi_i+2{\dot \bchi}_i \right) + {\ddot \bchi}_i  \right].
\end{equation}
\eqref{eq_detach_Newton} may be rewritten as
\begin{equation} \label{eq_detach_Newton2}
\mathbf{0} = \mathbf{F}_\mathrm{e}-\sum_{i=0}^n m_i \left( {\ddot {\mathbf{z}}}_i +g\mathbf{e}_3 \right). 
\end{equation}
Multiplying both sides of \eqref{eq_detach_Newton2} by $\Lambda^{-1}$ gives
\begin{equation} \label{eq_detach_Newton3}
\mathbf{0} = \tilde{\bGam}-\sum_{i=0}^n m_i \left( \Lambda^{-1} {\ddot {\mathbf{z}}}_i+g \bGam \right).
\end{equation}
Crossing both sides of \eqref{eq_detach_Newton3} by $\bchi_0$ and solving for $\tilde{\bGam} \times \bchi_0$ gives
\begin{equation} \label{eq_detach_Newton4}
\tilde{\bGam} \times \bchi_0 = \sum_{i=0}^n m_i \left( \Lambda^{-1} {\ddot {\mathbf{z}}}_i+g \bGam \right) \times \bchi_0.
\end{equation}
Setting $N=f_\mathrm{s}=0$ in \eqref{eq_rot_Newton3}, Newton's laws of angular motion about the ball's CM give
\begin{equation} \label{eq_detach_rot_Newton}
\bOm \times \inertia \bOm + \inertia \dot \bOm  = - \bchi_0 \times \tilde \bGamma -\sum_{i=1}^n \left(\bchi_i - \bchi_0 \right) \times m_i \left(\Lambda^{-1} {\ddot {\mathbf{z}}}_i +g \bGamma \right).   
\end{equation}
Plugging \eqref{eq_detach_Newton4} into \eqref{eq_detach_rot_Newton} gives
\begin{equation} \label{eq_detach_rot_Newton2}
\bOm \times \inertia \bOm + \inertia \dot \bOm  =  -\sum_{i=0}^n \bchi_i \times m_i \left(\Lambda^{-1} {\ddot {\mathbf{z}}}_i +g \bGamma \right).   
\end{equation}
Multiplying both sides of \eqref{eq2_ddot_z_i} by $\Lambda^{-1}$ and using \eqref{eq_ddot_z_GC} gives
\begin{equation} \label{eq_Lam_ddot_z_i}
\Lambda^{-1} {\ddot {\mathbf{z}}}_i = \frac{1}{M}\tilde \bGam-g\bGam -\sum_{j=0}^n \left(\frac{m_j}{M}-\delta_{ij}\right) \left[ \dot \bOm \times \bchi_j+ \bOm \times \left(\bOm \times \bchi_j+2{\dot \bchi}_j \right) + {\ddot \bchi}_j  \right].
\end{equation}
Plugging \eqref{eq_Lam_ddot_z_i} into \eqref{eq_detach_rot_Newton2} yields
\begin{equation} \label{eq_detach_rot_Newton3}
\bOm \times \inertia \bOm + \inertia \dot \bOm  =  -\sum_{i=0}^n \bchi_i \times m_i \left\{ \frac{1}{M}\tilde \bGam -\sum_{j=0}^n \left(\frac{m_j}{M}-\delta_{ij}\right) \left[ \dot \bOm \times \bchi_j+ \bOm \times \left(\bOm \times \bchi_j+2{\dot \bchi}_j \right) + {\ddot \bchi}_j  \right] \right\},   
\end{equation}
which simplifies to
\begin{multline} \label{eq_detach_rot_Newton4}
\left[ \inertia + \sum_{i=0}^n m_i \widehat{\bchi_i} \left\{\sum_{j=0}^n \left(\frac{m_j}{M}-\delta_{ij} \right) \widehat{\bchi_j} \right\}  \right] \dot \bOm  \\ = \inertia \bOm \times  \bOm  -\sum_{i=0}^n m_i \bchi_i \times \left\{ \frac{1}{M}\tilde \bGam -\sum_{j=0}^n \left(\frac{m_j}{M}-\delta_{ij}\right) \left[ \bOm \times \left(\bOm \times \bchi_j+2{\dot \bchi}_j \right) + {\ddot \bchi}_j  \right] \right\}.   
\end{multline}
Solving \eqref{eq_detach_rot_Newton4} for $\dot \bOm$ yields
\begin{multline} \label{eq_detach_dOm}
\dot \bOm = 
\left[ \inertia + \sum_{i=0}^n m_i \widehat{\bchi_i} \left\{\sum_{j=0}^n \left(\frac{m_j}{M}-\delta_{ij} \right) \widehat{\bchi_j} \right\}  \right]^{-1} \\ \left[ \inertia \bOm \times  \bOm  -\sum_{i=0}^n m_i \bchi_i \times \left\{ \frac{1}{M}\tilde \bGam -\sum_{j=0}^n \left(\frac{m_j}{M}-\delta_{ij}\right) \left[ \bOm \times \left(\bOm \times \bchi_j+2{\dot \bchi}_j \right) + {\ddot \bchi}_j  \right] \right\} \right].   
\end{multline}
The detachment dynamics for the ball are given by \eqref{eq_detach_dOm} and \eqref{eq_ddot_z_GC}.

\paragraph{Ball with Static Internal Structure}
By setting the number of point masses $n$ to 0, the detachment dynamics for the ball with static internal structure are readily obtained. Letting $n=0$, \eqref{eq_detach_dOm} and \eqref{eq_ddot_z_GC} simplify to
\begin{equation} \label{eq_detach_ball_static}
\begin{split}
\dot \bOm &= 
\inertia^{-1} \left[ \inertia \bOm \times  \bOm  + \tilde \bGam \times \bchi_0  \right] \\
\ddot {\mathbf{z}}_\mathrm{GC} &= \frac{1}{m_0}\mathbf{F}_\mathrm{e}-g\mathbf{e}_3-\Lambda \left[ \dot \bOm \times \bchi_0+ \bOm \times \left(\bOm \times \bchi_0 \right)  \right].
\end{split} 
\end{equation}

\paragraph{Ball with 1-d Parameterizations of the Point Mass Trajectories}
Plugging the formulas for $\bchi_i$, $\dot \bchi_i$, and $\ddot \bchi_i$ given in \eqref{eq_1d_param} into \eqref{eq_detach_dOm} and \eqref{eq_ddot_z_GC} yields the detachment dynamics for a ball with 1-d parameterizations of the point mass trajectories:
\begin{equation} \label{eq_detach_ball_1d}
\begin{split}
\dot \bOm &= 
\left[ \inertia + \sum_{i=0}^n m_i \widehat{\bzeta_i} \left\{\sum_{j=0}^n \left(\frac{m_j}{M}-\delta_{ij} \right) \widehat{\bzeta_j} \right\}  \right]^{-1} \\ &\hspace{10mm} \left[ \inertia \bOm \times  \bOm  -\sum_{i=0}^n m_i \bzeta_i \times \left\{ \frac{1}{M}\tilde \bGam -\sum_{j=0}^n \left(\frac{m_j}{M}-\delta_{ij}\right) \left[ \bOm \times \left(\bOm \times \bzeta_j +2 \dot \theta_j \bzeta_j^{\prime} \right) + \dot \theta_j^2 \bzeta_j^{\dprime} + \ddot \theta_j  \bzeta_j^{\prime}   \right] \right\} \right] \\
\ddot {\mathbf{z}}_\mathrm{GC} &= \frac{1}{M}\mathbf{F}_\mathrm{e}-g\mathbf{e}_3-\frac{1}{M}\Lambda \sum_{i=0}^n m_i \left[ \dot \bOm \times \bzeta_i+ \bOm \times \left(\bOm \times \bzeta_i +2 \dot \theta_i \bzeta_i^{\prime} \right) + \dot \theta_i^2 \bzeta_i^{\dprime} + \ddot \theta_i  \bzeta_i^{\prime}  \right].
\end{split} 
\end{equation}

\paragraph{Disk with 1-d Parameterizations of the Point Mass Trajectories} By using the results in Section 3.4 and Appendix C of \cite{Putkaradze2018dynamicsP} and in Appendix~\ref{app_disk}, \eqref{eq_detach_ball_1d} simplifies to give the detachment dynamics for a disk with 1-d parameterizations of the point mass trajectories:
\begin{equation} \label{eq_detach_disk_1d}
\begin{split} 
\ddot \phi &= \frac{ \sum_{i=0}^n m_i \left[F_{\mathrm{e},1} \left(\zeta_{i,1} \sin \phi + \zeta_{i,3} \cos \phi\right)-F_{\mathrm{e},3}\left(\zeta_{i,1} \cos \phi - \zeta_{i,3} \sin \phi \right)+\sum_{j=0}^n \left(m_j-M \delta_{ij}\right) V_{ij} \right] } {M d_2 - \sum_{i=0}^n m_i \left(m_i-M \right)\left(\zeta_{i,1}^2+\zeta_{i,3}^2\right) - 2 \sum_{i=0}^n \sum_{j=i+1}^n m_i m_j \left(\zeta_{i,1} \zeta_{j,1} +\zeta_{i,3} \zeta_{j,3}\right)} \\
\ddot {\mathbf{z}}_\mathrm{GC} &= \frac{1}{M} \begin{bmatrix} F_{\mathrm{e},1} - \sum_{i=0}^n m_i \left[\cos \phi \left(-\ddot \phi \zeta_{i,3}+Q_i  \right) - \sin \phi \left(\ddot \phi \zeta_{i,1}+P_i  \right) \right] \\ 0 \\ F_{\mathrm{e},3} -Mg - \sum_{i=0}^n m_i \left[\sin \phi \left(-\ddot \phi \zeta_{i,3}+Q_i  \right) + \cos \phi \left(\ddot \phi \zeta_{i,1}+P_i  \right) \right] \end{bmatrix},
\end{split} 
\end{equation}
where
\begin{equation}
\begin{split}
P_i &\equiv - {\dot \phi}^2 \zeta_{i,3}+ 2 \dot \phi {\dot \theta}_i \zeta_{i,1}^{\prime}+ {\dot \theta}_i^2 \zeta_{i,3}^{\dprime} + {\ddot \theta}_i \zeta_{i,3}^{\prime}  \\
Q_i &\equiv - {\dot \phi}^2 \zeta_{i,1} - 2 \dot \phi {\dot \theta}_i \zeta_{i,3}^{\prime} + {\dot \theta}_i^2 \zeta_{i,1}^{\dprime} + {\ddot \theta}_i \zeta_{i,1}^{\prime} \\
V_{ij} &\equiv \zeta_{i,1} P_j -\zeta_{i,3} Q_j.
\end{split}
\end{equation}

\paragraph{Disk with Static Internal Structure} By setting the number of point masses $n$ to 0, the detachment dynamics for the disk with static internal structure are readily obtained. Letting $n=0$, \eqref{eq_detach_disk_1d} simplifies to
\begin{equation} \label{eq_detach_disk_static}
\begin{split} 
\ddot \phi &= \frac{1}{d_2}\left[ F_{\mathrm{e},1} \left(\zeta_{0,1} \sin \phi + \zeta_{0,3} \cos \phi\right)-F_{\mathrm{e},3}\left(\zeta_{0,1} \cos \phi - \zeta_{0,3} \sin \phi \right) \right] \\
\ddot {\mathbf{z}}_\mathrm{GC} &=  \begin{bmatrix} \frac{F_{\mathrm{e},1}}{M} + \cos \phi \left(\ddot \phi \zeta_{0,3} + {\dot \phi}^2 \zeta_{0,1}  \right) + \sin \phi \left(\ddot \phi \zeta_{0,1}- {\dot \phi}^2 \zeta_{0,3}  \right)  \\ 0 \\ \frac{F_{\mathrm{e},3}}{M} -g + \sin \phi \left(\ddot \phi \zeta_{0,3} + {\dot \phi}^2 \zeta_{0,1}  \right) - \cos \phi \left(\ddot \phi \zeta_{0,1}- {\dot \phi}^2 \zeta_{0,3}  \right)  \end{bmatrix}.
\end{split} 
\end{equation}

\end{document}